\documentclass[a4paper,fleqn,usenatbib]{mnras}

\usepackage{graphicx}
\usepackage{mathptmx}
\usepackage{savesym}
\usepackage{amsmath}
\usepackage{txfonts}

\usepackage[T1]{fontenc}
\usepackage{ae,aecompl}

\usepackage{amssymb}	
\usepackage{gensymb}
\usepackage{longtable,lscape}

\title[Open clusters in AurOB2]{Open clusters in Auriga OB2\thanks{Partially based on observations collected at the Jacobus Kapteyn Telescope, the Telescopio Nazionale Galileo and the Nordic Optical Telescope (La Palma) and at the 1.50-m telescope at Observatoire de Haute Provence (CNRS), France.}}

\author[A. Marco and I. Negueruela]{
Amparo Marco$^{1}$$^{,}$$^{2}$\thanks{E-mail: amparo.marco@ua.es (AM)} and Ignacio Negueruela$^{1}$
\\
$^{1}$DFISTS, EPS, Universidad de Alicante, Carretera San Vicente del Raspeig s/n,  E03690, San Vicente del Raspeig, Spain\\
$^{2}$Department of Astronomy, University of Florida, 211 Bryant Space Science Center, Gainesville, FL 32611, USA\\
}

\date{Accepted 2016 March 14. Received 2016 March 11; in original form 2016 January 7}

\pubyear{2016}

\begin{document}
\label{firstpage}
\pagerange{\pageref{firstpage}--\pageref{lastpage}}
\maketitle

\begin{abstract}
We study the area around the \ion{H}{ii} region Sh~2-234, including the young open cluster Stock~8, to investigate the extent and definition of the association Aur OB2 and the possible role of triggering in massive cluster formation. We obtained Str\"{o}mgren and $J,H,K_{S}$ photometry for Stock~8 and Str\"{o}mgren photometry for two other cluster candidates in the area, which we confirm as young open clusters and name Alicante~11 and Alicante~12. We took spectroscopy of $\sim 33$ early-type stars in the area, including the brightest cluster members. We calculate a common distance of $2.80^{+0.27}_{-0.24}$~kpc for the three open clusters and surrounding association. We derive an age 4\,--\,6~Ma for Stock~8, and do not find a significantly different age for the other clusters or the association. The star LS\,V~$+34\degr$23, with spectral type O8\,II(f), is likely the main source of ionization of Sh~2-234. We observe an important population of pre-main sequence stars, some of them with disks, associated with the B-type members lying on the main-sequence. We interpret the region as an area of recent star formation with some residual and very localized ongoing star formation. We do not find evidence for sequential star formation on a large scale. The classical definition of Aur~OB2 has to be reconsidered, because its two main open clusters, Stock~8 and NGC~1893, are not at the same distance. Stock~8 is probably located in the Perseus arm, but other nearby \ion{H}{ii} regions whose distances also place them in this arm show quite different distances and radial velocities and, therefore, are not connected.
\end{abstract}




\section{Introduction}

Stellar associations are large and loose, comoving stellar groups of low-density stars often associated with smaller open cluster groups. OB associations contain O and/or early B-type stars. They must be young ($\leq30$~Ma) because they are unstable against Galactic tidal forces. Therefore, most of their low-mass members are still found in the pre-main sequence (PMS) phase. They are excellent targets for detailed studies of the initial mass function and the star formation history because they represent a place where the star formation process has been completed recently. Their space distribution also provides useful information about the spiral structure of the Galaxy \citep{blaauw1964,preibisch2007}.

OB associations are the natural outcome of star-formation processes in giant molecular clouds (GMCs) in most Milky Way environments. These large clouds demonstrate hierarchical structure in both time and space, resulting in the formation of star clusters that are not bound gravitationally to one another, but share a common distance. A well-studied example of such processes is the W3 region, which contains several small embedded clusters surrounding the optically visible cluster IC~1795 \citep{bik12,roman15,kiminki15}. A more dispersed population of early-type stars seems to connect it with IC~1805, the ionizing cluster of W4, which includes a few early O-type stars.  The nearby cluster IC~1848, ionizing W5, is also likely connected. Overall stars traditionally considered to belong to Cas~OB6 cover more than 6 degrees on the sky. Other examples of large star-forming regions containing moderately massive clusters include the G305 region, centred on the Danks~1 and~2 clusters \citep[ and references therein]{hindson13} and the Carina Nebula \citep[e.g.][]{preibisch11}
. Other GMCs, such as W33 \citep{messineo15} or W51, lack the central massive clusters, and will very likely evolve into dispersed associations, similar to Cyg~OB2 \citep{wright14}. 

The  Perseus  arm  has  been  proposed  as  one  of  the  two main spiral  arms  of  the  Galaxy
\citep{churchwell2009}.
It emerges from behind obscuring clouds in the Local arm around $l=75\degr$ and over a significant fraction of the northern sky, it is clearly delineated by early-type luminous stars \citep[as first relised by][]{morgan53}. Between $l=100\degr$ and $140\degr$, the Perseus arm contains both large active star-forming regions, 
and more evolved associations, with ages between $\sim10$ and $\sim20$~Ma. Among them, some are dispersed and contain small clusters (Cas OB2/OB4/OB5/OB7), while other have massive central clusters, such as the twin clusters NGC~869/884 at the core of Per~OB1 or NGC~663 and NGC~654, defining Cas~OB8.
However, beyond $l=140\degr$ stellar and molecular tracers of the Perseus arm suddenly become very rare. Cloud complexes identified with the Perseus arm are seen again in the third galactic quadrant (already at $l\approx215\degr$), but its stellar tracers are still scarce \citep{vazquez2008}. Partly because of this absence of tracers, we still lack a complete picture of the extent of the Perseus Arm towards the Third Quadrant, and the connection between the Local Feature and the major spiral design (\citealt{vazquez2008}; see a recent discussion in \citealt{foster15}). Recently, \citet{choi2014} have used trigonometric parallaxes for water masers in massive star-forming regions to trace the Perseus Arm in the second and third quadrants.
They find a number of tracers of the Perseus arm around $l=190\degr$, with distances clustered around 2~kpc, but again no tracers are found between $l=140\degr$ and $l=180\degr$.
Over this longitude range, there are a number of young open clusters with distances around $\sim4$~kpc at $l=150\degr$ and $\sim5$~kpc at $l\sim180\degr$ \citep[and references therein]{negueruela2003}. They seem to delineate a poorly-populated spiral arm, but their distance is closer than expected for the position of the continuation of the Cygnus (+II) arm in models such as those of  \citet{vallee2015}

To help casting light on these issues, the aim of this work is to clarify the extent and definition of what has traditionally known as association Aur OB2. Aur~OB2 was identified as a distant and compact association lying behind Aur~OB1, with boundaries $l=172\degr$-$174\degr$ and $b=-1\fdg8$ to $b=+2\fdg0$ \citep{humphreys1978}. The star-forming open cluster NGC~1893, containing five O-type stars, was considered the core of this association, but most modern studies based on accurate photometry \citep{marco2001, fitzsimmons1993} and spectroscopy \citep{negueruela2007} agree on a distance around 5~kpc for NGC~1893, incompatible with the distance determined to other presumed members \citep{humphreys1978}. Visually, IC~410, the \ion{H}{ii} region illuminated by NGC~1893, seems to form part of a single complex of illuminated clouds together with IC~417 and IC~405. This is just an illusion, as IC~405 is a completely unrelated nebula, associated to a foreground ($d\approx450$~pc) cloud illuminated by the runaway star AE~Aur, believed to have been ejected from the Orion Cloud \citep[and references therein]{france04}. IC~417, also known as Sh2-234, is the \ion{H}{ii} region illuminated  by the open cluster Stock~8. Its Galactic coordinates are $l=173\fdg37$ and $b=-0\fdg18$. In this paper, we present a photometric and spectroscopic study of Stock~8, the diffuse population of OB stars generally assigned to Aur~OB2, and two other young open clusters that we identify in its vicinity.

There have been a few previous studies of Stock~8 and surroundings. However, the extent of the cluster itself has not been clearly defined. The name Stock~8 has been traditionally given to a strong stellar concentration that seems to be embedded in the nebula associated with Sh~2-234. \citet{mm71} carried out an spectroscopic and $UBV$ photoelectric photometry study of 11 bright stars in its surroundings, finding a common distance for Stock~8 and some of the OB stars in its vicinity (a few arcminutes to the west; see Fig.~\ref{todo}), namely, a distance modulus ($DM$) of $12.36$ magnitudes ($2\,965$\,pc). 
\citet{malysheva1990} used $UBV$ photographic photometry of $\sim66$ stars brighter than $V=16$ inside an angular diameter of $20\arcmin$ from Stock~8, and obtained a $DM$ of 11.39 magnitudes ($\sim 1897$\,pc) and an age of $\sim12$ Ma. \citet{jose2008} took $UBVI_{\textrm{c}}$ CCD photometry of the compact cluster and found a cluster radius of $\sim6\arcmin$, a variable reddening within the cluster region, going from $E(B-V)=0.40$ to $0.60$~mag, and a distance of $2.05\pm0.10$ kpc. They detected a significant number of young stellar objects (YSOs) inside the cluster and in a Nebulous Stream towards the east side of the cluster (see Fig.~\ref{todo}). The YSOs lying in the Nebulous Stream were found to be younger than the stars in Stock~8. They conclude that the morphology of the region seems to indicate that the ionization/shock front caused by the ionizing sources located inside and west of Stock~8 has not reached the Nebulous Stream and the star formation activity in both regions may be independent. 

In this paper, we study a much larger area of $\sim40\arcmin \times 40\arcmin$, where we find Stock~8 and two candidates to young clusters, as well as a large number of early-type stars spread over the whole area. We aim to determine accurately the distance to all these objects and analyze how the stellar formation process has developed. The paper is organized as follows. In Section~\ref{data} we present the photometric and spectroscopic data used to make the subsequent analysis. This analysis, including the determination of spectral types for individual stars, and the reddening, distance and age for the three clusters studied, is developed in Section~\ref{results}. In Section~\ref{discusion} we comment the impact of our results on the history of star formation and the position in the Milky Way of Stock~8 and surroundings. Finally, we enumerate our conclusions.

\section{Observations and data}
\label{data}

\subsection{Optical photometry}
\label{opt_phot}

We obtained Str\"{o}mgren photometry of three fields in the area of study with the 1-m Jacobus Kapteyn Telescope (JKT) at the Roque de los Muchachos Observatory (La Palma, Spain) during a run in 2003 February, 9\,--\,16. The telescope was equipped with the $2048 \times 2048$ SITe1 chip CCD and the four Str\"{o}mgren $uvby$ and the narrow and wide $H\beta$ filters. The camera covers a field of $10\arcmin \times 10\arcmin$ and has a pixel scale of $0\farcs33$ $pixel^{-1}$.

The first field was the area traditionally assigned to Stock~8, which was observed on the nights of 2003 February, 9\,--\,10.  For each frame, we obtained 7 series of different exposure times in each filter to achieve accurate photometry for a broad magnitude range. The central position for the cluster and the exposure times used are presented in Table~\ref{tab1a} and Table~\ref{tab2a}, respectively 

The other two fields observed were centred on two stellar aggregates that we considered open cluster candidates. Subsequently to our observations, these open clusters have appeared in the literature as cluster candidates with the names of FSR~777 \citep{froebrich2007} and Kronberger~1  \citep{kronberger2006}. These two objects are confirmed as young open clusters here and named Alicante~11 and Alicante~12, respectively. The frames, taken on the nights of 2003 February, 15\,--\,16, were centred on the coordinates displayed in Table~\ref{tab1a}. We obtained 7 series of different exposure times for Alicante~11 and 3 series for Alicante~12. The exposure times used are presented in Table~\ref{tab2a}. Since Stock~8 is partially embedded in dark nebulosity, exposure times for this field were longer than those in the other two fields. Our intention was to reach an absolute magnitude that would allow us to trace the whole sequence of B types. As will be seen later, this was effectively achieved.

\begin{table}
\caption{Clusters observed from the JKT in February 2003. The top panel contains the target fields, while the bottom panel includes the clusters that contain the photometric standards. The coordinates are those provided by the WEBDA database: {\tt http://www.univie.ac.at/webda/}\citep{netopil2012}, except for the two newly defined clusters.}
\label{tab1a}
\centering
\scalebox{0.9}{
\begin{tabular}{lll}
\hline\hline
\noalign{\smallskip}
Name&RA(J2000)&Dec(J2000)\\
\noalign{\smallskip}
\hline
\noalign{\smallskip}
Stock~8 & 05h 28m 08.0s & $+34\degr\,25\arcmin\,42\farcs0$ \\
Alicante~11& 05h 27m 36.2s & $+34\degr\,45\arcmin\,19\farcs0$ \\
Alicante~12& 05h 28m 20.3s & $+34\degr\,47\arcmin\, 13\farcs5$ \\
\hline
\hline
NGC~2244&06h 31m 55.0s &  $+04\degr\,56\arcmin\, 30\farcs0$ \\
NGC~2169&06h 08m 24.0s &  $+13\degr\,57\arcmin\, 54\farcs0$ \\
NGC~1502& 04h 07m 40.4s &  $+62\degr\,20\arcmin\, 59\farcs1$ \\
NGC~1039& 20h 23m 10.6s &  $+40\degr\,46\arcmin\, 22\farcs4$ \\
NGC~~869&02h 19m 04.4s &  $+57\degr\,08\arcmin\, 07\farcs8$ \\
NGC~~884&02h 22m 00.6s &  $+57\degr\,08\arcmin\, 42\farcs1$ \\
\noalign{\smallskip}
\hline\hline
\smallskip\\
\end{tabular}}
\end{table}

\begin{table}
\caption{Log of the optical photometric observations taken at the JKT in February 2003.}
\label{tab2a}
\centering
\scalebox{0.85}{
\begin{tabular}{c c c}
\hline\hline
\multicolumn{3}{c}{Exposure times(s)}\\
\hline
Filter & Long times & Short times \\
\noalign{\smallskip}
\hline
\noalign{\smallskip}
\multicolumn{3}{c}{Stock~8}\\
\hline
\noalign{\smallskip}
$u$ & 1800 & 400  \\
$v$ & 500 &175 \\
$b$ & 200 &100  \\
$y$ & 150 & 50 \\
$H\beta_{\textrm{n}}$&1800&400\\
$H\beta_{\textrm{w}}$&200&100\\
\noalign{\smallskip}
\hline\hline
\noalign{\smallskip}
\multicolumn{3}{c}{Alicante~11}\\
\noalign{\smallskip}
\hline
\noalign{\smallskip}
$u$ & 450 & 100  \\
$v$ & 180 &40 \\
$b$ & 100 &20  \\
$y$ & 50 & 8 \\
$H\beta_{\textrm{n}}$&400&100\\
$H\beta_{\textrm{w}}$&60&15\\
\noalign{\smallskip}
\hline\hline
\noalign{\smallskip}
\multicolumn{3}{c}{Alicante~12}\\
\hline
\noalign{\smallskip}
$u$ & 450 & $-$  \\
$v$ & 180 & $-$ \\
$b$ & 100 &$-$  \\
$y$ & 50 &$-$  \\
$H\beta_{\textrm{n}}$&400&$-$\\
$H\beta_{\textrm{w}}$&60&$-$\\
\noalign{\smallskip}
\hline\hline
\noalign{\smallskip}
\multicolumn{3}{c}{NGC~2244}\\
\noalign{\smallskip}
\hline
\noalign{\smallskip}
$u$ & 35 & 20  \\
$v$ & 30 &20 \\
$b$ & 15 &10  \\
$y$ & 20 & 5 \\
$H\beta_{\textrm{n}}$&25&$-$\\
$H\beta_{\textrm{w}}$&10&$-$\\
\noalign{\smallskip}
\hline\hline
\noalign{\smallskip}
\multicolumn{3}{c}{NGC~2169}\\
\noalign{\smallskip}
\hline
\noalign{\smallskip}
$u$ & 50 &$-$   \\
$v$ & 30 &$-$ \\
$b$ & 20 &$-$  \\
$y$ & 20 & 10 \\
$H\beta_{\textrm{n}}$&50&$-$\\
$H\beta_{\textrm{w}}$&25&$-$\\
\noalign{\smallskip}
\hline\hline
\noalign{\smallskip}
\multicolumn{3}{c}{NGC~1502}\\
\noalign{\smallskip}
\hline
\noalign{\smallskip}
$u$ & 400&$-$   \\
$v$ & 200 &$-$ \\
$b$ & 150 &$-$  \\
$y$ & 30& 10 \\
$H\beta_{\textrm{n}}$&350&$-$\\
$H\beta_{\textrm{w}}$&30&$-$\\
\noalign{\smallskip}
\hline\hline
\noalign{\smallskip}
\multicolumn{3}{c}{NGC~1039}\\
\noalign{\smallskip}
\hline
\noalign{\smallskip}
$u$ & 50&$-$   \\
$v$ & 30 &$-$ \\
$b$ & 20 &$-$  \\
$y$ & 20& 5 \\
$H\beta_{\textrm{n}}$&50&$-$\\
$H\beta_{\textrm{w}}$&25&5   \\
\noalign{\smallskip}
\hline\hline
\noalign{\smallskip}
\multicolumn{3}{c}{NGC~869 and NGC~884}\\
\noalign{\smallskip}
\hline
\noalign{\smallskip}
$u$ & 600&$-$  \\
$v$ &350&$-$  \\
$b$ &200&$-$   \\
$y$ &100&$-$  \\
$H\beta_{\textrm{n}}$&600&$-$\\
$H\beta_{\textrm{w}}$&120&$-$\\
\noalign{\smallskip}
\hline
\end{tabular}}
\end{table}

Standard stars were observed throughout the run in the clusters NGC~869, NGC~884, NGC~1039, NGC~1502, NGC~2169 and NGC~2244, using the exposure times suitable to obtain good photometric values for all stars selected as standards in these clusters. The central positions for the clusters and the exposure times used are presented in Table~\ref{tab1a} and Table~\ref{tab2a}, respectively. If any of the selected standards turned out to be saturated, we repeated the observations with a shorter exposure time. 

We reduced the frames for all clusters with {\sc iraf}\footnote{{\sc iraf} is distributed by the National Optical Astronomy Observatories, which are operated by the Association of Universities for Research in Astronomy, Inc., under cooperative agreement with the National Science Foundation} routines
for the bias and flat-field corrections. Photometry was obtained by
point-spread function (PSF) fitting using the {\sc daophot} package
\citep{stetson1987} provided by {\sc iraf}. The apertures used are of the order of 
the full width at half maximum. In this case, we used a value of 6 pixels for each image in all filters. In order to construct the PSF empirically,
we automatically selected bright stars (typically 25 stars). After this,
we reviewed the candidates and we discarded those that did not reach the
best conditions for a good PSF star. Once we had the list of PSF
stars ($\approx 20$), we determined an initial PSF by fitting the best
function between the 5 options offered by the PSF routine inside
{\sc daophot}. We allowed the PSF to be variable (of order 2) across the
frame to take into account the systematic pattern of PSF variability
with position on the chip. 

We needed to perform an aperture correction of 26, 25 and 30 pixels for the $u$, $v$ and $b$ filters, respectively. The aperture correction for the rest of the filters: $y$, $H\beta_{\textrm{n}}$ and $H\beta_{\textrm{w}}$ was of 28 pixels.
The atmospheric extinction corrections were performed using the {\sc ranbo2} program, which implements the method described by \citet{manfroid1993}. Finally, we obtained the instrumental magnitudes for all stars.

The selection of standard stars has been explained in \citet{marco2013}. In this work, we used two new open clusters that fulfilled the requirements to provide standard stars. The list of newly adopted standard stars and their photometric
data to be used in the transformations of CCD Str\"omgren photometry of open clusters are given in Table~\ref{tab2}.

\begin{table*}
\centering
\caption{New adopted standard stars with their cataloged values and spectral types taken from the literature}
\label{tab2}
\begin{tabular}{ccccccc}
\hline
\noalign{\smallskip}
Star&$V$&$b-y$&$m_{1}$&$c_{1}$&$\beta$&Spectral Type\\
\noalign{\smallskip}
\hline
\noalign{\smallskip}
\multicolumn{7}{c}{NGC~2244}\\
\noalign{\smallskip}
\hline
\noalign{\smallskip}
114 &  7.590  &   0.207 &  $-0$.048&   $-0$.081 & 2.608 &O8.5\,V \\
\noalign{\smallskip}
\hline
\noalign{\smallskip}
\multicolumn{7}{c}{NGC~2169}\\
\noalign{\smallskip}
\hline
\noalign{\smallskip}
11 & 10.600	& 0.084  & 0.065 &  0.541 & 2.698 & B8\,V  \\ 
15 & 11.080	& 0.130 & 0.109 &  0.944 &2.864  &B9.5\,V	\\
18 & 11.800	& 0.115 & 0.105 &  0.912 &2.872  &B9.5\,V	\\
\noalign{\smallskip}
\hline
\end{tabular}
\begin{center}
\begin{list}{}{}
\item[]The data are taken from \citet{crawford1975} and \citet{johnson1953} for NGC~2244 and \citet{perry1978} for NGC~2169. Spectral types are taken from \citet{walborn1971} for NGC~2244, and \citet{perry1978} for NGC~2169.
\end{list}
\end{center}               
\end{table*}

With the standard stars, we transformed the instrumental magnitudes to the
standard system using the {\sc photcal} package inside {\sc iraf}. We implemented the following $uvby$ transformation equations, after \citet{crawford1970a} and the $\beta$ transformation equation after \citet{crawford1966}:

\begin{eqnarray}
V& = &(-4.937+A)-0.110(b-y)+y_{i}\\
&&\pm0.009\pm0.035\nonumber \\
(b-y)& = &(0.231+B)+1.037(b-y)_{i}\\
&&\pm0.002\pm0.013\nonumber \\
m_{1}& = &(-0.668+C)+0.923m_{1_{i}}-0.159(b-y)\\
&&\pm0.047\pm0.051\pm0.031\nonumber\\
c_{1}& = &(0.016+D)+1.020c_{1_{i}}-0.202(b-y)\\
&&\pm0.011\pm0.011\pm0.030\nonumber\\
\beta& = &(1.198+E)+0.703\beta_{i}\\
&&R^2=0.97\nonumber\\
\nonumber\end{eqnarray}

where each coefficient is given with the error resulting from the transformation. The values of A, B, C, D and E represent zero points whose values are different for each night. These values are provided in Table~\ref{tab3}.

\begin{table}
\caption{Values of the constants A, B, C, D and E for each night. \label{tab3}}
\centering
\begin{tabular}{cccccc}
\hline\hline
\noalign{\smallskip}
Night&A&B&C&D&E\\
\noalign{\smallskip}
\hline
\noalign{\smallskip}
20030209& 18.079&0.174&0.816&0.431&2.105\\
20030210& 18.036&0.173&0.818&0.429&2.104\\
20030215& 16.574&0.432&0.259&0.805&2.049\\
20030216& 13.784&0.805&0.156&0.478&1.687\\
\noalign{\smallskip}
\hline
\end{tabular}
\end{table}

\begin{figure*}
\resizebox{18 cm}{!}{\includegraphics[angle=0]{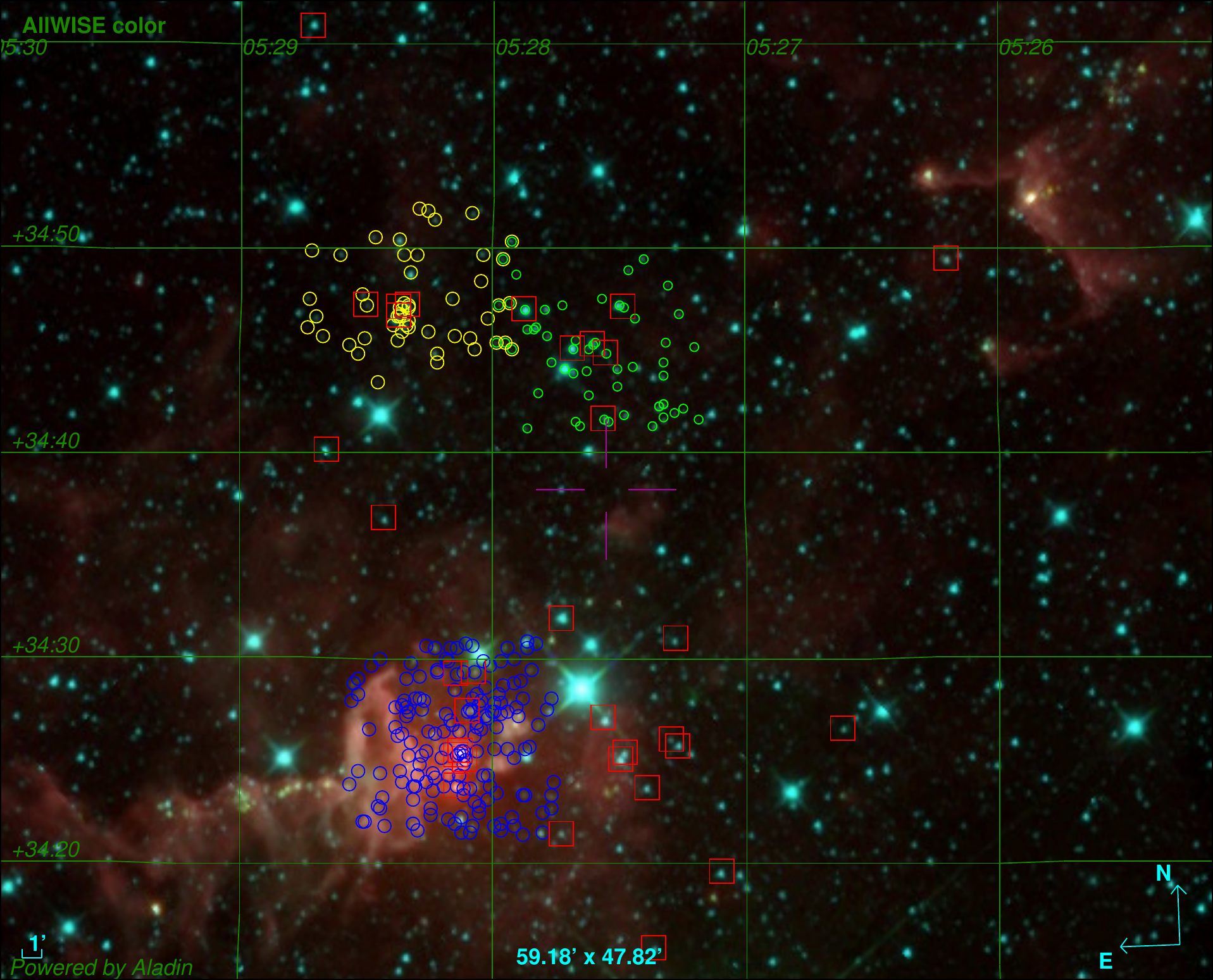}
}
\centering
\caption{Map of the area showing the full field studied and the three areas for which we obtained Str\"omgren photometry. The image has been downloaded with Aladin and represents a false-colour composite of WISE bands. The pinkish regions represent extended dust emission. Red squares indicate stars with spectra. Circles in blue, green and yellow are stars with optical photometry in the three areas (Stock~8, Alicante~11 and Alicante~12). The Nebulous Stream of \citet{jose2008} can be seen just to the east of Stock~8. The cluster candidate CC~14 is the small clump of infrared sources located where the stream meets the grid line corresponding to R.A. = 05h 29m. The size of field is $59\farcm18 \times 47\farcm82$. North is up and east is left. }
\label{todo} 
\end{figure*}

\begin{figure*}
\resizebox{18 cm}{!}{\includegraphics[angle=0]{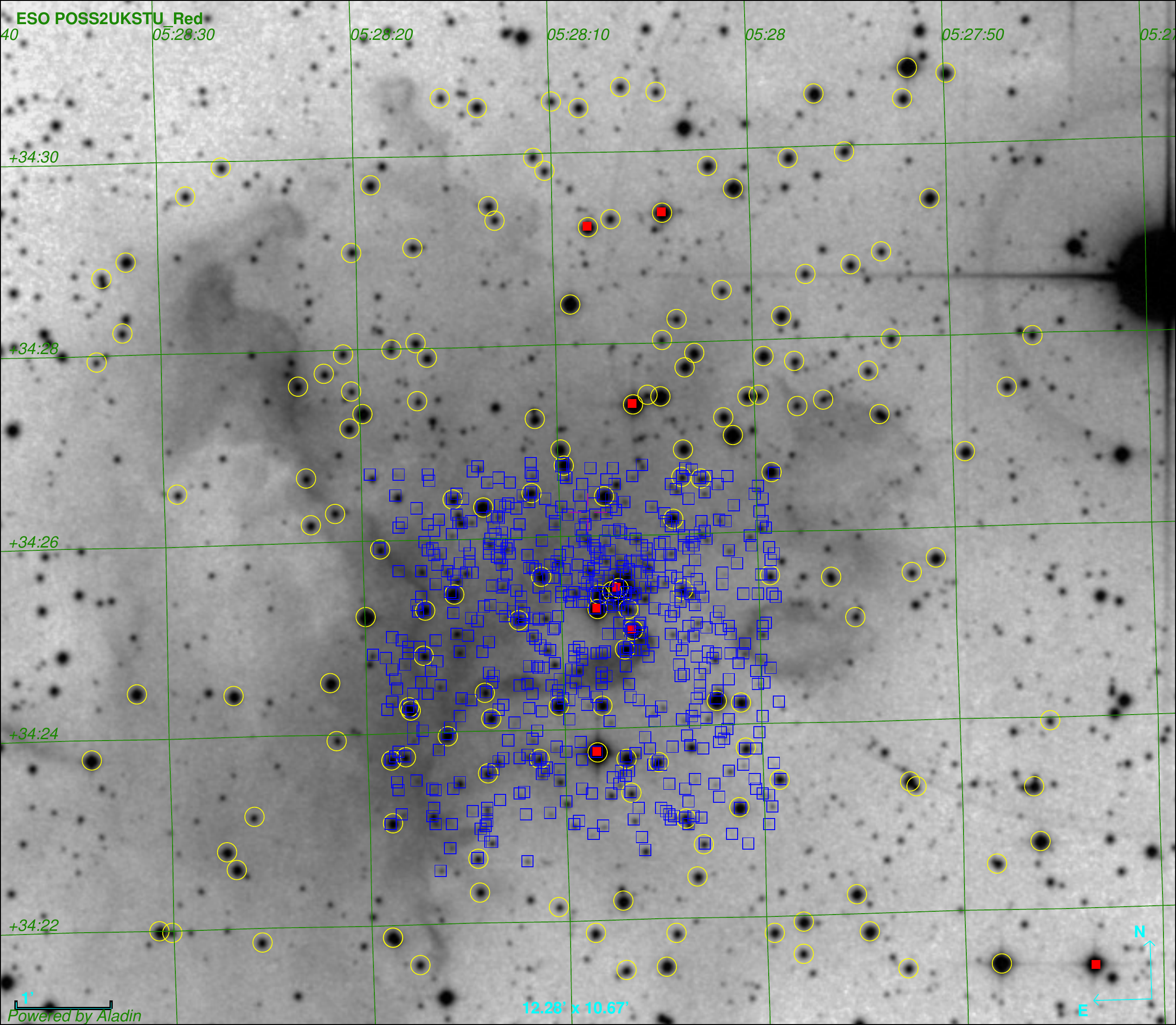}
}
\centering
\caption{Finding chart for stars with photometry in the cluster Stock~8. The image, downloaded from Aladin, is a DSS2 red digitization. The dark areas thus map nebular H$\alpha$ emission. Red solid squares represent stars with spectra. Circles in yellow are stars with optical photometry and blue rectangles are stars with TNG near-IR photometry. Numbers and coordinates in J(2000) are listed in Tables~\ref{coorSTOCK8} and~\ref{Near_IR}. The size of field is $12\farcm28 \times 10\farcm67$. North is up and east is left. }
\label{STOCK8} 
\end{figure*}

\begin{figure*}
\resizebox{18 cm}{!}{\includegraphics[angle=0]{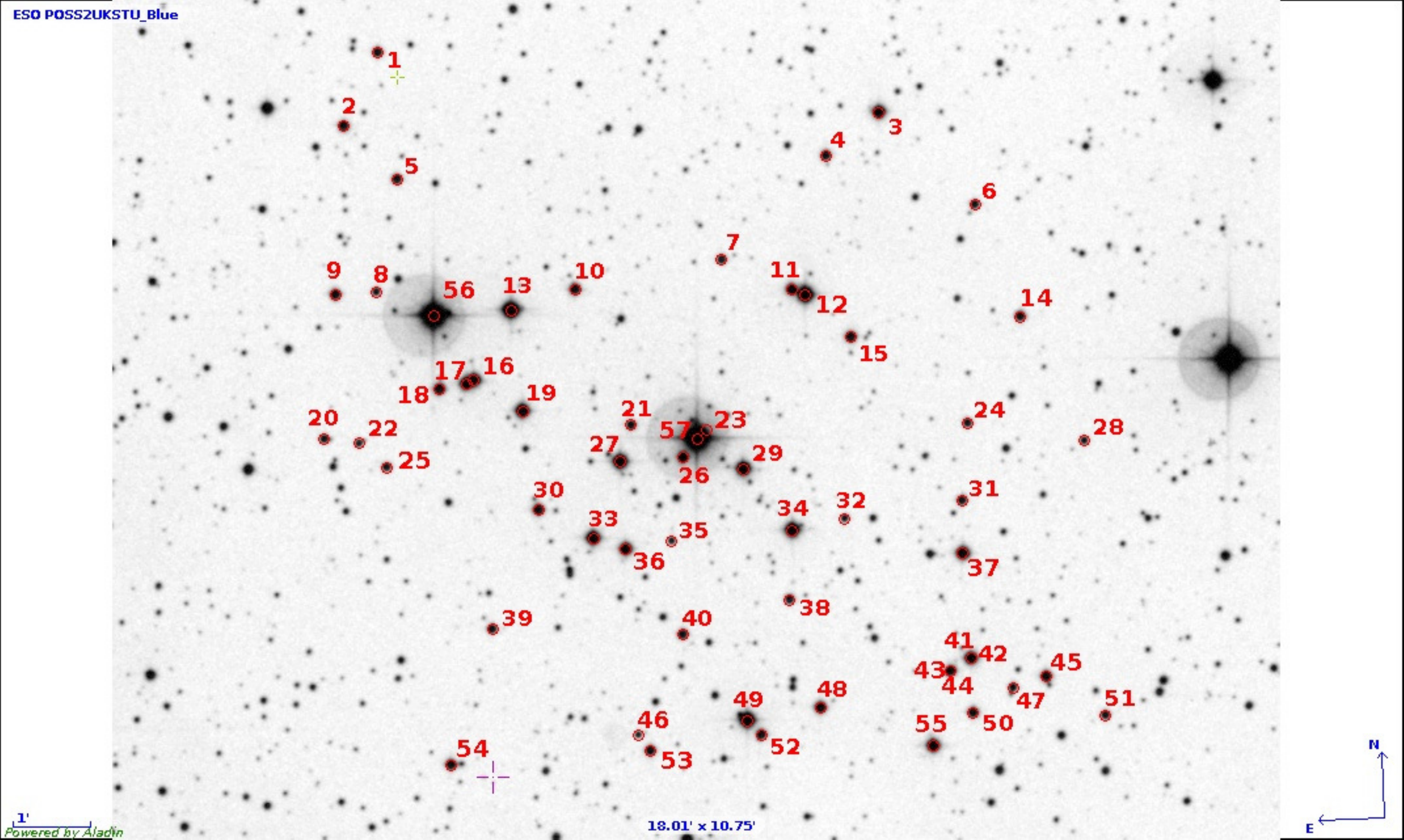}
}
\centering
\caption{Finding chart for stars with photometry in Alicante~11. The image, provided by Aladin, is a DSS2 blue digitization. Numbers and coordinates in J(2000) are listed in Table~\ref{coor-Alicante11}. The size of the field is $18\farcm01 \times 10\farcm75$. North is up and east is left.\label{Alicante11}}
\end{figure*}

\begin{figure*}
\resizebox{18 cm}{!}{\includegraphics[angle=0]{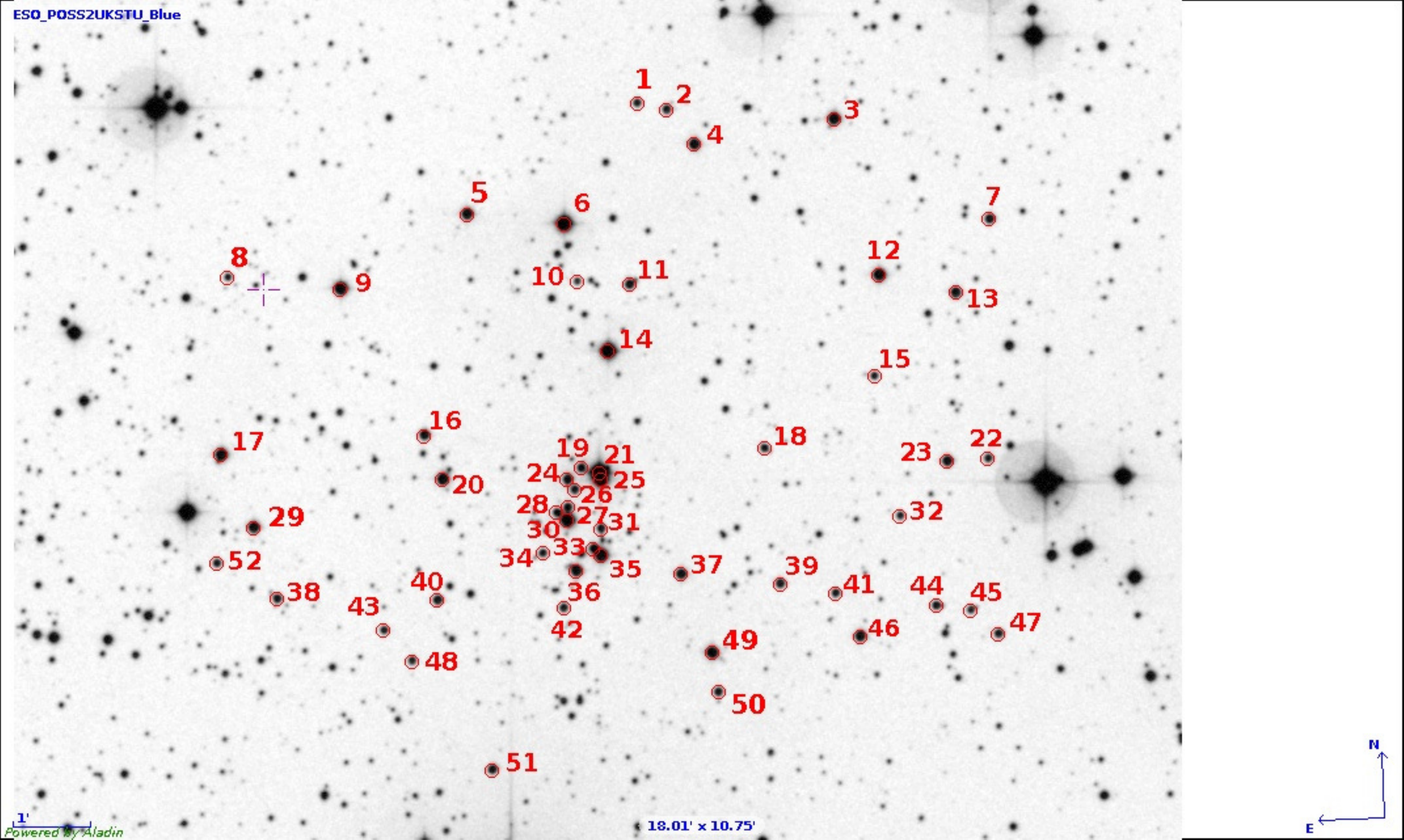}
}
\centering
\caption{Finding chart for stars with photometry in Alicante~12. The image, provided by Aladin, is a DSS2 blue digitization. Numbers and coordinates in J(2000) are listed in Table~\ref{coor-Alicante12}. The size of field is $18\farcm01 \times 10\farcm75$. North is up and east is left.\label{Alicante12}}
\end{figure*}

The number of stars that can be detected in all filters is limited by
the long exposure time in the $u$ filter. We selected for the analysis all stars with good photometry (photometric errors $\le 0.05$~mag) in all six filters. We identify these stars on the images in Figures~\ref{STOCK8}, \ref{Alicante11} and~\ref{Alicante12} for each field observed.

We have obtained $uvby\beta$ CCD photometry for 163 stars in the cluster Stock~8, reaching a magnitude limit $V\approx16$. In Table~\ref{coorSTOCK8}, we list their coordinates in $J2000$. The second column cross-references with the near-infrared (near-IR) photometry, indicating their number in Table~\ref{Near_IR}, for those stars that have values in our near-IR photometry. For stars with no measurements in Table~\ref{Near_IR}, the last columns show $JHK_{s}$ photometry from the 2MASS
catalog \citep{Skrutskie2006}. The $uvby\beta$ CCD photometry is given in Table~\ref{strom-stock8}, where $N$ is the number of measurements for each magnitude, colour or index and $\sigma$ is the error assigned to each value. The error assigned is the standard deviation for stars with more than one measurement and the photometric errors from {\sc daophot} for stars with only one measurement, with the errors in the colour and indices calculated through error propagation. In Fig.~\ref{STOCK8} we indicate with yellow open circles stars with optical photometry. The open blue squares are stars that we could detect in our near-IR photometry. Solid red squares are stars for which we obtained spectra.
In Table~\ref{coor-Alicante11} and Table~\ref{coor-Alicante12} we list the number and the coordinates in J2000 for stars in the Alicante~11 and the Alicante~12 fields with optical photometry. In these tables we show their $JHK_{s}$ photometry from the 2MASS
catalog \citep{Skrutskie2006} too. In Table~\ref{strom-Alicante11} and Table~\ref{strom-Alicante12} we present the resulting values for $V$, $b-y$, $m_{1}$, $c_{1}$ and $\beta$ for stars in Alicante~11 and Alicante~12, together with the number of measurements for each magnitude, colour or index and the error assigned for each value. These errors are calculated as above. The designation of each star is given by the number indicated on the images (Fig.~\ref{Alicante11} and Fig.~\ref{Alicante12}). We obtained optical photometry for 57 stars in Alicante~11 and for 52 stars in Alicante~12.  

\subsection{Near-IR photometry} 

Near-IR $JHK$ images were obtained in the area of the open cluster Stock~8 using NICS (Near Infrared Camera Spectrometer) on the Telescopio Nazionale Galileo (TNG) in the La Palma observatory on 14 May 2011. NICS was equipped with a HgCdTe Hawaii $1024 \times 1024$ array, providing a pixel scale of $0\farcs25$ pixel and a field of view of $4\farcm2 \times 4\farcm2$ in its large field (LF) mode. The observations were carried out under photometric conditions.

These near-IR observations consisted of series of several integrations repeated at dithered positions on the detector. The total effective exposure, resulting from the co-addition of the jittered exposures, is 60~s for long exposures and 15~s for short exposures in all filters  

The standard reduction of near-IR data includes the dark subtraction and flat-field correction of individual frames
that are median-combined to build a sky image. The sky image is subsequently subtracted from the individual
frames that are then shifted and finally stacked into one single image. 
The procedure for obtaining the photometry from the reduced frames was the same as described in Section~\ref{opt_phot} for the optical images.

The next step is to transform these instrumental magnitudes to the 2MASS magnitude system \citep{Skrutskie2006}. For this, we selected those stars in our field having $JHK_{{\rm S}}$ magnitudes in the 2MASS catalogue with photometric errors smaller than 0.03 magnitudes in every filter (about 80 stars). A linear transformation was carried out between instrumental and 2MASS magnitudes. No colour term was needed, since a simple shift in zero point
results in a good transformation. This was checked by plotting the transformed magnitudes against 2MASS magnitudes for all the stars with 2MASS magnitudes in the field, finding that the best fit corresponds to a straight line of slope 1 in all three filters. 

We have near-IR photometry for 589 stars. In Table~\ref{Near_IR}  we list the number of each star, their RA and DEC coordinates in J2000, and the value of $J$, $H$ and $K_{{\rm S}}$ with the photometric error for each magnitude. There is only one measurement for each filter. The completeness limits in the 2MASS standard system are $J\sim18$, $H\sim17$ and $K_{{\rm S}}\sim18$.

Astrometric referencing of our images was made using positions of a number of stars from the 2MASS catalog. The
PSFs of these stars are not corrupted by the CCD over-saturation effects. We used {\sc iraf} tasks {\tt ccmap/cctran} for the astrometric
transformation of the image. Formal rms uncertainties of the astrometric fit for our images are $<0\farcs10$ in both right ascension and declination.

\subsection{Spectroscopy} 

Spectra of the brightest stars in the area were taken with the {\em Aur\'{e}lie}
spectrograph on the 1.52-m telescope at the Observatoire de Haute
Provence (OHP) during two dedicated runs on
2002 January 18\,--\,22 and 2002 February 25\,--\,28 (when only the
first night was useful because of the weather). The
spectrograph was equipped with grating \#3 (600 ln mm$^{-1}$) and
the Horizon 2000 $2048\times1024$ EEV CCD camera (see \citealt{gil94}
for a description of the instrument). In the classification region,
this configuration gives a dispersion of 0.22\,\AA/pixel (resolving
power of approximately 7\,000), covering a wavelength range of $\approx
440$\,\AA. It is, therefore, necessary to observe two wavelength regions
to cover the classical classification region.

For this programme, the two regions selected (identified as Position 1 and~2 in Table~\ref{tab:ohp}) were centred at
$\lambda=4175$\,\AA\ and $\lambda=4680$\,\AA. In principle, all objects were intended to be observed in both 
regions, resulting in coverage in the $\lambda\lambda 3950 - 4900$\,\AA\ range, with a small gap ($\sim 60\,$\,\AA) around $\lambda 4425$\,\AA. No
strong photospheric lines are found in the gap, but the strong diffuse
interstellar line (DIB) at $\lambda 4428$\,\AA, which is a good
indicator of the reddening, is lost.
The complete log of observations at the 1.52-m OHP is given
in Table~\ref{tab:ohp}.

These spectroscopic data were reduced with the {\em Starlink}
packages {\sc ccdpack} \citep{Draper2000} and {\sc figaro}
\citep{Shortridge1997} and analyzed using {\sc figaro} and {\sc dipso}
\citep{Howarth1998}. 

We took more spectra with the 2.6-m Nordic Optical Telescope (NOT, La Palma, Spain) on the nights of 2011 September, 1\,--\,2. The telescope was equipped with the imager and spectrograph Andaluc\'{i}a Faint Object Spectrograph and Camera (ALFOSC). Spectra of some of the brightest stars in Stock~8, Alicante~11 and Alicante~12 were obtained with ALFOSC. In spectroscopic mode, we used the grism \#16 combined with a $1\arcsec$ slit to obtain intermediate resolution spectroscopy. Grism \#16  covers the 3500\,--\,5060\:\AA{} range with a nominal dispersion of 0.8\,\AA/pixel. The resolving power for this configuration is $R\sim1\,000$. The log of observations at the NOT is given in Table~\ref{tab:not}. 

These spectroscopic data were reduced using {\sc iraf} routines for the bias and flat-field corrections. The extraction of the spectra was done using the {\sc apall} package inside {\sc apextract} in {\sc twodspec}. Once we had the spectra of targets and arcs, we used the package {\sc identify} to perform the wavelength calibration. We used ThAr arc lamp spectra, taken between the exposures. The rms for the wavelength solution is $\approx0.2$ pixels.

\begin{table*}
\caption{Stars observed from the OHP 1.52-m. The first column indicates the volume and number in the Luminous Stars catalog or the name in the Henry Draper Catalog. The second and third columns indicate the dates when the two spectral positions were observed and (between brackets) the exposure time in seconds. The derived spectral types are given in the fourth column. Spectral types marked with a '*' are less secure than the average, because of poor signal to noise or presence of double lines. References for the photometric values in column 7 are: (1) \citet{hiltner1956} and (2) \citet{mm71}. 
\label{tab:ohp}}
\begin{center}
\begin{tabular}{lcccccccc}
\hline
\hline
\noalign{\smallskip}
Name& Position 1 & Position 2& Spectral & $V$ &  $(B-V)$ & Reference & $DM\pm0.5$&$K_{S}$\\
Number& & & Type &  & &&&\\
\noalign{\smallskip}
\hline\hline
\noalign{\smallskip}
LS\,V $+34\degr$22$^{1}$ & 18/01 (600) & 19/01 (700)& O7.5\,V&8.58 &0.23 &2 & 11.5  &$7.89\pm0.02$\\
LS\,V $+34\degr$15 & 18/01 (750) & 19/01 (900)& B1\,V&$10.02\pm0.05$ &0.20 &1 &11.6&$9.37\pm0.02$\\
LS\,V $+34\degr$16 & 18/01 (1000) & 19/01 (1000)& B1\,V$^{*}$&$-$&$-$&$-$&$-$&$9.22\pm0.02$\\
LS\,V $+34\degr$18 & 18/01 (750) & 19/01 (1200)& O9.5\,V$^{*}$& 10.01&0.67&2&11.0&$8.04\pm0.02$    \\
LS\,V $+34\degr$21 & 18/01 (1200) & 19/01 (1500)&O9\,V$^{*}$&10.70 &0.94&2&10.9&$7.88\pm0.02$   \\
LS\,V $+34\degr$29$^{2}$ & 20/01 (1000) & 25/02 (1200)&O9.7\,IV & 8.89& 0.18   &2&12.0&$8.32\pm0.02$\\ 
LS\,V $+34\degr$31$^{3}$ & 20/01 (1200) & 25/02 (1200)& B0.5\,V$^{*}$& 9.87&0.19    &2&12.0&$9.23\pm0.02$\\
LS\,V $+34\degr$36$^{4}$& 18/01 (600) & 19/01 (750) & O8\,V(n)&8.78&0.26&1&11.4&$8.02\pm0.03$\\
LS\,V $+34\degr$23$^{5}$& 20/01 (750) & 25/02 (900) & O8\,II(f) &8.04&0.32&1 &12.0&$6.99\pm0.02$\\
LS\,V $+34\degr$25$^{6}$& 20/01 (750) & 25/02 (1200) & O9.5\,V + B0.2\,V&$-$&$-$&$-$& &$7.59\pm0.02$\\
LS\,V $+34\degr$11$^{7}$& 18/01 (900) & 19/01 (750)& B0.5\,V&9.70&0.17&2&11.9&$9.27\pm0.02$\\
HD 281147 & 18/01 (900) & 19/01 (1200)& B1\,V&$-$&$-$&$-$&$-$&$9.63\pm0.02$\\
\noalign{\smallskip}
\hline
\end{tabular}
\begin{list}{}{}
\item[]$^{1}$ HD~35619 = Alicante~11-57; $^{2}$ BD~$+34\degr$1054 = ST93; $^{3}$ BD~$+34\degr$1056 = ST120; $^{4}$ BD~$+34\degr$1058; $^{5}$ HD~35633; $^{6}$ HD~35652 = Alicante~11-56 (Eclipsing binary IU~Aur); $^{7}$ HD~281150
\end{list}
\end{center}
\end{table*}

\begin{table*}
\caption{Classification spectra of stars in the area of Stock~8. These stars were observed with the NOT.  References
for the photometric values in column 7 are: (2) \citet{mm71} and (3) \citet{mayer1964}.  \label{tab:not}}
\begin{center}
\begin{tabular}{lcccccccc}
\hline\hline
\noalign{\smallskip}
Star&LS&Spectral Type&Exposure Time(s)&$V$&$(B-V)$&Reference&$DM\pm0.5$&$K_{S}$  \\
\noalign{\smallskip}
\hline
\noalign{\smallskip}
Alicante~11-12& V $+34\degr$17&	B2\,V&250&$-$ &$-$ &$-$ &$-$&$10.16\pm0.02$\\						
Alicante~12-21& V $+34\degr$33&	B2\,V&300&$-$&$-$&$-$&$-$&$9.66\pm0.03$	\\	
Alicante~12-24&$-$&B7\,V	& 300&$-$&$-$&$-$&$-$&$12.82\pm0.03$    \\					  
Alicante~12-20&$-$&B9\,V&300&$-$&$-$&$-$&$-$&$11.81\pm0.02$\\		
Alicante~11-29& V $+34\degr$19&	B2\,V&400&$-$&$-$&$-$&$-$&$10.77\pm0.02$\\
Alicante~11-49& V $+34\degr$20&B1.5\,V&400&$-$&$-$&$-$&$-$&$10.29\pm0.02$\\  
$-$& V $+34\degr$14& B1\,V& 300&$-$&$-$&$-$&$-$&$10.19\pm0.02$\\   
ST101&$-$&A3\,V&400&$-$&$-$&$-$&$-$&$11.39\pm0.01$\\				  
ST97& V $+34\degr$30&	B1\,V&400&-&-&-&-&$10.45\pm0.02$	\\					  
ST61& V $+34\degr$28$^{1}$	&	B1\,V&350&10.92&0.27&3&12.3&$10.03\pm0.02$	\\					  
ST23& V $+34\degr$27	&	B2\,IV&400&11.66&0.33&3&13.0&$10.70\pm0.02$	\\					  
ST25&$-$&B5\,V	&400&$-$&$-$&$-$&$-$&$11.63\pm0.02$		\\
$-$& V $+34\degr$35	&	B0.7\,V&200&$-$&$-$&$-$&$-$&$9.75\pm0.02$	\\				  
$-$& V $+34\degr$12	&	B1\,V&300&$10.91$&$0.30$&2&12.2&$10.01\pm0.02$		\\					  
$-$& V $+34\degr$13	&	B2\,IV&	400&$-$&$-$&$-$&$-$&$10.92\pm0.02$	\\					  
Alicante~11-27&$-$&	B8\,V&300&$-$&$-$&$-$&$-$&$8.06\pm0.03$	\\					  
Alicante~12-30&$-$	&	B2\,V&	400&$-$&$-$&$-$&$-$&$10.69\pm0.02$	\\					  
$-$& V $+34\degr$8&B4\,III	&300&$-$&$-$&$-$&$-$&$10.57\pm0.02$		\\
$-$& V $+34\degr$24&B1\,V	&350&11.24&0.40&3&12.2&$10.05\pm0.02$	\\
$-$& V $+34\degr$10$^{2}$	& B0.2\,V&200& 9.43&0.16&2&11.7&$9.01\pm0.02$\\
$-$& V $+34\degr$37$^{3}$	&O9.7\,V&200&9.22&0.19&2&11.5  &$8.80\pm0.02$\\
\noalign{\smallskip}
\hline
\end{tabular}
\begin{list}{}{}
\item[]$^{1}$ BD $+34\degr$1053; $^{2}$ HD 281151; $^{3}$ BD $+34\degr$1059
\end{list}
\end{center}
\end{table*}			  

\section{Results}
\label{results}
		  
\subsection{Spectral classification}

To characterize the morphology of the whole region, we observed spectroscopically almost all the stars in the area that have been classified as early-type stars in the literature. 
The stars observed cover a region of approximately $40\arcmin \times 40\arcmin$ that includes the 3 open clusters studied and a very significant diffuse population scattered over most of the region. These stars are represented by open red squares in Fig.~\ref{todo} and listed in Tables~\ref{tab:ohp} and~\ref{tab:not}. In these tables all the stars are listed by their identifier in the LS V \citep{hardorp1965} catalog except for HD~281147, which is missing in this catalog. Cross-correlation with our numbering system is provided for stars with photometry. Other more usual identifiers of the brightest stars, such as HD numbers, are also given.

\begin{figure}
\resizebox{\columnwidth}{!}{\includegraphics[angle=-90]{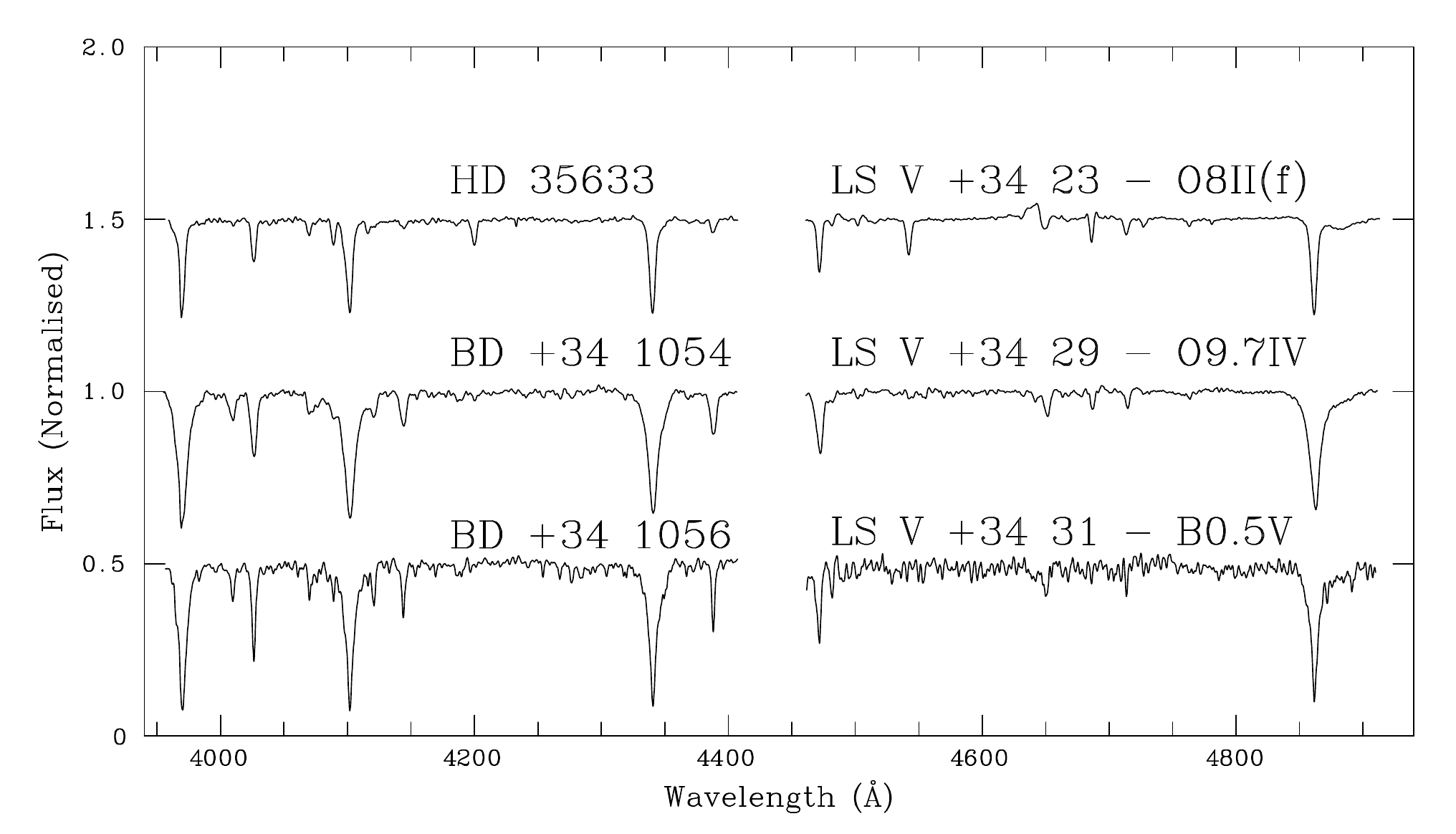}
}
\centering
\caption{OHP spectra of the two brightest stars in Stock~8, and the nearby HD~35633 that is, very likely, the main source of ionization in IC~417. The small gap is due to the two different settings explained in the text.
\label{stock_spec}} 
\end{figure}

The stars observed, according to their distribution, can be grouped as follows: 
\begin{itemize}
\item The two brightest stars illuminating the \ion{H}{ii}
region, at the core of Stock 8, LS\,V$+34\degr$29 (ST93) and LS\,V$+34\degr$31 (ST120), and a set of 5 stars within the limits given for Stock~8 by \citet{jose2008}: ST23, ST25, ST61, ST97 and ST101

\item A group of 12 stars spread to the west of Stock8 that have been assumed
to be somehow connected to Stock 8 by previous authors: LS\,V $+34\degr$23, LS\,V $+34\degr$12, 
LS\,V $+34\degr$21, LS\,V $+34\degr$13,  LS\,V $+34\degr$14, LS\,V $+34\degr$11, LS\,V $+34\degr$16, LS\,V $+34\degr$18, LS\,V $+34\degr$15, LS\,V $+34\degr$24, LS\,V $+34\degr$10 and LS\,V $+34\degr$8

\item Two stars to the north of Stock 8, clearly detached
from the cluster,  LS\,V $+34\degr$35 and LS\,V $+34\degr$36

\item A more distant star to the NW, HD 281147 and another isolated star to the north of Alicante~12: LS\,V $+34\degr$37

\item Four stars in the core of Alicante~12: 20, 21, 24 and 30

\item Six stars in the core of Alicante~11: 12, 27, 29, 49, 56 and 57

\end{itemize} 

The spectra obtained were used for spectral classification by comparison to spectra of MK standard stars observed at similar resolution, following the standard classification procedures of \citet{walborn1990}. In a second step, we checked internal consistency by comparing all the spectra obtained amongst
themselves. In principle, the much higher resolution of the OHP spectra allows a much more accurate classification, but in some cases the signal-to-noise ratio in the 4000\,--\,4500~\AA\ region is very low, resulting in inaccurate classifications, marked as such in Table~\ref{tab:ohp}. Spectra observed with ALFOSC can be considered accurate to $\pm1$ subtype.

\begin{figure}
\resizebox{\columnwidth}{!}{\includegraphics[angle=-90]{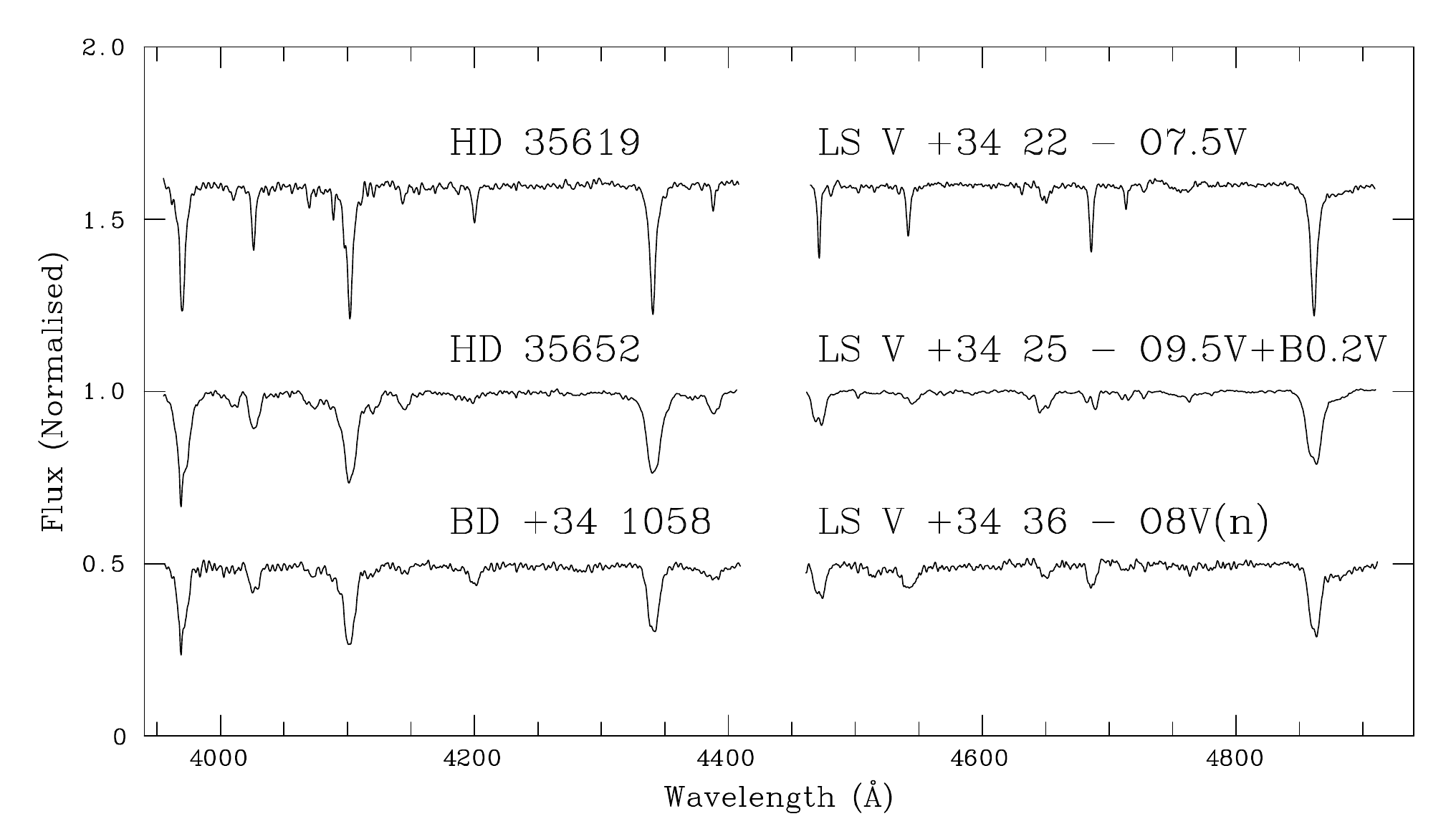}
}
\centering
\caption{OHP spectra of other O-type stars in the area. HD~35619 and HD~35652 are part of Alicante~11. BD~$+34\degr$1058 lies north of Stock~8.
\label{ostars}} 
\end{figure}

All the stars in the LS catalog \citep{hardorp1965}, as expected, turn out to be of early type. The only object that does not belong to the OB group is LS\,V~$+34\degr$8 that, at B4\,III, is a bit too late for the catalog, and too old to belong to the same population as the rest of the stars in the region, even though its $DM$ (calculated from the 2MASS data) is readily compatible. All the stars observed within the clusters have spectral types compatible with membership except for two early-type foreground objects, ST101 in Stock~8, and Alicante~11-27. Noteworthy aspects of some of the spectra are discussed below:

\begin{itemize}
\item LS\,V +34$\degr$29 (ST~93) is the brightest star in Stock~8. The spectrum suggests it is a moderately fast rotator, but the lines appear very shallow (Fig.~\ref{stock_spec}). Formally, the criterion \ion{Si}{iii}~4553\AA $\simeq$ \ion{He}{ii}~4542\AA\ defines spectral type O9.7. The relative weakness of $\ion{He}{ii}~4686$\AA\ indicates then a luminosity class IV. However, the \ion{Si}{iv} lines are quite weak (\ion{Si}{iv}~4116\AA\ is not even seen). All metallic lines look too weak and shallow, and this suggests that this object is really a spectroscopic binary whose spectral type results from the combination of a late-O and an early-B star.
\item LS\,V $+34\degr$23 has a spectral type O8, rather earlier than previously assigned (Fig.~\ref{stock_spec}). It had been classified as B0.5\,IV by \citet{morgan53}, most likely as a consequence of misidentification. The \ion{N}{iii} 4634--41--42 complex is in emission, while
\ion{He}{ii}~$\lambda4686$\AA\ is very weak and flanked by emission
components. The object, therefore, presents a rather high
luminosity. Just short of being a supergiant according to the criteria of \citet{sota11}, we classify it as O8\,II(f), while noting the complete absence of any absorption \ion{N}{iii} lines. If the star is nitrogen deficient, it could probably have a higher luminosity.
\item LS\,V $+34\degr$25 (Alicante~11-56) is a well known spectroscopic binary (IU~Aur), which has received several very discrepant classifications. Though the two components are not very well resolved in the Position 1 spectrum, they are clearly separated in the Position
2 one (Fig.~\ref{ostars}), and we derive spectral types of O9.5\,V and approximately B0.2\,V for
the two components, based mainly on the relative strengths of the
\ion{He}{ii}~4686\AA\ and \ion{C}{iii}~4650\AA\
lines. These values are in surprisingly good agreement with the masses
$M_{1}=21.3\,M_{\sun}$ and $M_{2}=14.4\,M_{\sun}$ derived by
\citet{dre94}. \citet{dre94} also identify a third body in this
system, with a mass $M_{2}=17.8\,M_{\sun}$, which accounts for
$\sim20$\% of the light. Since a third spectrum is not seen,
\citet{dre94} support the idea that the third body is a close binary
consisting of two B2\,--\,B3 stars. 
\item LS\,V $+34\degr$22 (Alicante~11-57) is the brightest star in Alicante~11. Based on standard criteria, we classify it as O7.5\,V. \citet{sota14} give a more accurate classification O7.5\,V((f))z, based on higher-quality spectra.

\item LS\,V $+34\degr$36, classified as O8nn by \citet{morgan55} does not show very broad lines in our spectrum (Fig.~\ref{ostars}). The object, though, is a double-lined spectroscopic binary. The combined spectrum has spectral type O8\,V,  but all the lines are broad and asymmetric, clearly showing the presence of a second companion. Given the width of \ion{He}{ii}~4542\AA, the two components are O-type stars. Since \ion{He}{ii}~4686\AA\ is abnormally weak, at least one of them could be of higher luminosity.
\item HD~281147 is given as O8nn in Simbad, after \citet{garmany84}. This is almost certainly due to a confusion with  BD $+34\degr$1058, which is the object observed by {\it IUE}. Our spectrum gives a a spectral type B1\,V.
\item The Position~1 spectra of LS\,V~$+34\degr$16, LS\,V~$+34\degr$18, and LS\,V~$+34\degr$21 are little more than noise, and therefore their spectral types are based on a very limited spectral range, and must not be considered accurate.
\end{itemize}

The distribution of spectral types in the region is shown in Fig~\ref{spectral_image}. No obvious pattern emerges.

\subsection{HR diagrams}

\subsubsection{Optical photometry for Stock~8}
\label{ana1stock8}
\begin{figure*}
\centering
\resizebox{\columnwidth}{!}{\includegraphics[clip]{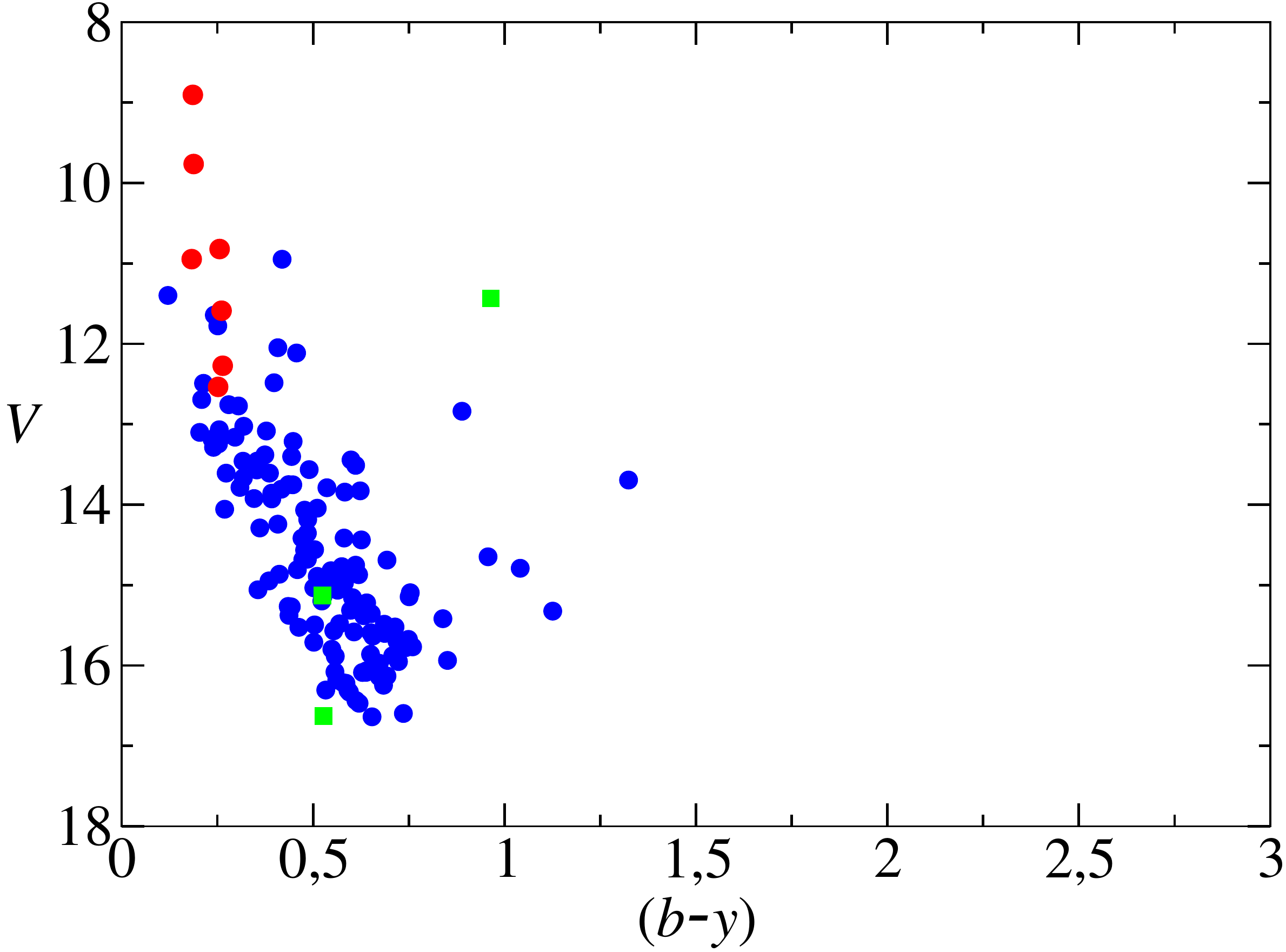}
}
\resizebox{\columnwidth}{!}{\includegraphics[clip]{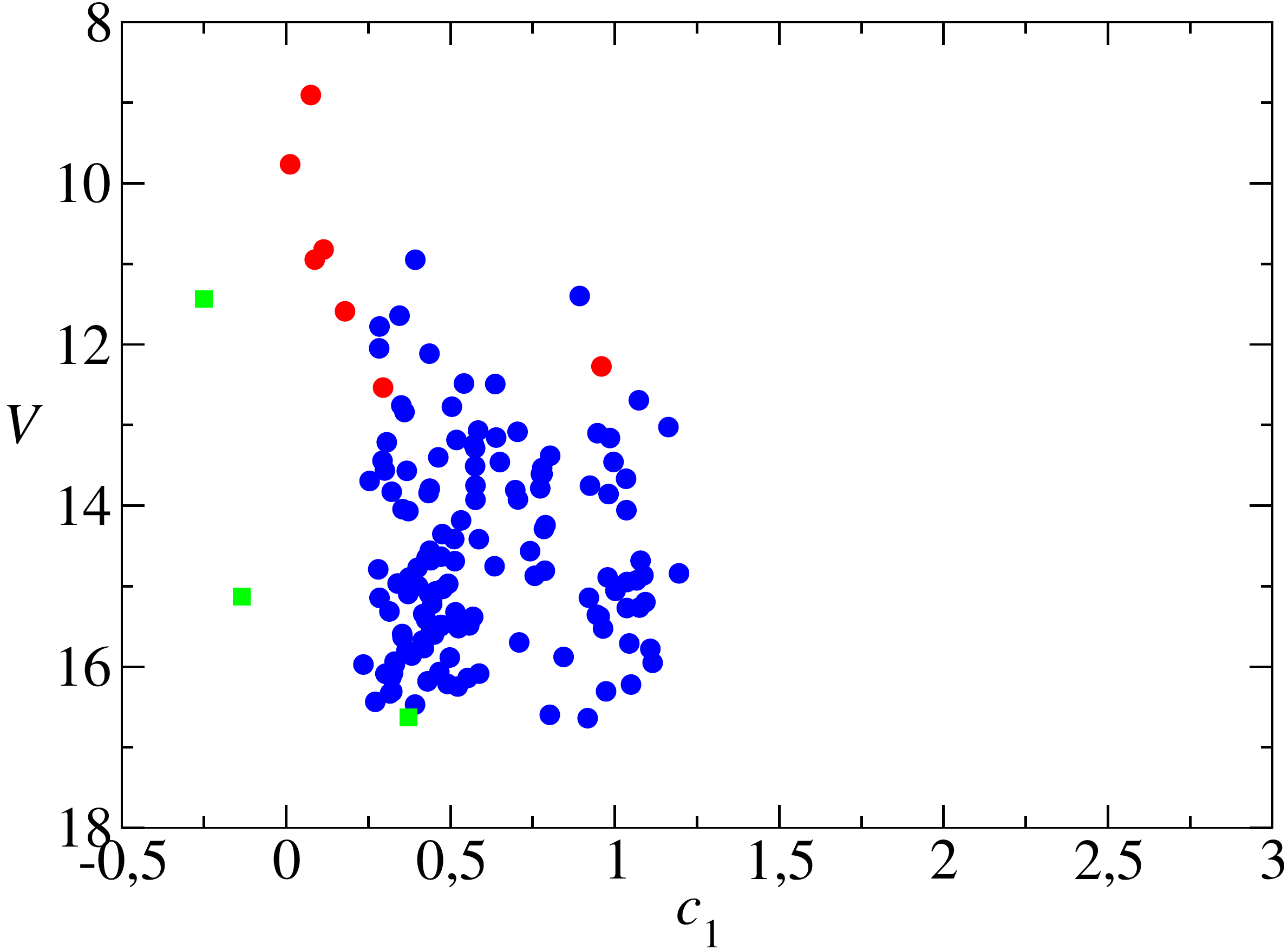}
}
\caption{\textit{Left}: $V/(b-y)$ diagram for all stars in Stock~8. Red circles represent stars spectroscopically observed and green squares, non-member stars\label{Vby_stock8}. \textit{Right}: $V/c_{1}$ diagram for all stars in Stock~8. Red circles represent stars spectroscopically observed and green squares non-member stars.\label{Vc1_stock8}}
\end{figure*}

As a first step, we analyzed the $uvby\beta$ CCD photometry obtained for the cluster Stock~8 (Table~\ref{strom-stock8}) to determine the physical parameters of the cluster: reddening, distance and age.
Initially, we plotted the diagrams $V/(b-y)$ and $V/c_{1}$ for all the stars in the field. In Fig.~\ref{Vby_stock8} red circles are stars with spectra and green squares are stars that appear misplaced in both diagrams and, therefore, were removed for the following analysis. 

The next step in the analysis is the estimate of membership for the stars measured in the field. Given the depth of our observations, and the presence of O-type stars confirmed by spectroscopy, most cluster members in our sample must be B-type stars. Now, we can calculate the reddening-free indices $[m_{1}]$, $[c_{1}]$ and $[u-b]$, where:

\begin{eqnarray}
&[m_1]&=m_1+0.32(b-y)\\
&[c_1]&=c_1-0.20(b-y)\\
&[u-b]&=[c_1]+2[m_1]
\end{eqnarray}

\begin{figure*}
\centering
\resizebox{\columnwidth}{!}{\includegraphics[clip]{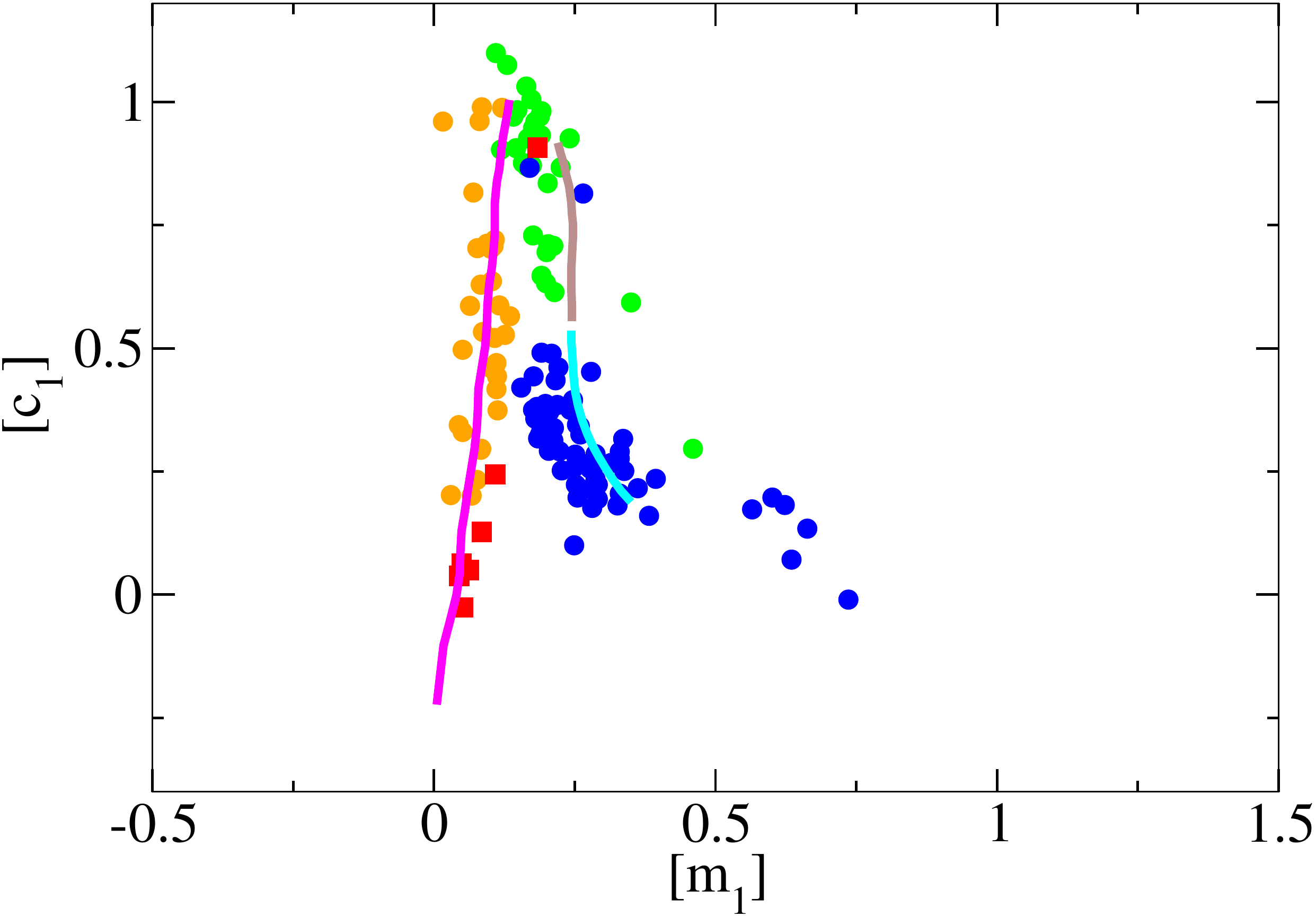}
}
\resizebox{\columnwidth}{!}{\includegraphics[clip]{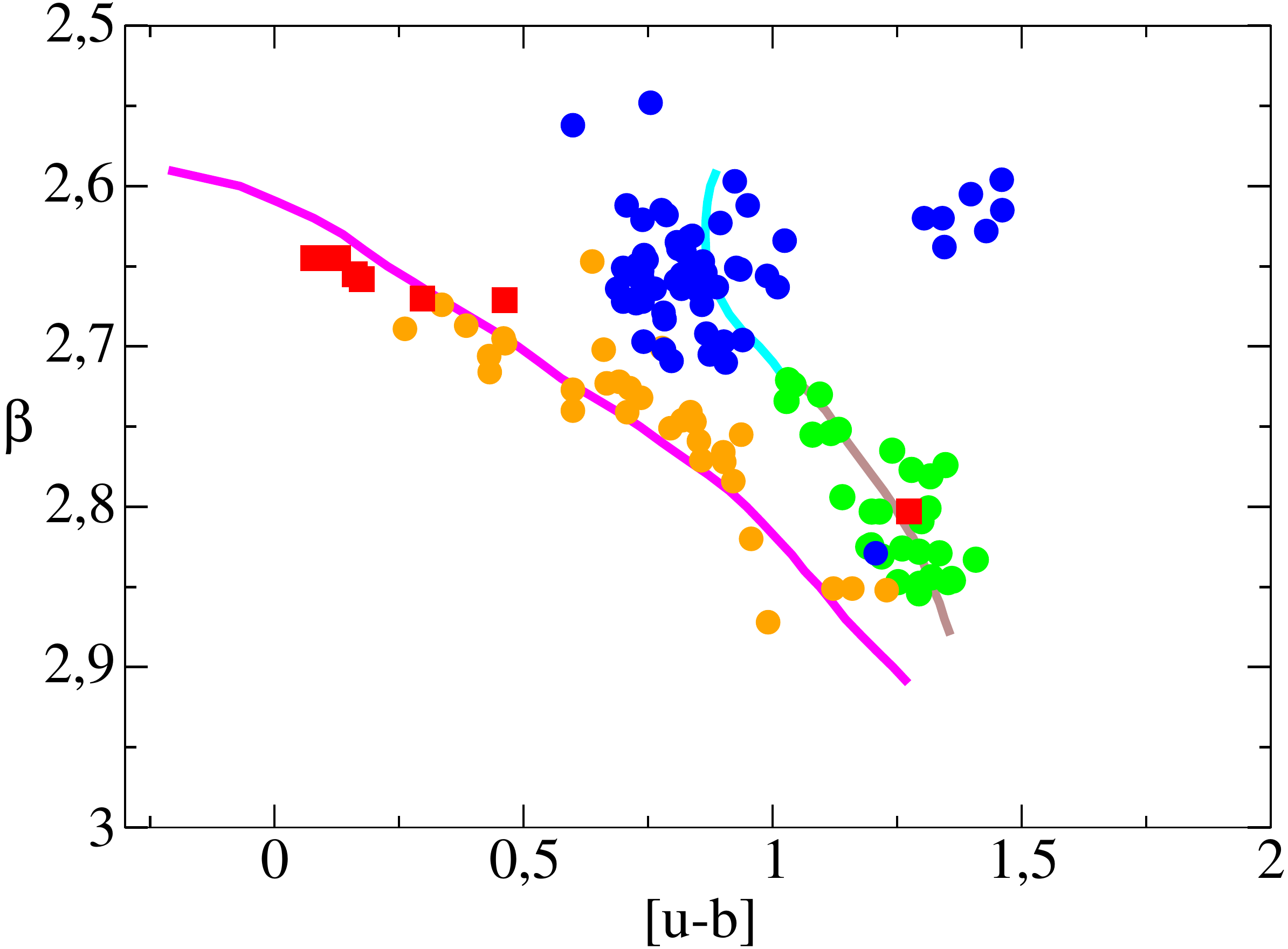}
}
\caption{\textit{Left}: The $[c_{1}]$-$[m_{1}]$ diagram for the stars observed in the open cluster Stock~8. The solid lines represent the average loci of field B-type (magenta line), A-type (brown line) and F-type (cyan line) main-sequence stars \citep{perry1987}. \label{m1c1_stock8}. \textit{Right}: The $\beta$-$[u-b]$ diagram for the stars observed in the open cluster Stock~8. The colour convention is as in the left panel, with the average loci of field stars also taken from \citet{perry1987}. In both diagrams, red squares correspond to spectroscopically observed stars. Orange circles are stars falling close to the position of B-type stars in the two diagrams; green dots are A-type stars, and blue dots are identified as F-type stars}.\label{betaub_stock8}
\end{figure*}

We plot the $[c_{1}]$-$[m_{1}]$ and $\beta$-$[u-b]$ diagrams for all stars chosen in the $V$-$(b-y)$ and $V$-$c_{1}$ diagrams (see Fig.~\ref{m1c1_stock8}). We can divide the stars according to their spectral type. In this figure we can see that the stars spectroscopically classified (red squares) as B-type stars fall on the B-type star standard line, confirming the validity of this relationship. The only star classified as A3\,V falls on the A-type star standard line. We can observe that the majority of the stars are in the same region in both diagrams. Therefore, using these standard relationships, we assign a spectral type for each star. After this, we use those stars falling on the B-type branch in both diagrams to make the following analysis. 

\begin{figure*}
\resizebox{18 cm}{!}{\includegraphics[angle=0]{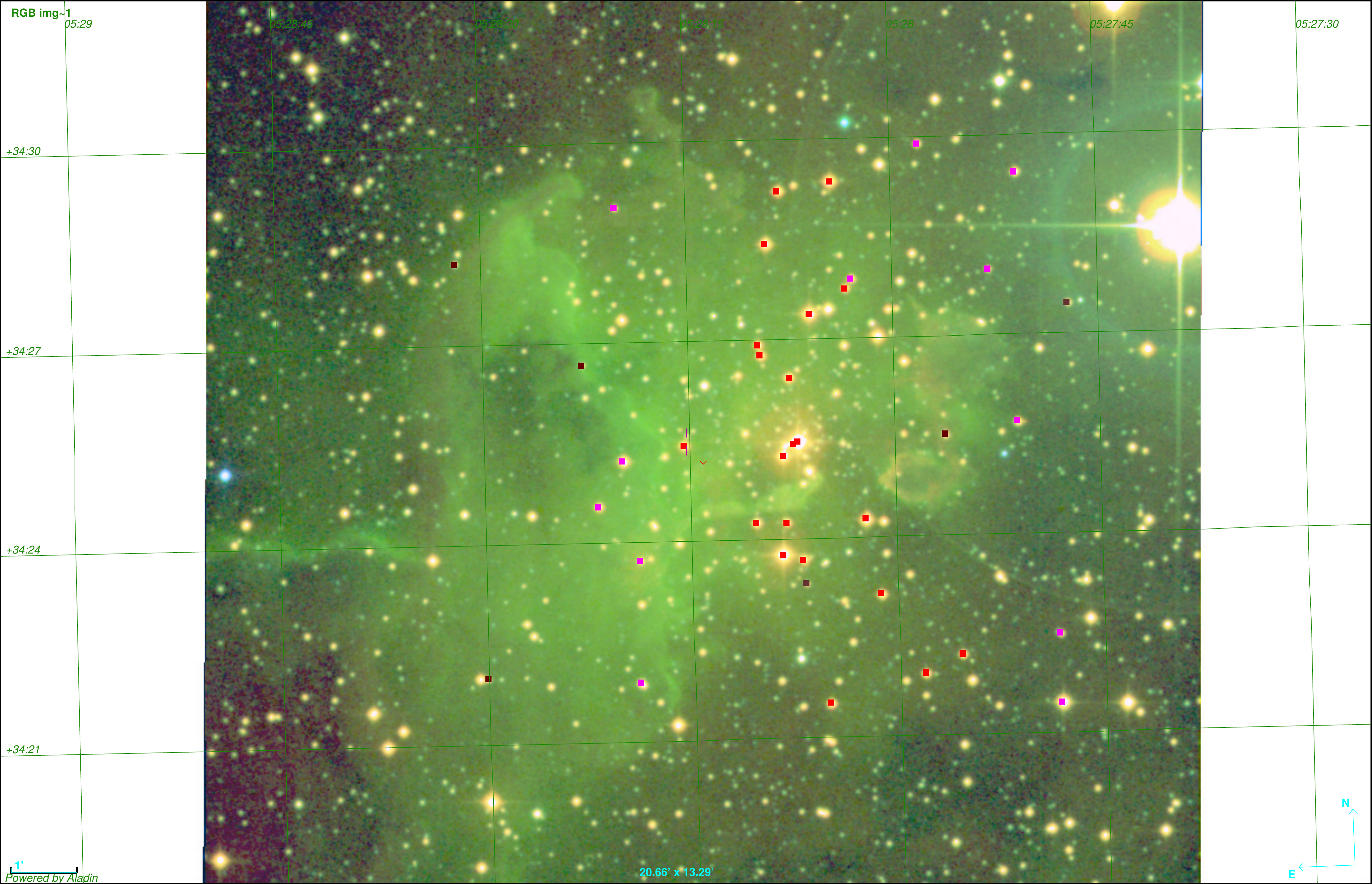}
}
\centering
\caption{Reddening map for B-type stars with optical photometry in the cluster Stock~8. The image, downloaded from Aladin, is a false-colour composite of the three DDS bands, where $R$ (and, hence H$\alpha$) is represented by green colours. Red solid squares represent stars with low reddening [\textbf{$E(b-y)=0.3-0.4$}]. Pink solid squares are stars with moderately reddening [\textbf{$E(b-y)=0.5-0.6$}] and black solid squares are stars with high reddening [\textbf{$E(b-y)=0.7-0.8$}]. North is up and east is left. \label{STOCK8_reddening}} 
\end{figure*}

Firstly, we calculate individual reddenings. We follow the procedure described
by \citet{crawford1970b}: we use the observed $c_{1}$ to
predict the first approximation to $(b-y)_{0}$ with the expression:
$(b-y)_{0}$ = $-0.116$ + $0.097$$c_{1}$. Then we calculate:
$E(b-y)$ = $(b-y)-(b-y)$$_{0}$ and use $E(c_{1})$ = $0.2$$E(b-y)$
to correct $c_{1}$ for reddening $c_{0}$ = $c_{1}-E(c_{1})$. The intrinsic
colour $(b-y)_{0}$ is now calculated by replacing $c_{1}$ with $c_{0}$ in the
above equation for $(b-y)_{0}$ . Three iterations are enough to
reach convergence in the process. Naturally, this procedure only results in physically meaningful values for B-type stars. 
As we can see in the maps (see Fig.~\ref{todo} and Fig.~\ref{STOCK8}), the area of Stock~8 is still embedded in the parental cloud and therefore we expect to find an important degree of differential reddening amongst members. $E(b-y)$ spans values between $0.3$ and $0.8$. In Fig.~\ref{STOCK8_reddening} we plot their surface distribution that can be compared with the extent of the cloud. Most of the stars with relatively low reddening (marked in red) fall along a narrow strip, almost vertical, that seems to coincide with a gap in the dark cloud. Stars wih higher reddening concentrate almost exclusively close to the illuminated rim of the cloud.

With the aid of these individual values, we calculate the intrinsic colour $(b-y)_{0}$, index $c_{0}$, and magnitude $V_{0}$ of the 38 likely B-type members. The values of $E(b-y)$, $c_{0}$ and $V_{0}$ for these likely B-type members are shown in Table~\ref{ebystock8}. 

\begin{table}
\centering
\caption{Values of $E(b-y)$, $c_{0}$, and $V_{0}$ for likely B-type members in Stock~8.\label{ebystock8}}
\begin{tabular}{lccc}
\hline
\hline
\noalign{\smallskip}
Star&$E(b-y)$&$c_{0}$&$V_{0}$\\
\noalign{\smallskip}
\hline
ST14	&	0.46	&	0.70	&	12.28\\
ST21	&	0.53	&	0.36	&	11.14\\
ST23	&	0.37	&	0.11	&	10.01\\
ST25	&	0.35	&	0.23	&	11.05\\
ST28	&	0.46	&	0.98	&	13.30\\
ST34	&	0.35	&	0.21	&	10.29\\
ST35	&	0.78	&	0.55	&	12.34\\
ST45	&	0.54	&	0.98	&	12.86\\
ST46	&	0.46	&	0.48	&	11.95\\
ST50	&	0.40	&	0.62	&	12.20\\
ST60	&	0.68	&	0.50	&	11.84\\
ST61	&	0.37	&	0.04	&	9.24\\
ST69	&	0.32	&	0.71	&	12.23\\
ST71	&	0.32	&	0.52	&	11.69\\
ST72	&	0.76	&	0.69	&	12.62\\
ST80	&	0.38	&	0.43	&	11.14\\
ST86	&	0.62	&	0.31	&	11.12\\
ST89	&	0.77	&	0.96	&	12.49\\
ST91	&	0.37	&	0.28	&	11.17\\
ST93	&	0.30	&	0.02	&	7.61\\
ST96	&	0.51	&	0.18	&	9.87\\
ST97	&	0.30	&	0.03	&	9.67\\
ST104	&	0.59	&	0.18	&	11.03\\
ST110	&	0.43	&	0.62	&	11.22\\
ST111	&	0.36	&	0.70	&	12.25\\
ST112	&	0.33	&	0.28	&	10.22\\
ST119	&	0.57	&	0.33	&	12.23\\
ST120	&	0.31	&	$-$0.05	&	8.44\\
ST124	&	0.32	&	0.51	&	11.87\\
ST128	&	0.71	&	0.45	&	13.03\\
ST134	&	0.32	&	0.58	&	11.77\\
ST142	&	0.55	&	0.96	&	12.57\\
ST146	&	0.43	&	0.69	&	11.74\\
ST148	&	0.75	&	0.40	&	12.91\\
ST151	&	0.56	&	0.81	&	12.72\\
ST152	&	0.38	&	0.57	&	11.84\\
ST159	&	0.31	&	0.46	&	11.86\\
ST163	&	0.31	&	0.51	&	11.97\\
\noalign{\smallskip}
\hline
\end{tabular}
\end{table}

\subsubsection{Determination of the distance for Stock~8}
\label{ana2stock8}

\begin{figure*}
\centering
\resizebox{\columnwidth}{!}{\includegraphics[clip]{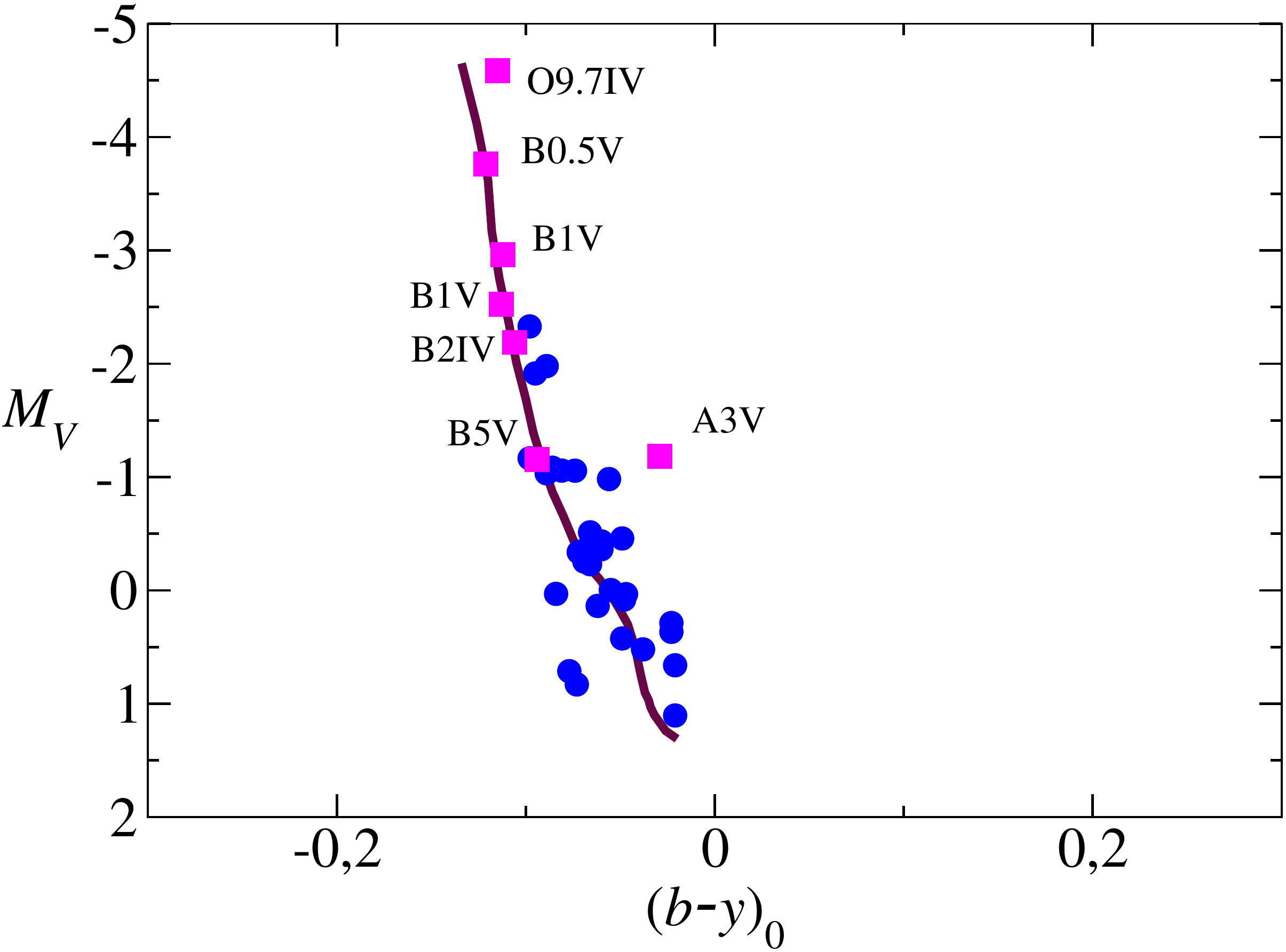}
}
\resizebox{\columnwidth}{!}{\includegraphics[clip]{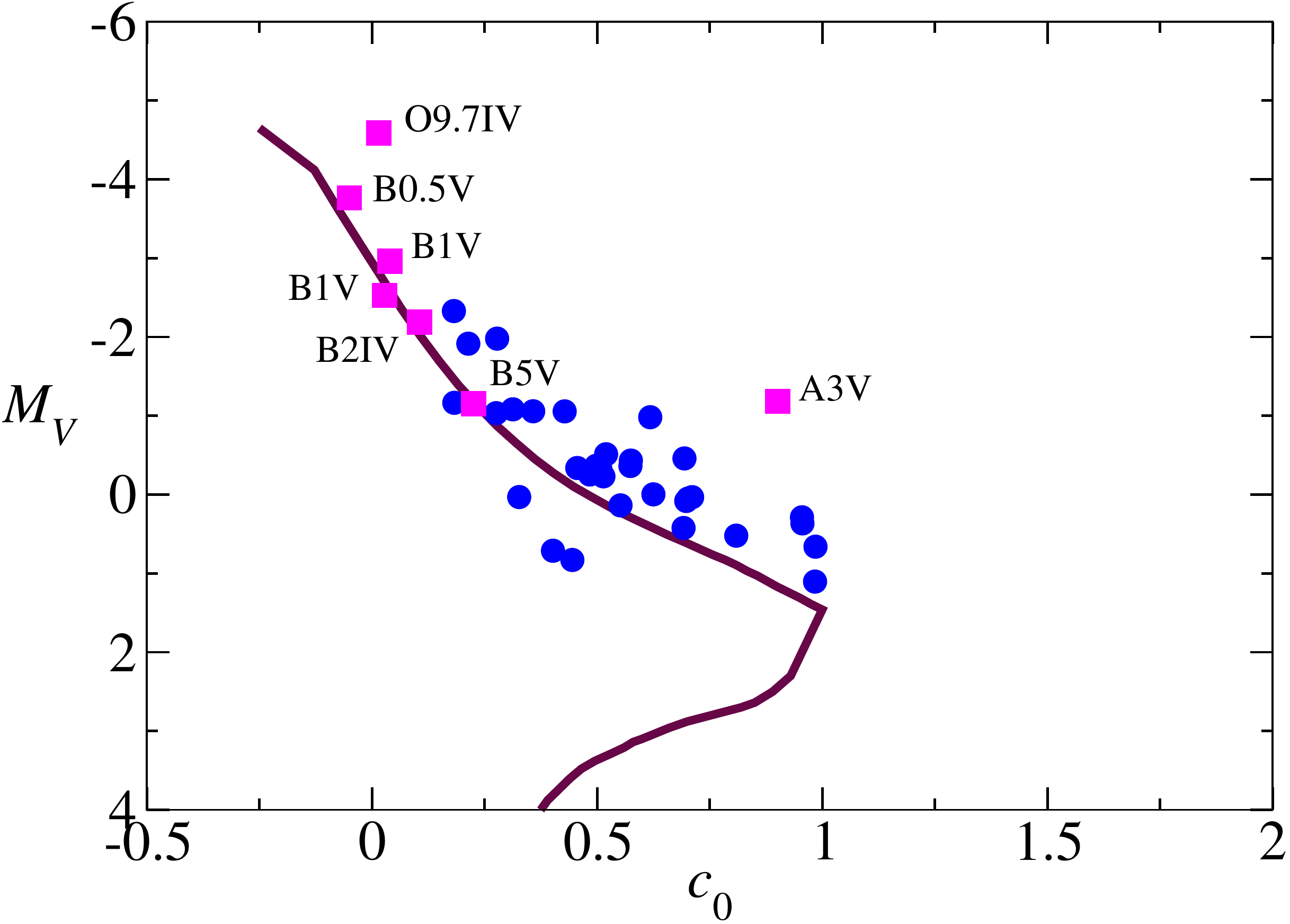}
}
\caption{\textit{Left}: Absolute magnitude $M_{V}$ against intrinsic colour $(b-y)_{0}$ for Stock~8. Solid magenta squares are stars with spectra. The solid line represents the ZAMS from \citet{perry1987}\label{Mvby0_stock8}. \textit{Right}: Absolute magnitude $M_{V}$ against intrinsic index $c_{0}$ for Stock~8. Solid magenta squares are stars with spectra. The solid line represents the ZAMS from \citet{perry1987}.\label{Mvc0_stock8}}
\end{figure*}

We estimate the $DM$ to Stock~8
by fitting the observed $V_{0}$ .vs. $(b-y)_{0}$ zero-age main sequence (ZAMS) and $V_{0}$ .vs. $c_{0}$ ZAMS to the mean calibrations of
\citet{perry1987}. We fit the ZAMS as a lower envelope for the majority of members, deriving a best fit distance modulus of $V_{0}-M_{V}$ = $12.2\pm0.2$ (the error indicates the uncertainty in positioning the theoretical
ZAMS and its identification as a lower envelope; see Fig~\ref{Mvby0_stock8}). This $DM$ corresponds to a distance of $2.80^{+0.27}_{-0.24}$ kpc.

We can check the validity of the distance modulus adopted by eye fitting, by comparing the absolute magnitude $M_{V}$ obtained for the stars with spectral classification and their values in the calibration from \citet{turner1980}. We can say that all values are compatible within errors, except for the A3\,V star that is considered a non-member.

\subsubsection{Optical photometry for Alicante~11}
\label{ana1Alicante11}

\begin{figure*}
\centering
\resizebox{\columnwidth}{!}{\includegraphics[clip]{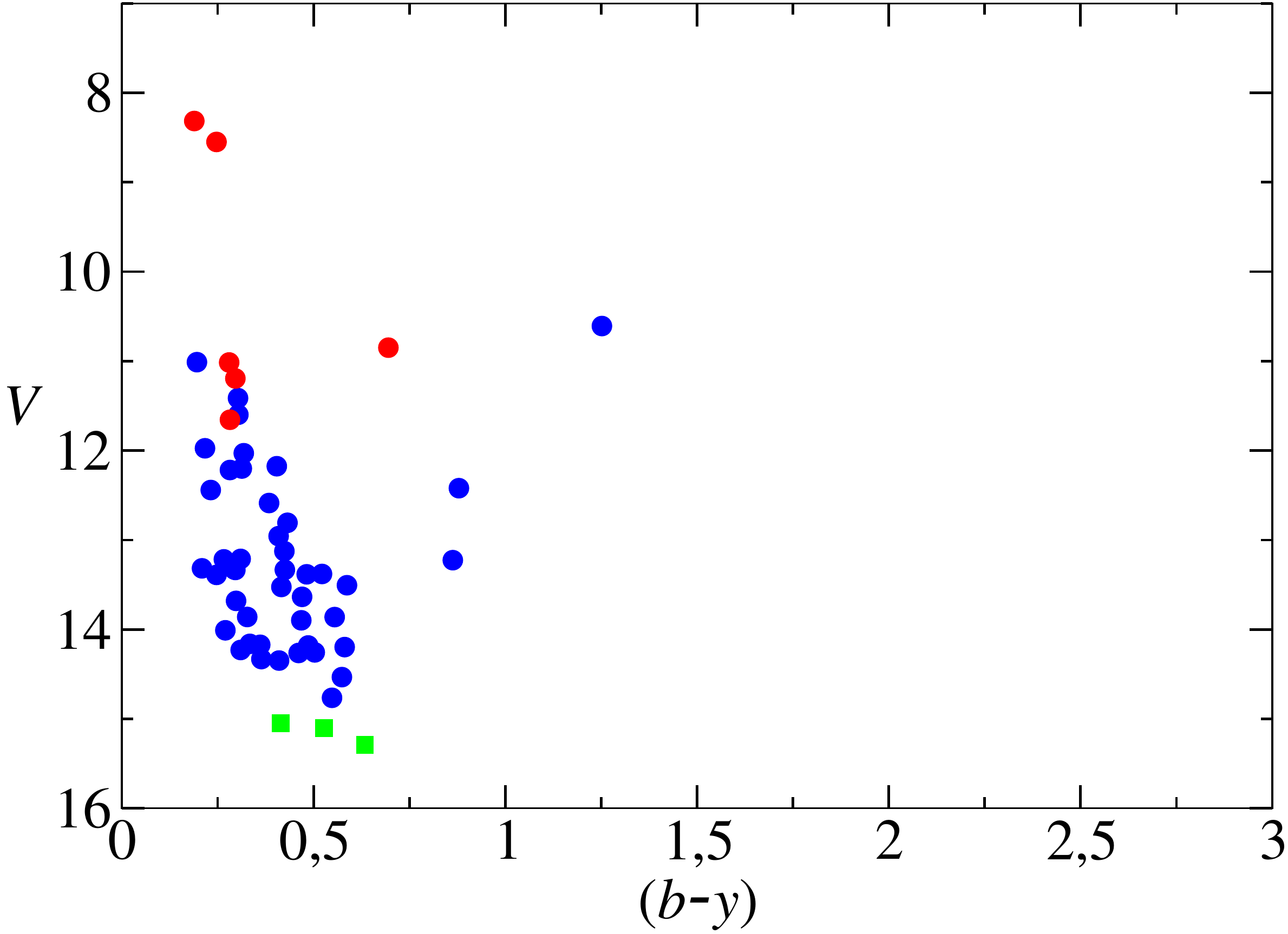}
}
\resizebox{\columnwidth}{!}{\includegraphics[clip]{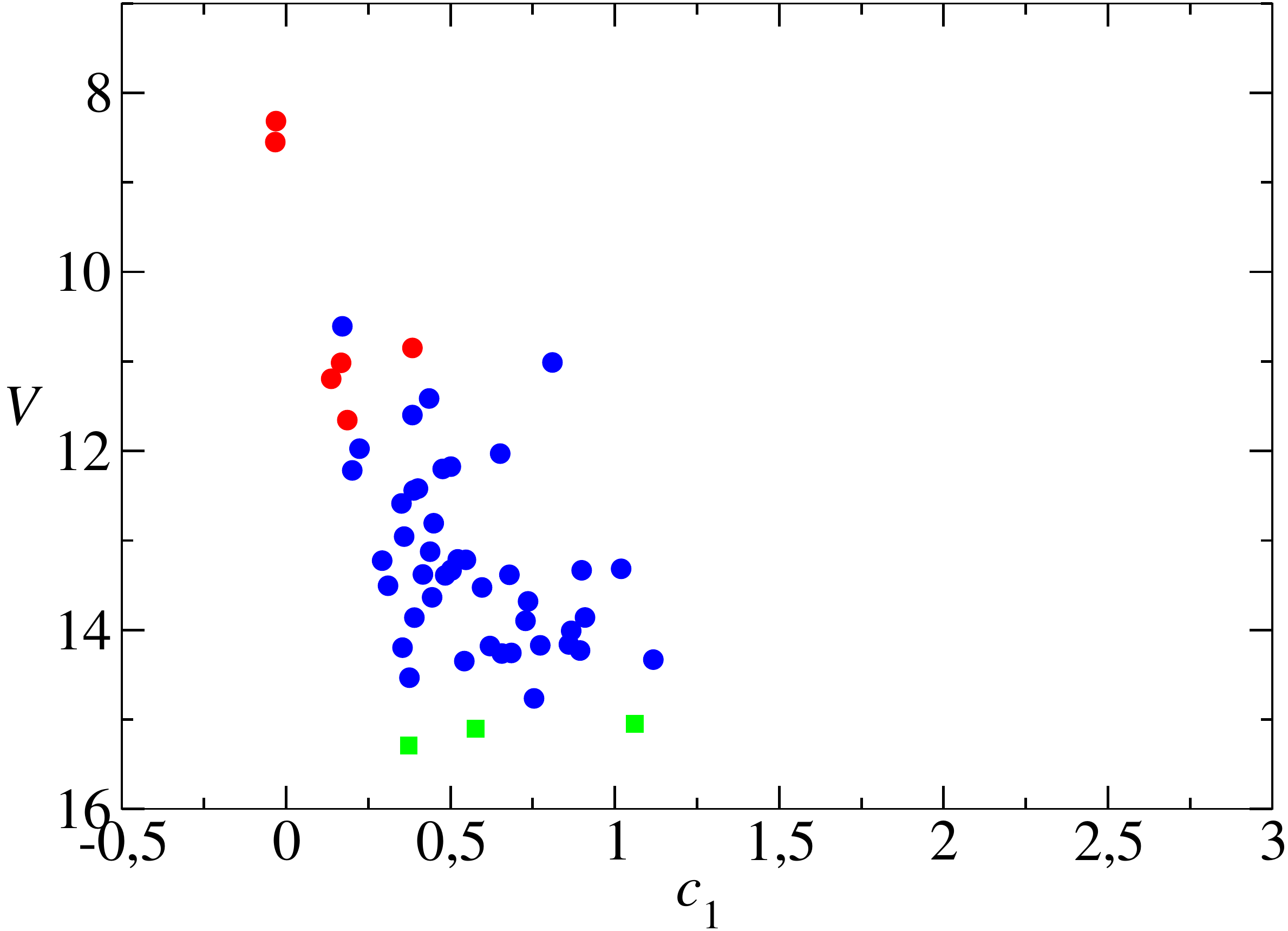}
}
\caption{\textit{Left}: $V/(b-y)$ diagram for all stars in Alicante~11. Red circles represent stars spectroscopically observed and green squares non-member stars\label{VbyAlicante11}. \textit{Right}: $V/c_{1}$ diagram for all stars in Alicante~11. Red circles represent stars spectroscopically observed and green squares non-member stars.\label{Vc1_Alicante11}}
\end{figure*}

We used the same procedure described in Sections~\ref{ana1stock8} and~\ref{ana2stock8} to analyze the photometric data from  Table~\ref{strom-Alicante11}. Firstly, we plotted the $V/(b-y)$ and $V/c_{1}$ diagrams for stars in Alicante~11, where the red points are stars with spectra and green squares are stars that do not ocuppy the expected position in any of the diagrams. These outliers are considered field population and therefore, not used in the subsequent analysis (see Figure~\ref{VbyAlicante11}). Secondly, we selected B-type stars using the $[c_{1}]$-$[m_{1}]$ and $\beta$-$[u-b]$ diagrams. All the early-type stars spectroscopically classified fall on the correct position in both diagrams, except Alicante~11-27 (B8\,V), that we consider a non-member. After this, using the procedure described
by \citet{crawford1970b}, we calculated the individual reddenings $E(b-y)$ for 28 stars that we label as likely members. As seen in Figure~\ref{todo} and Figure~\ref{Alicante11}, the region of Alicante~11 is outside the gas and dust cloud that covers the area where Stock~8 lies. Because of this, stars in this region present rather lower values of reddening compared to those of Stock~8. Most of them have values of $E(b-y)$ around 0.4, which is the same $E(b-y)$ that we found for the stars in Stock~8 falling along the narrow strip that seems to coincide with a gap in the dark cloud. 

With the aid of these individual values, we calculated
the intrinsic colour $(b-y)_{0}$, index $c_{0}$ and magnitude $V_{0}$ of the 28 likely B-type members. The values of $E(b-y)$, $c_{0}$ and $V_{0}$ for these likely members are shown in Table~\ref{EbyAlicante11}.

\begin{table}
\centering
\caption{Values of $E(b-y)$, $c_{0}$ and $V_{0}$ for likely B-type members in Alicante~11.\label{EbyAlicante11}}
\begin{tabular}{lccc}
\hline
Star&$E(b-y)$&$c_{0}$&$V_{0}$\\
\hline
Alicante~11-2	&	0.46	&	0.81	&	11.34\\
Alicante~11-3	&	0.39	&	0.40	&	10.52\\
Alicante~11-5	&	0.36	&	0.84	&	12.30\\
Alicante~11-6	&	0.56	&	0.57	&	11.83\\
Alicante~11-12	&	0.39	&	0.09	&	9.35\\
Alicante~11-15	&	0.37	&	0.43	&	11.75\\
Alicante~11-17	&	0.32	&	0.32	&	11.08\\
Alicante~11-19	&	0.32	&	0.16	&	10.61\\
Alicante~11-21	&	0.35	&	0.83	&	12.74\\
Alicante~11-22	&	0.60	&	0.63	&	12.17\\
Alicante~11-23	&	0.39	&	0.31	&	9.92\\
Alicante~11-25	&	0.52	&	0.55	&	12.01\\
Alicante~11-26	&	0.32	&	0.42	&	12.00\\
Alicante~11-29	&	0.39	&	0.11	&	9.99\\
Alicante~11-30	&	0.34	&	0.48	&	11.77\\
Alicante~11-31	&	0.52	&	0.62	&	11.65\\
Alicante~11-37	&	0.39	&	0.12	&	10.56\\
Alicante~11-38	&	0.41	&	0.69	&	12.41\\
Alicante~11-39	&	0.37	&	0.79	&	12.55\\
Alicante~11-40	&	0.31	&	0.81	&	12.69\\
Alicante~11-41	&	0.48	&	0.26	&	10.54\\
Alicante~11-42	&	0.48	&	0.50	&	11.45\\
Alicante~11-45	&	0.38	&	0.45	&	11.57\\
Alicante~11-48	&	0.51	&	0.35	&	10.60\\
Alicante~11-49	&	0.41	&	0.06	&	9.45\\
Alicante~11-53	&	0.35	&	0.67	&	12.18\\
Alicante~11-56	&	0.31	&	$-$0.09	&	6.96\\
Alicante~11-57	&	0.37	&	$-$0.11	&	6.94\\
\hline
\end{tabular}
\end{table}

\subsubsection{Determination of the distance for Alicante~11}
\label{ana2Alicante11}

\begin{figure*}
\resizebox{\columnwidth}{!}{\includegraphics[clip]{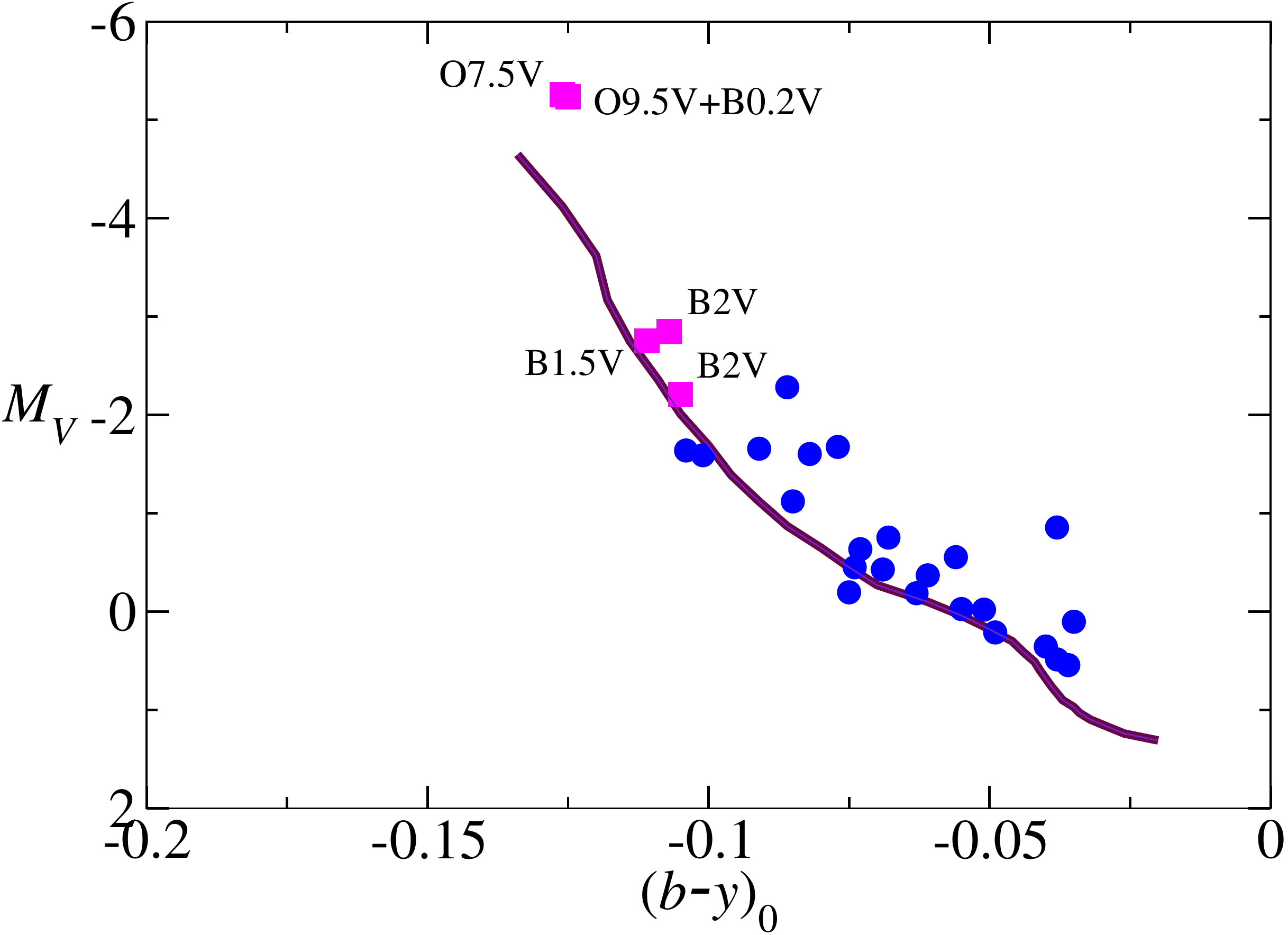}
}
\resizebox{\columnwidth}{!}{\includegraphics[clip]{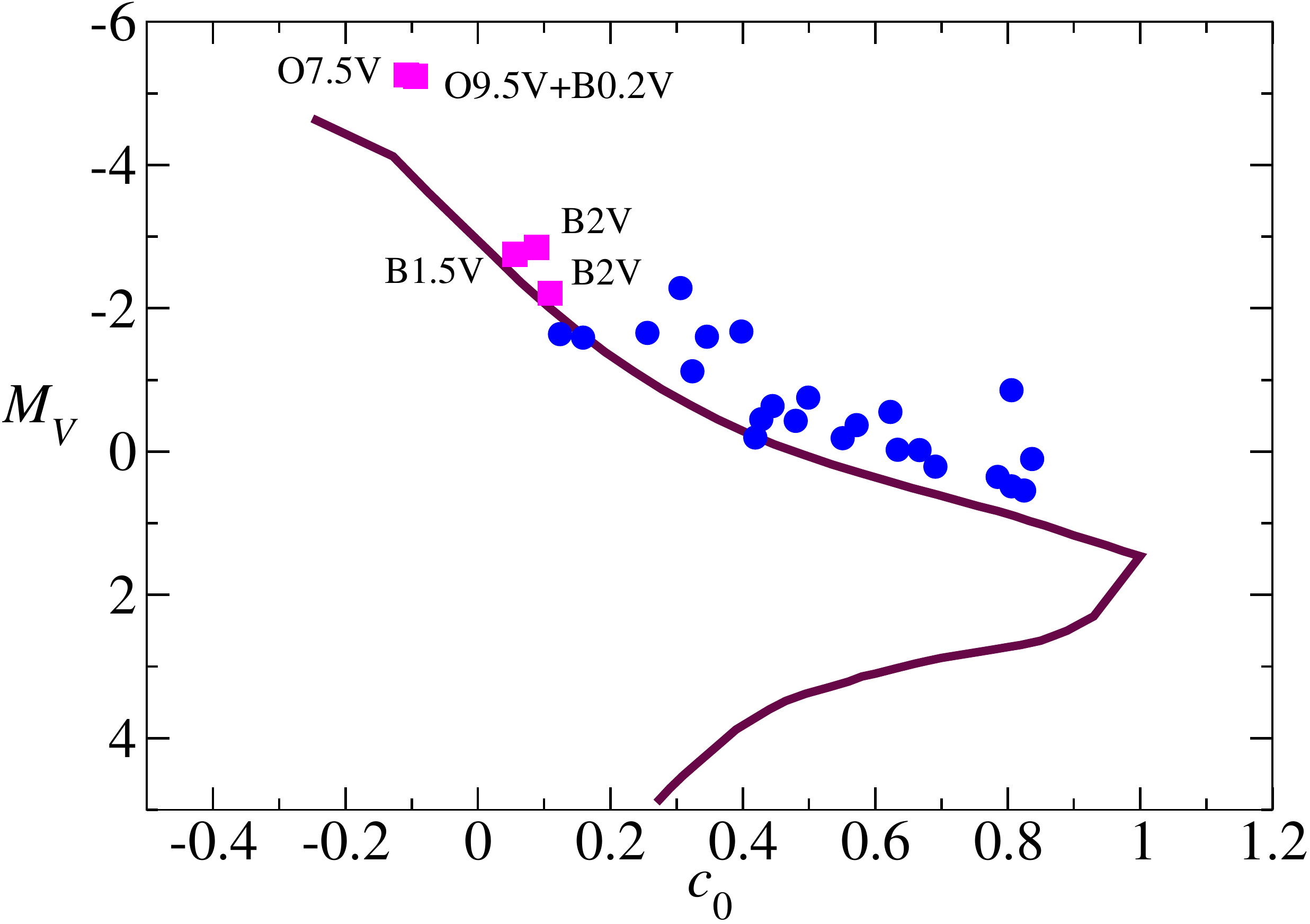}
}
\caption{\textit{Left}: Absolute magnitude $M_{V}$ against intrinsic colour $(b-y)_{0}$ for Alicante~11. Solid magenta squares are stars with spectra. The solid line represents the ZAMS from \citet{perry1987}\label{Mvby0Alicante11}. \textit{Right}: Absolute magnitude $M_{V}$ against intrinsic index $c_{0}$ for Alicante~11. Solid magenta squares are stars with spectra. The solid line represents the ZAMS from \citet{perry1987}.\label{Mvc0_Alicante11} }
\end{figure*}

We estimated the distance modulus to Alicante~11
by using the same procedure as for Stock~8. The fit of the ZAMS as a lower envelope for the majority of members in the $V_{0}$ .vs. $(b-y)_{0}$ and $V_{0}$ .vs. $c_{0}$ diagrams (see Fig~\ref{Mvby0Alicante11}) gives a distance modulus of $V_{0}-M_{V}$ = $12.2\pm0.2$ ($2.80^{+0.27}_{-0.24}$ kpc). We can then check the validity of the distance modulus by comparing the absolute magnitude $M_{V}$ obtained for the stars with spectral classification. All of them give values compatible within errors with the calibration of \citet{turner1980}, except for the binary star IU~Aur, which, as discussed above, is likely a complex system with no less than four OB stars. The absolute magnitude that we obtain for IU~Aur assuming the distance to the cluster is totally compatible with the configuration outlined by \citet{dre94}.

\subsubsection{Optical photometry for Alicante~12}

\begin{figure*}
\centering
\resizebox{\columnwidth}{!}{\includegraphics[clip]{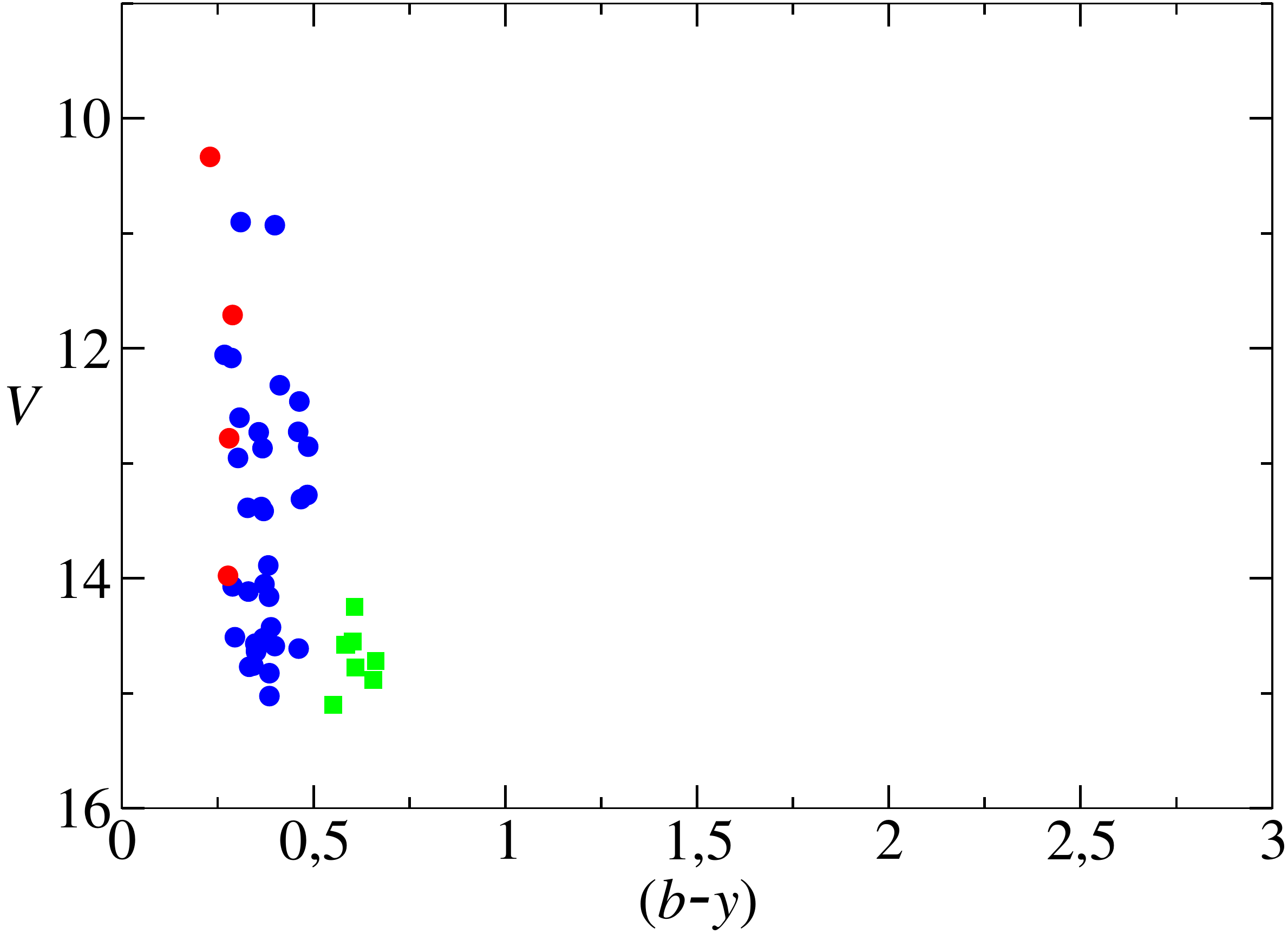}
}
\resizebox{\columnwidth}{!}{\includegraphics[clip]{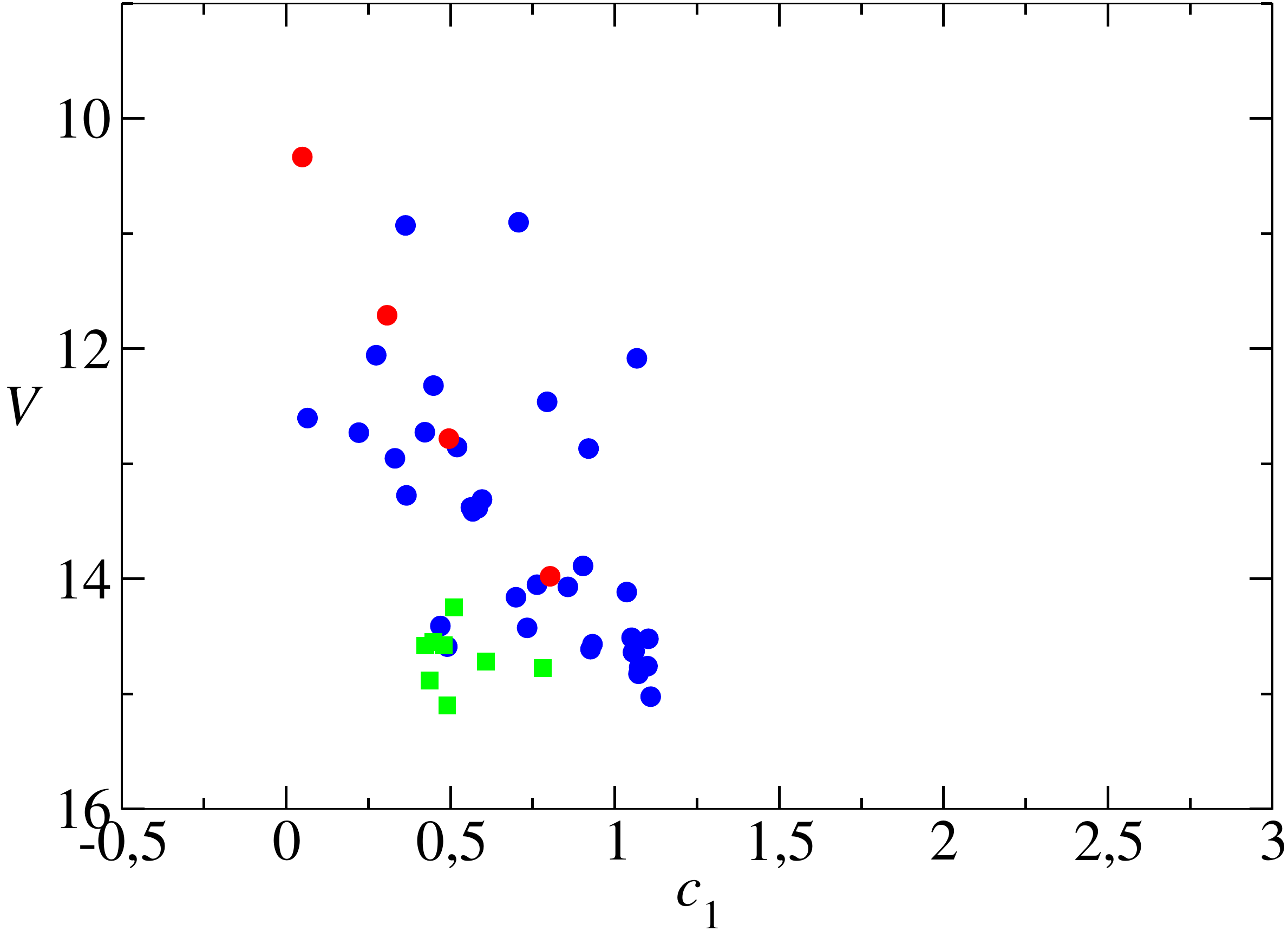}
}
\caption{\textit{Left}: $V/(b-y)$ diagram for all stars in Alicante~12. Red circles represent stars spectroscopically observed and green squares non-member stars.\label{VbyAlicante12}. \textit{Right}: $V/c_{1}$ diagram for all stars in Alicante~12. Red circles represent stars spectroscopically observed and green squares non-member stars.\label{Vc1_Alicante12}}
\end{figure*}

We performed the same analysis done in Sections~\ref{ana1stock8} and~\ref{ana2stock8} for Stock~8 and Sections~\ref{ana1Alicante11} and~\ref{ana2Alicante11} for Alicante~11. We plot the $V/(b-y)$ and $V/c_{1}$ diagrams for stars in Alicante~12, where the red points are stars with spectra and green squares are stars considered non-members by applying the same criteria as in the other clusters (see Figure~\ref{VbyAlicante12}). We selected the B-type members from the $[c_{1}]$-$[m_{1}]$ and $\beta$-$[u-b]$ diagrams, checking that their $V$ values corresponded to their spectral types and, then, we calculated their individual $E(b-y)$. We find that the average value is $E(b-y)=0.4$, similar to the value for the stars in Alicante~11. We can observe in Figure~\ref{todo} and Figure~\ref{Alicante12} that the cluster is not inside the parental cloud. Finally, we calculate the deredenned values $(b-y)_{0}$, $c_{0}$ and $V_{0}$ of the 17 likely B-type members (displayed in Table~\ref{EbyAlicante12}). 

\begin{table}
\centering
\caption{Values of $E(b-y)$, $c_{0}$ and $V_{0}$ for likely B-type members in Alicante~12}
\label{EbyAlicante12}
\begin{tabular}{lccc}
\hline
Star&$E(b-y)$&$c_{0}$&$V_{0}$\\
\hline
Alicante~12-3	&	0.46	&	0.13	&	10.75\\
Alicante~12-13	&	0.54	&	0.49	&	11.01\\ 
Alicante~12-21	&	0.35	&	-0.02	&	8.84 \\
Alicante~12-24	&	0.32	&	0.74	&	12.60\\ 
Alicante~12-30	&	0.38	&	0.23	&	10.07\\
Alicante~12-31	&	0.38	&	0.86	&	12.93\\
Alicante~12-32	&	0.66	&	0.65	&	11.93\\ 
Alicante~12-33	&	0.42	&	0.68	&	12.23\\ 	
Alicante~12-35	&	0.36	&	0.20	&	10.49\\ 	
Alicante~12-36	&	0.39	&	0.25	&	11.26\\ 
Alicante~12-37	&	0.44	&	0.48	&	11.52\\ 
Alicante~12-40	&	0.33	&	0.79	&	12.66\\ 
Alicante~12-41	&	0.36	&	1.03	&	13.21\\ 
Alicante~12-42	&	0.37	&	0.99	&	13.04\\ 
Alicante~12-45	&	0.73	&	0.46	&	11.57\\ 	  
Alicante~12-46	&	0.40	&	0.50	&	11.69\\ 	  	  
Alicante~12-51	&	0.44	&	0.61	&	12.27\\	 
\hline
\end{tabular}
\end{table}

\subsubsection{Determination of the distance for Alicante~12}

\begin{figure*}
\resizebox{\columnwidth}{!}{\includegraphics[clip]{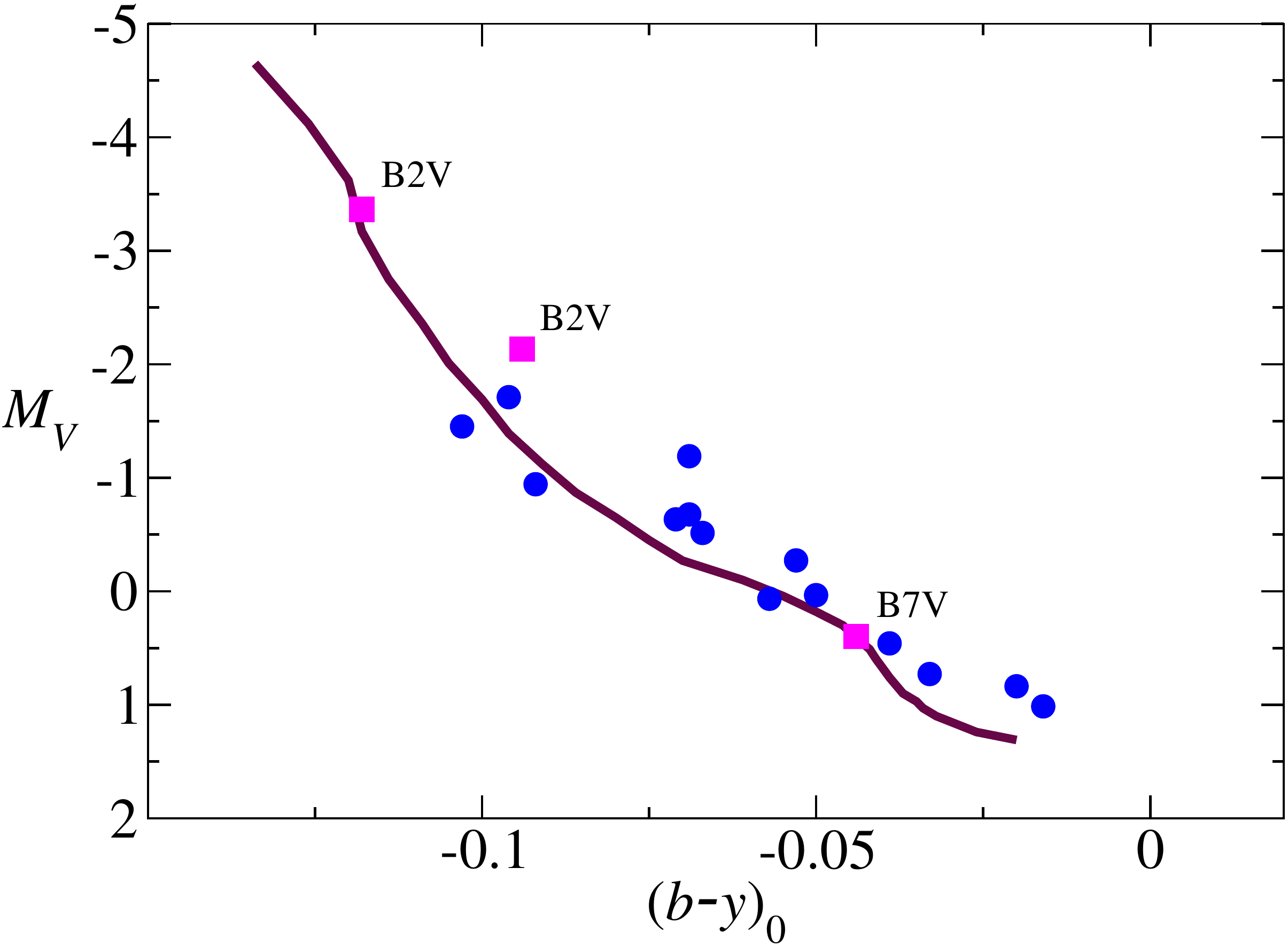}
}
\resizebox{\columnwidth}{!}{\includegraphics[clip]{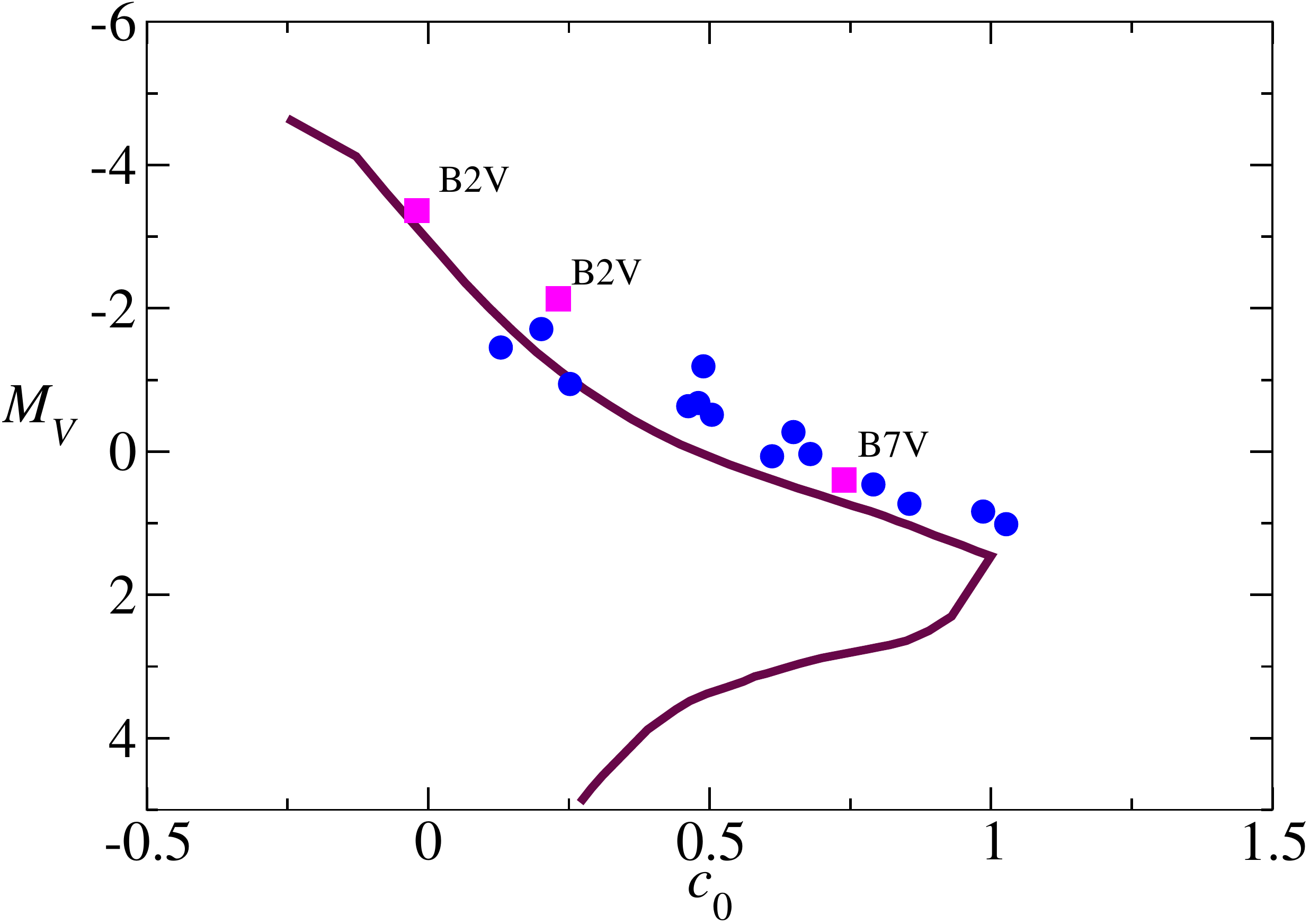}
}
\caption{\textit{Left}: Absolute magnitude $M_{V}$ against intrinsic colour $(b-y)_{0}$ for Alicante~12. Solid magenta squares are stars with spectra. The solid line represents the ZAMS from \citet{perry1987}\label{Mvby0_Alicante12}. \textit{Right}: Absolute magnitude $M_{V}$ against intrinsic index $c_{0}$ for Alicante~12. Solid magenta squares are stars with spectra. The solid line represents the ZAMS from \citet{perry1987}.\label{Mvc0_Alicante12}}
\end{figure*}

As with the other two clusters, we estimated the distance modulus to Alicante~12
by fitting the ZAMS as a lower envelope in the deredenned photometric diagrams (see Fig~\ref{Mvby0_Alicante12}). For this cluster, we also derive a best fit distance modulus of $V_{0}-M_{V}$ = $12.2\pm0.2$ ($2.80^{+0.27}_{-0.24}$ kpc).
Again, we can check the validity of the distance modulus, by comparing the absolute magnitude $M_{V}$ obtained for the stars with spectral classification to the calibration of \citet{turner1980}. All stars are compatible within errors, except for the star classified as B9\,V that we do not consider a member. 

\subsubsection{A common distance for Stock~8, Alicante~11 and Alicante~12}

\begin{figure}
\resizebox{\columnwidth}{!}{\includegraphics[clip]{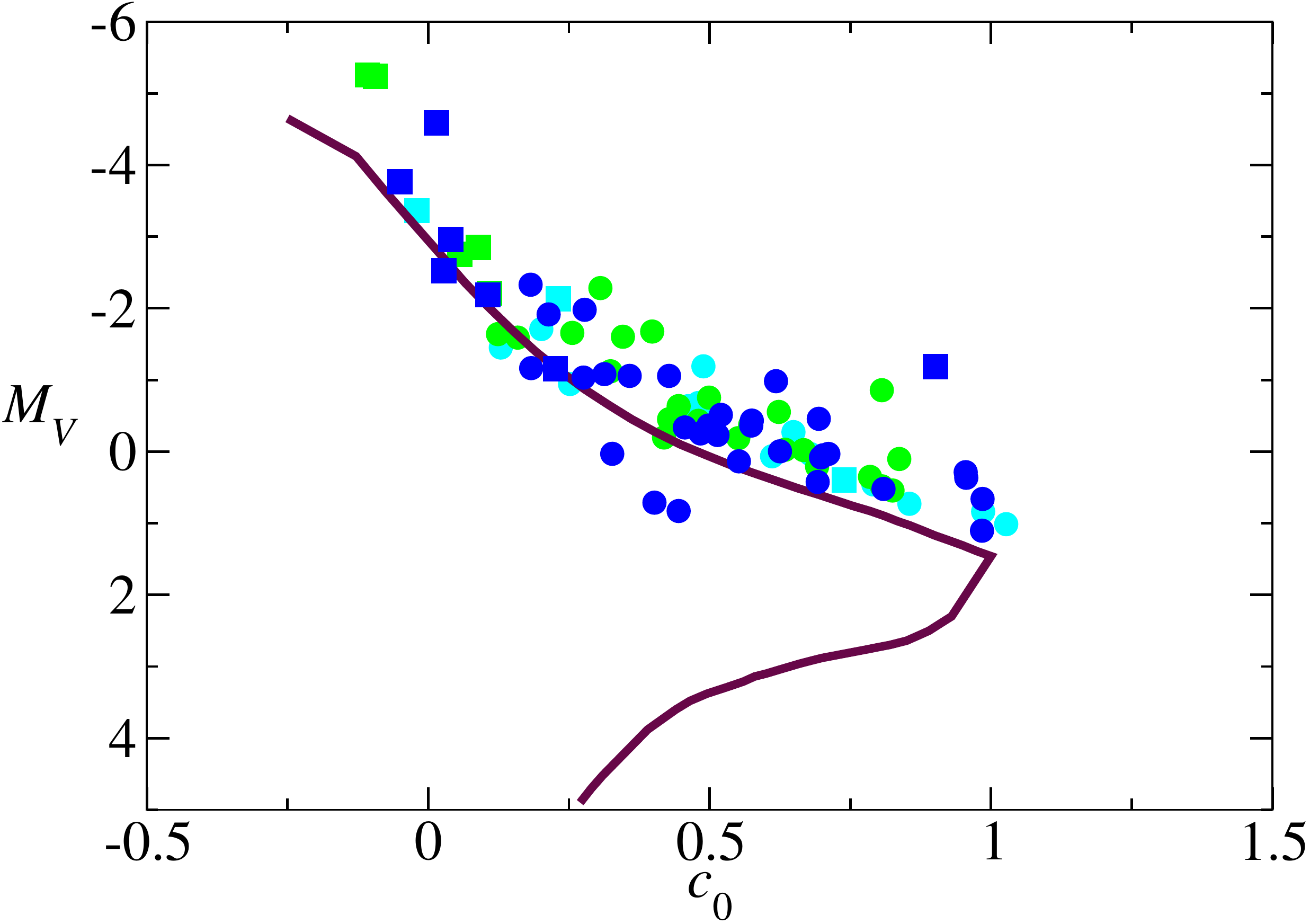}
}
\caption{Absolute magnitude $M_{V}$ against intrinsic index $c_{0}$ for Stock~8 (blue), Alicante~11 (green) and Alicante~12 (cyan). Squares are stars with spectra. The thick line represents the ZAMS from \citet{perry1987}.\label{Mvc0_todos}}
\end{figure}

We have obtained optical photometry for the three different regions indicated in Figure~\ref{todo}. The first region corresponds to the known open cluster Stock~8 that is still partially embedded in its parental cloud. The two other regions are approximately $20\arcmin$ north of Stock~8 and not associated with the nebulosity. Str\"{o}mgren photometry allows us to study the population of these three regions and to determine their distances with accuracy. 
We have calculated a distance for Stock~8 of $2.80^{+0.27}_{-0.24}$ kpc using 38 likely early type members. In the other two regions, we find clear sequences of early type stars that define two new clusters that had not been studied before, and we name Alicante~11 and Alicante~12. We have found 28 likely B-type members for Alicante~11 and 17 likely B-type members for Alicante~12, and we have estimated the same distance as for Stock~8. Their coordinates are shown in Table~\ref{tab1a}. The typical $E(b-y)$ values for B-type members in the two new clusters are very similar, and also similar to those found for the stars placed in a hole in the cloud surrounding Stock~8. This concordance suggests that the two new open clusters are fully detached from their parental cloud, and therefore, the reddening in their directions is entirely caused by intervening foreground material. In Figure~\ref{Mvc0_todos}, we plot the absolute magnitude $M_{V}$ against intrinsic index $c_{0}$ for all the stars considered members in Stock~8 (blue), Alicante~11 (green) and Alicante~12 (cyan). Squares are stars with spectra. We can see that all of them are placed at the same distance modulus of $V_{0}-M_{V}$ = $12.2\pm0.2$, corresponding to a distance of $2.80^{+0.27}_{-0.24}$ kpc

\subsubsection{Bright stars with spectra. Determination of their distance modulus}
\label{brightdistance}

In addition to the three open clusters, there is a large, diffuse population of OB stars scattered over the whole area. Many of them have $UBV$ photometry available in the literature, and all of them have $JHK_{\textrm{S}}$ photometry from 2MASS. We used their spectral types to calculate their distance moduli and check if they are compatible with the common $DM$ that we find for  Stock~8, Alicante~11 and Alicante~12 ($V_{0} -M_{V}$ = $12.2\pm0.2$). For objects with $UBV$ photometry, we utilized the calibration of intrinsic colours from \citet{fitzgerald1970} and the calibration of average absolute magnitude against spectral type from \citet{turner1980}. The results are shown in Tables~\ref{tab:ohp} and~\ref{tab:not}.
 
We find that essentially all the stars with $UBV$ photometry have a distance modulus compatible within errors with the value obtained for the three  open clusters studied. Generally, calibrations are assumed to have an intrinsic dispersion around $\pm0.5$ magnitudes. Possible exceptions would be LS\,V $+34\degr$18, LS\,V $+34\degr$21, and LS\,V $+34\degr$36. For the latter, we have already mentioned that the spectral type is simply the average of two components, one of which is likely more luminous, and so the discrepancy in $DM$ is only apparent. The other two are marked as the less reliable spectral types, as their spectra are very poor. In many cases early-type stars are binaries or multiple systems, and then the spectral type derived from a single spectrum of moderate resolution can be an average of the spectral types of the components. Therefore, we do not think that any of these stars is ruled out as a member of the association, while LS\,V $+34\degr$36 is very likely a member, and one of the main sources of ionizing photons in the area.

For the stars without $UBV$ photometry, we can use the $(J-K)_{\textrm S}$ colour to check that the extinction is not higher than that of members of Alicante~11 and~12, and the $K_{\textrm{S}}$ magnitude to check that their brightness is not incompatible with its spectral type at the $DM$ of the clusters. Thus, we can conclude that all of the isolated stars seen in the area are compatible with the common distance of all three clusters, and therefore they form an OB association extending over at least the region studied, i.e., over $\sim 50\arcmin$ corresponding to $\sim 40$~pc at 2.80~kpc.

\subsubsection{2MASS diagram}
Most of the stars observed lie in areas completely devoid of not only H$\alpha$ nebulosity, but also dust emission (see Fig.~\ref{todo}). The obvious exception is the region of the open cluster Stock~8, which lies in a typical \ion{H}{ii} ``blister'' on the wall of a molecular cloud (see Fig.~\ref{STOCK8}). This area, as can be seen in Fig.~\ref{todo}, also shows strong dust emission. It is therefore important to use infrared wavelengths to detect and study the population hidden by nebulosity. To this purpose, we have used our own TNG near-IR photometry, combined with the 2MASS catalog (see Tables~\ref{coorSTOCK8} and~\ref{Near_IR} for Stock~8).

\begin{figure}
\resizebox{\columnwidth}{!}{\includegraphics[clip]{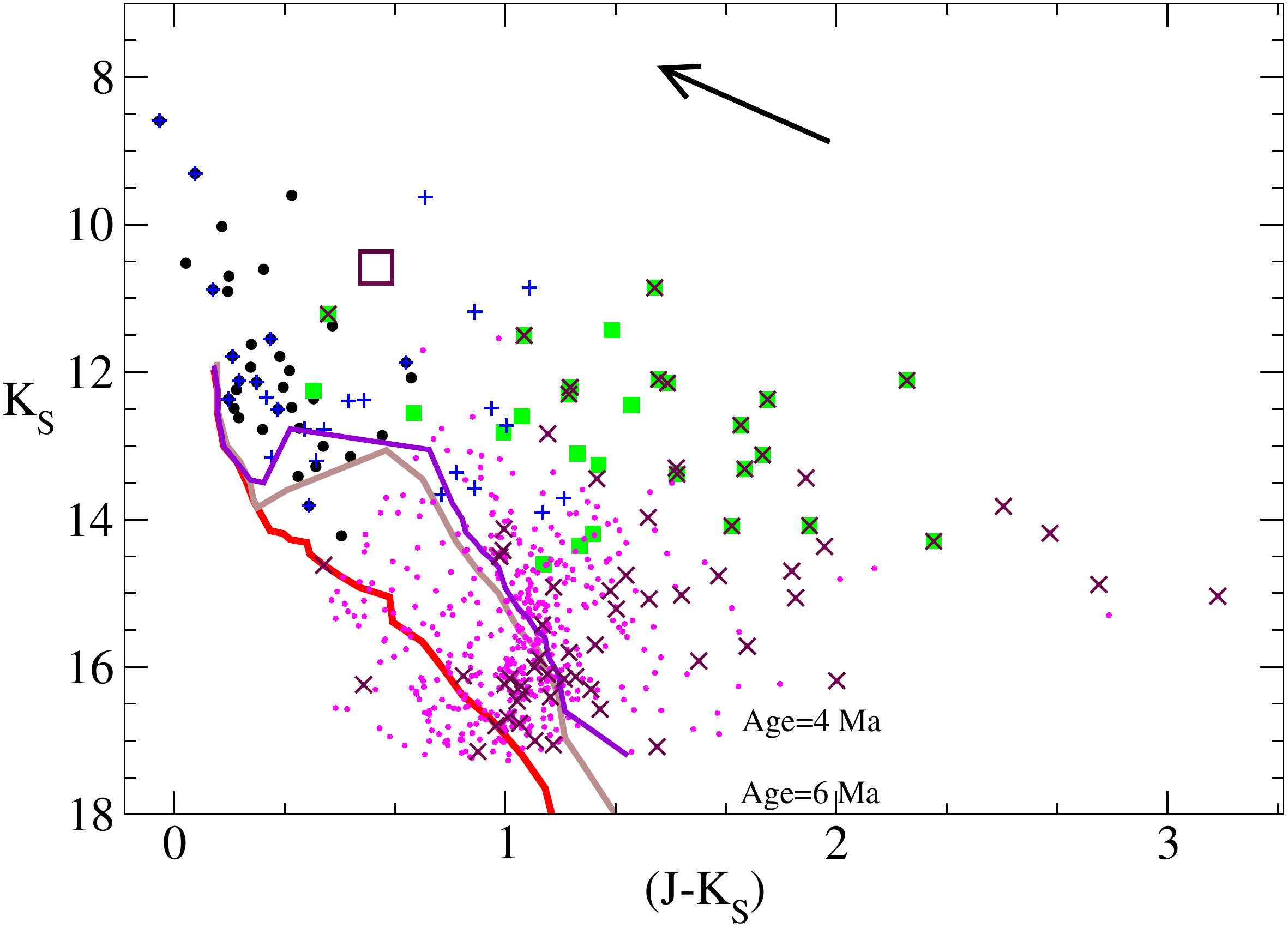}
}
\caption{$K_{\textrm{S}}$ magnitude against $(J-K_{\textrm{S}})$ colour for Stock~8. See the text for a detailed explanation of the symbols. The arrow on the top is the standard reddening vector $A_{K_{\textrm{S}}}=0.67\cdot(J-K_{\textrm{S}})$ of length $(J-K_{\textrm{S}})=0.5$~mag.\label{IR_stock8}}
\end{figure}

Initially, we plot the $K_{\textrm{S}}$ magnitude against the $(J-K_{\textrm{S}})$ colour in Fig.~\ref{IR_stock8}. Since we have already determined a distance modulus of $V_{0}-M_{V}$ = $12.2\pm0.2$ and a minimum $E(b-y)=0.4$ with the optical photometry, we use these values to draw the ZAMS. We use isochrones from \citet{siess2000} with $Z=0.020$ and no overshooting. The ZAMS extends from spectral type M6 until B3. The PMS isochrones for ages $4 \times 10^6$ and $6 \times 10^6$ years are also plotted in Figure~\ref{IR_stock8}. We calculated $E(J-K_{S})=0.3$ using the standard relationships: $E(b-y)=0.74E(B-V)$ and $E(J-K_{S})=0.476E(B-V)+0.007(E(B-V)^2)$. In Fig.~\ref{IR_stock8}, the likely B-type members (from the optical analysis) are represented as black solid circles, while the pink dots are stars with TNG photometry. We can observe that there is clearly a PMS star population associated with the B-type likely members. We can see a gap between $K_{\textrm{S}}$ of 14 and 15 mag, where there are no stars and so we fit the PMS isochrones to the stars on the top of this gap. From their position, we conclude that all the PMS stars formed in a single process of star formation between $4\times 10^6$ and $6 \times10^6$ years ago. We have no stars already leaving the upper-main sequence (because of evolution) and, therefore, we cannot use post-main-sequence isochrones to estimate the age of the cluster. Taking into account the position of the PMS isochrones, we can interpret that the pink dots with $(J-K_{S})<1$ are foreground stars and the rest of the stars are the PMS population of the cluster together with some field contamination that we estimate below. 

To characterize some properties of the PMS stars in Stock~8, we used the Wide-field Infrared Survey Explorer (WISE) catalog \citep{wright2010, jarret2011}. WISE mapped the entire sky simultaneously in four infrared (IR) bands centred at $3.4$, $4.6$, $12$, and $22\:\mu$m (W1, W2, W3, and W4, respectively). We selected those stars that have measurements in W1 and W2 with a precision better than $0.05$. We built the $(K_{\textrm{S}}-W1)$ against $(W1-W2)$ and $(H-K_{\textrm{S}})$ against $(W1-W2)$ diagrams to classify the stars as either objects with disk or diskless, following the criteria described in \citet{koenig2012} and \citet{koenig2014}. We cannot utilise diagrams based on the W3 or W4 bands, because dust emission from the molecular cloud dominates these bands for all sources in this area.

The results of this classification are represented in Fig.~\ref{IR_stock8}. Diskless stars are blue plusses, while stars with disks are green squares. We can observe that the majority of the diskless stars are B-type likely members already on the main sequence or stars not considered members (field stars). All green squares are placed on the right side of the diagram. This is only natural, as they should have an excess in the $(J-K_{\textrm{S}})$ colour arising from the disk. We have to emphasize that the limiting magnitude reached by WISE corresponds in most cases to stars with $K_{\textrm{S}}\approx14$ in this area. Moreover, WISE images have much worse spatial resolution than our IR images. For these two reasons, most of the stars seen in our IR diagram have no WISE counterpart, either because they are too faint or because of confusion. The WISE objects with disks therefore must represent a population of moderately massive PMS stars ($M\geq3M_{\sun}$).

To complement this analysis, we can use the $(J-H)-(H-K_{\textrm{S}})$ diagram, or equivalently the IR $Q$ reddening-free index. Stars with disks or strong infrared excess have negative values of $Q$ \citep{negueruela2007}. The use of this index to identify different types of red luminous stars is discussed by \citet{messineo2012} and \citet{gonzalez2015}. Red giants have $Q\geq0.35$. In Fig.~\ref{IR_stock8}, we plot with crosses stars having $Q<-0.1$. Most of the stars classified as PMS stars with disk from the WISE analysis are also selected by this criterion. In adition, this population extends to fainter magnitudes, covering the same range of $(J-K_{\textrm{S}})$. We interpret them as lower mass PMS objects.

The degree of contamination by background stars is very low. There are very few stars with $Q>0.35$ in our sample. Almost all of them are very faint, with $K_{\textrm{S}}\approx16$\,--\,17~mag. They must therefore be field red dwarfs. In Fig.~\ref{IR_stock8}, we show with a large empty square the approximate position of unreddened red clump stars at the distance of the cluster, using typical photometric intrinsic parameters \citep{alves2000,cabrera2005}. If we project this locus along the reddening vector (indicated on the top of Fig.~\ref{IR_stock8}), we can see that very few stars are compatible with being background red giants. Indeed, most of the stars in this part of the diagram are PMS stars with disk according to the WISE criteria. This is not surprising, as we are looking at a distant cluster in the direction of the Anticenter. The background contamination must thus be low.

\section{Discussion}
\label{discusion}

\begin{figure*}
\resizebox{18 cm}{!}{\includegraphics{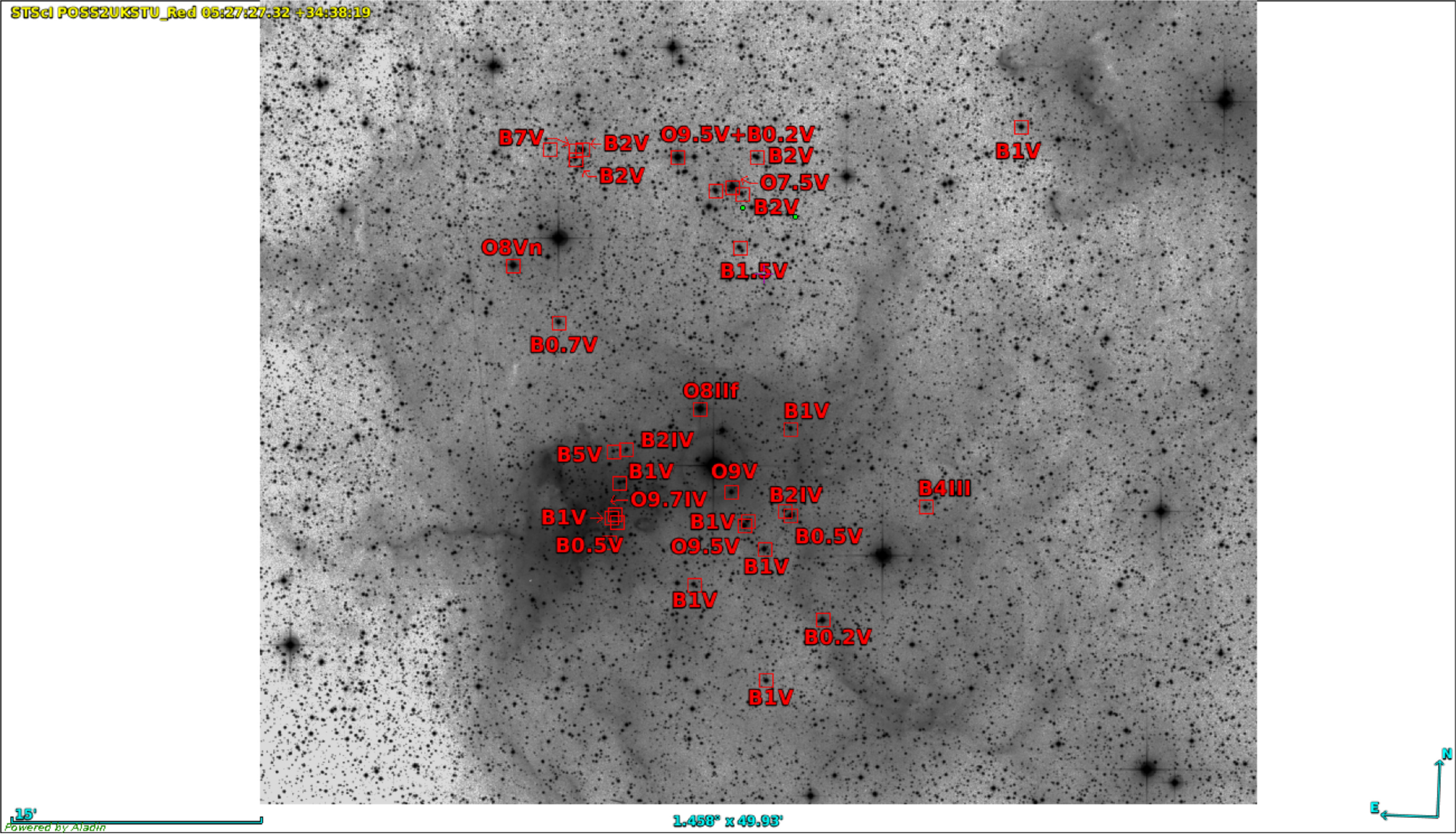}
}
\centering
\caption{Finding chart showing the spectral type distribution for stars in the region. The image is provided from Aladin. Spectral types are listed in Tables~\ref{tab:ohp} and~\ref{tab:not}.The size of field is $1.00\degr \times 49\farcm93$. North is up and east is left.}
\label{spectral_image} 
\end{figure*}

We have studied a region with a size of $\sim40\arcmin \times 40\arcmin$ located close to the known open cluster Stock~8. Much of this area shows evidence for dust emission. We provide spectral types for almost all the cataloged early-type stars in the region, many of which are part of a diffuse population scattered over the field. Our spectral types confirm that all of them are early-type stars (see Tables~\ref{tab:ohp} and~\ref{tab:not}). The spectral types range between O7.5\,V and B4\,III. The surface distribution of all the OB stars in the region can be seen in Fig~\ref{spectral_image}. The population can be roughly divided in two clumps. The first one, to the north, comprising Alicante~11 and Alicante~12 and a few other stars, lies in the area devoid of dust emission (Fig.~\ref{todo}) and with weak H$\alpha$ emission (Fig.~\ref{spectral_image}). The second one, on the southern half of the field, is formed by Stock~8 and a large population of stars scattered to the west of the cluster. This second group lies in an area showing evidence for diffuse dust emission and stronger H$\alpha$ emission. The spectral types observed, however, do not evidence any difference in age between the two groups, with the earliest spectral type found in the northern clump.

We have obtained a common distance of $2.80^{+0.27}_{-0.24}\:$kpc ($DM=12.2\pm0.2$) for the three open clusters in the area studied (Stock~8, Alicante~11 and Alicante~12) and an age between $4-6\:$Ma for Stock~8. These values are compatible with the previous estimation of the distance ($DM=12.4$) for Stock~8 by \citet{mm71}. They show  rather worse agreement with the values of $DM=11.4$ of \citet{malysheva1990} and $d= 2.05^{+0.1}_{-0.1}\:$kpc ($DM=11.6$) of \citet{jose2008}. The discrepancy with \citet{malysheva1990} is likely due to her derivation of an age of 12~Ma (at least double than the value that we find, 4\,--\,6~Ma). Her analysis, based on photographic photometry, cannot be so accurate as Str\"{o}mgren photometry combined with spectroscopy. The presence of several O-type stars in the area and the PMS population in Stock~8 rule out such a high age. In the case of \citet{jose2008}, the discrepancy can have two procedural reasons. The main one is the value adopted for the reddening in calculating the distance. The reddening is variable because of the presence of the parental cloud, and for this reason, we have used individual values for B-type members. The second reason is our choice of the ZAMS as a lower envelope to the position of cluster stars. Conversely, their fit in their Fig.~9 is almost an upper envelope to the early-type stars (perhaps, again, because of their value for the reddening). Given the width of the main sequence, this difference alone can account for 0.3 or 0.4~mag in the $DM$ derived.

\subsection{Morphology of the region}

The most remarkable feature in the region is the bright H$\alpha$ and dust shell lying immediately to the east of Stock~8 (Figures~\ref{todo} and~\ref{STOCK8_reddening}). This shell, which is also bright in radio \citep{jose2008}, must be ionized by some of the OB stars in its vicinity. \citet{jose2008} suggest that LS\,V $+34\degr$29, the brightest star in Stock~8, can emit enough ionizing photons to account for the photoionisation of the whole shell. In spite of this, they argue that the O-type stars LS\,V~$+34\degr$18 and LS\,V~$+34\degr$21 can contribute to the ionizing flux. However, we note that these two stars are much more reddened than any other OB star in the area (compare their $V$ and $K_{\textrm S}$  in Table~\ref{tab:ohp}), and seem to be partially immersed in dust in the WISE image. 

Our new spectral type for LS\,V~$+34\degr$23 (HD~35633), O8\,II(f), undoubtedly means that this is the main source of ionizing photons in the area. The morphology of the shell is perfectly compatible with this (see Figures~\ref{todo} and~\ref{STOCK8_reddening}). This star was classified as B0\,IV by \citet{morgan53}. This classification is almost certainly due to a misidentification. \citet{sergio14} classify it as O7.5\,II(f)(n) in good agreement with our value. They measure a $v\sin i \approx 170\:\rm{km}\,\rm{s}^{-1}$. LS\,V~$+34\degr$23 is the only star in the area that is clearly evolved away from the ZAMS\footnote{As commented before, LS\,V $+34\degr$29 (BD~$+34\degr$1054) has spectral type O9.7\,IV (see Table~\ref{tab:ohp}), but this is very likely a composite spectrum.}. As such, it provides the only estimate of the age of the cluster from the massive stars. Comparison to typical spectral types in very young open clusters suggests an age of $\sim 4-5$ Ma. This value matches well the age obtained by using the PMS isochrones in Fig.~\ref{IR_stock8}.

A second feature of high interest is the apparent filamental structure extending to the east of the \ion{H}{ii} shell that \citet{jose2008} call the Nebulous Stream. This structure contains a stream of embedded sources and an obscured cluster known as CC~14 (see Figure~\ref{todo}). Even though \citet{ivanov05} studied the possibility that this is a massive and distant cluster, \citet{jose2008} conclude that it is a small embedded cluster, associated to the stream of young objects. They also find that all these YSOs are younger than Stock~8. Indeed, the images suggest that the filament is not directly associated to Stock~8. The wide-field H$\alpha$ image (see Fig.~\ref{spectral_image}) shows a bright filament of \ion{H}{ii} emission that is probably marking a photoionisation front that appears almost perpendicular to our line of sight in projection. The only star that can ionize this filament is LS\,V $+34\degr$36 (BD~$+34\degr$1058). This star, situated to the north of Stock~8 and in relative isolation, is an SB2 with integrated spectral type O8\,V that probably contains a moderately-luminous O-type star.

As discussed in Sect.~\ref{brightdistance}, the hypothesis that all the OB stars that we have observed are at the same distance is perfectly consistent with observations. Unfortunately, there is no CCD optical photometry available in the literature for the area to the west of the cluster containing a large number of early-type stars to study if there is an underlying population of stars accompanying the concentration of 6 OB stars in an area $\sim3\arcmin\times4\arcmin$ (LS\,V $+34\degr$11, 13, 15, 16, 18 and 21), which could represent a diffuse cluster similar to Alicante~11. However, observation of 2MASS images and simple star counts in the 2MASS catalog suggest that, even though most of the OB stars are outside the area covered by our photometry, most of the intermediate-mass stars are inside the area covered. This configuration is very reminiscent of that found in the very young open cluster NGC~1893, which also sports a number of OB stars at some distance of the cluster core, in a region where there are essentially no intermediate-mass stars \citep{negueruela2007}. 



\subsection{History of star formation in the region}

\citet{kang2012} have suggested the presence of a large-scale star-formation structure in this area. They present the results of \ion{H}{i}~21\:cm-line observations to explore the nature of the high-velocity \ion{H}{i} gas at $l\approx173\degr$. They designated this feature as Forbidden Velocity Wing (FVW) 172.8+1.5. Since this direction lies very close to the Galactic Anticentre, the Galactic rotation curve predicts very low radial velocities, and high-velocity components are not expected. FVW 172.8+1.5 seems to be associated with the \ion{H}{ii} complex G173+1.5, interpreted as one the largest star-forming regions in the outer Galaxy.  \citet{kang2012} consider that the complex is composed by two groups of Sharpless \ion{H}{ii} regions and a surrounding population of OB stars, distributed along a radio continuum loop of size $4.4\degr\times3.4\degr$. The geometry of the area (in two dimensions) can be seen in their Fig.~5. 

Two groups of \ion{H}{ii} regions can be seen in the area studied by \citet{kang2012}. To the northeast, the \ion{H}{ii} regions Sh~2-231, Sh~2-232, Sh~2-233 and Sh~2-235 have been claimed to be associated with a single molecular cloud. The distances determined by the spectrophotometry of the exciting stars of the individual \ion{H}{ii} regions are between $2.3$~kpc and $1.0$~kpc, while their radial velocities are: $-18.1\pm0.9$, $-23\pm0.5$, $-18.4\pm0.5$ and $-18.8\pm1.7$, respectively \citep[from the catalogue of][]{blitz1982}. \citet{kang2012} adopt a distance of 1.8~kpc, after \citet{evans1981}. However, a recent comprehensive study of the extinction in this area \citep{straizys2010} finds a lower distance of 1.3~kpc for the complex (though these authors suggest that Sh~2-231 may not belong to the complex, but rather be a background object with a much higher distance). Since this distance is compatible with the estimate for Aur~OB1 found by \citet{humphreys1978} from a few isolated OB stars, \citet{straizys2010} identify the complex as the core of Aur~OB1. To the southwest, the \ion{H}{ii} regions Sh~2-234 and Sh~2-237 are associated with the open clusters Stock~8 and NGC~1931, respectively.

\citet{kang2012} assume that all the high-velocity structures in the area are connected, concluding that the \ion{H}{i}-line feature FVW 172.8+1.5 is well correlated with a radio continuum loop, and the two seem to trace an expanding shell. The expansion velocity of the shell would be $55\:{\rm km}\,{\rm s}^{-1}$ (well beyond the velocities allowed by Galactic rotation) and the kinetic energy of the shell would then be $2.5\times10^{50}$~erg, suggesting that it represents an old ($\sim 0.33$~Ma) supernova remnant produced inside this complex. They propose that the progenitor belonged to a stellar association near the center of the shell, and that this association could have triggered the formation of the OB stars currently exciting the \ion{H}{ii} regions on both edges of the shell. The \ion{H}{ii} complex G173+1.5 would then be an excellent example of sequential star formation over several stellar associations. 

Our data do not seem to provide support for this interpretation. A close connection between Stock~8 and NGC~1931 seems unlikely, in spite of their proximity on the sky. The radial velocities of the associated molecular clouds are $-13.4\pm0.7\:{\rm km}\,{\rm s}^{-1}$ and $-4.3\pm0.7\:{\rm km}\,{\rm s}^{-1}$, respectively \citep[as measured by CO velocities in][]{blitz1982}. \citet{fich1990} observed the radial velocity and line width of H$\alpha$ emission from 284 objects listed in Galactic \ion{H}{ii} regions catalogs. The values measured for Sh 2-234 and Sh 2-237 were $-14.3\pm0.5\:{\rm km}\,{\rm s}^{-1}$ and $+1.4\pm0.4\:{\rm km}\,{\rm s}^{-1}$, respectively. Differences of a few ${\rm km}\,{\rm s}^{-1}$ are regularly observed between the cold and ionized components \citep{fich1990}, as is the case for Sh~2-237. However, the values of radial velocities from both measurements for Sh~2-234 (Stock~8) match within errors. This does not suggest the presence of perturbed gas, as would be expected if star formation was triggered by a shock wave. Moreover, the geometry of the shell, as discussed in the previous section, clearly demonstrates that it is ionized by stars to its west, and this is very difficult to reconcile with a shock wave coming from the east as the triggerer of star formation.

Independently of these measurements, the main difficulty with the scenario proposed by \citet{kang2012} is the lack of cataloged OB stars inside the giant shell. If the supernova happened 0.33~Ma ago, it cannot have triggered the formation of a cluster that is $\ga4$~Ma old. \citet{kang2012} suggest that the progenitor of the supernova was part of a population that might have triggered large-scale star formation, but there are very few stars that could belong to this population. The only O-type star that has a $DM$ perhaps compatible is the multiple system LY~Aur \citep{mayer2013}.

We have searched the catalog of \citet{wouterloot1989} for other molecular clouds that could be associated to Sh2-234. \citet{wouterloot1989} looked for  CO$(1-0)$ emission in the direction of 1302 IRAS sources with colours of star forming regions. The only measurement in the vicinity of Stock~8 is IRAS~05274+3345, which is located $\sim27\arcmin$ to the southwest of NGC~1931 and $\sim50\arcmin$ to the southeast of Stock~8. This compact \ion{H}{ii} region is associated with a small young embedded cluster, containing perhaps two massive stars \citep{chen2005}. Its radial velocity of $-3.90\pm0.20\:{\rm km}\,{\rm s}^{-1}$ is identical within errors to that of NGC~1931, and quite different from that of Sh2-234. Therefore, given the small separation in the sky, it seems that NGC~1931 and IRAS~05274+3345 are physically connected, but Stock~8 is not. The distance to NGC~1931 has recently been found to be 2.3~kpc by two different groups \citep{pandey2013,lim2015}. The difference in radial velocity and the morphology of WISE images is compatible with a slightly longer distance for Stock~8.

The age that we find for Stock~8 is relatively high for a cluster still associated with nebulosity. It is also remarkable that most of the O-type stars in the area are not directly associated with the nebulosity. In particular, Alicante~11, which contains at least two O-type stars, including the earliest spectral type in the region, is completely devoid of any nebulosity. These characteristics suggest that star formation has essentially stopped in this area. The \ion{H}{ii} region that we see represents the last remnant of a large cloud, illuminated by the O-type stars that formed from this cloud, and Stock~8 is emerging from the parental cloud as it photo-dissociates. This scenario is supported by the radio-continuum and CO images of the area presented by \citet{kang2012}, which show that the molecular cloud associated with Stock~8 is very small compared to the GMC associated with IC~410, or even that connected to Sh~2-232\,--\,235.

Remarkably, the embedded population seems to have the same age as the OB population -- with the possible exception of objects in the Nebulous Stream. This is in contrast to the situation observed in W3, or in NGC~1893, where massive PMS stars (Herbig Be stars) are spatially segregated from main-sequence objects. Therefore the area surrounding Stock~8 does not seem to have fostered sequential star formation.

\subsection{Position in the Milky Way}

The current understanding of Galactic structure towards the Anticentre and in the third quadrant is very poor, in spite of many important recent developments. Several models of the structure and dynamics of the Galaxy have been published in the past few years, based on improved information \citep[e.g.][]{bobylev2010, mcmillan2010, honma2012,vallee2015}. In particular, \citet{reid2014} have 
used more than 100 trigonometric parallaxes and proper motions for water and methanol masers associated with high-mass star-forming regions, measured by Very Long Baseline Array (VLBA), VLBI Exploration of Radio Astrometry (VERA) and the European VLBI Network (EVN), to fit axially symmetric models of the Milky Way.
In their model (see their Fig.~1, and also \citealt{choi2014}), the Perseus arm lacks tracers between $\sim135\degr$ and $\sim170\degr$. \citet{choi2014} identify a single tracer close to Stock~8, the small embedded cluster IRAS~05168+3634 ($l=170\fdg7$), with a parallax distance of $1.9\pm0.2$~kpc and a radial velocity of $v_{\textrm{LSR}}=-15.5\pm1.9\:{\rm km}\,{\rm s}^{-1}$ \citep{sakai2012}. These values seem compatible with the Sh~2-231\,--\,235 complex, which is located $\sim4\degr$ away. \citet{reid2014} also identify a number of tracers of the Perseus arm between $l=184\degr$ and~$193\degr$, all of them with distances $d\la2$~kpc, slightly shorter than their best fit for the arm, which gives a distance of $\sim2$~kpc around $l\sim180\degr$. The typical width of the spiral arm should be around 0.5~kpc on each side of its mid-point.

With these characteristics, our distance to Stock~8 is just compatible with a location in the Perseus arm, within errors. We have to caution, however, that the simple picture of Galactic structure based on four main arms is not shared by many authors. The Orion arm, as observed by \citet{xu2013} is a major structure. According to some authors, it extends for many kpc towards $l=240\degr$\,-\,$250\degr$, intersecting the Perseus and even the Outer arm \citep{vazquez2008, costa2015}. On the other hand, the Perseus arm is very poorly traced beyond $l=193\degr$. All the possible tracers given by \citet{reid2014} lie at very large distances from their fit. Conversely, a large number of stellar tracers seem to lie between the Orion and the Perseus arm all the way between $l=180\degr$ and $l=240\degr$, perhaps suggesting a more fluffy structure.

\subsection{Is there an association Aur~OB2?}

As mentioned in the Introduction, in the classical list of OB associations by \citet{humphreys1978}, Aur~OB2 was described as composed by the open clusters Stock~8 and NGC~1893, together with some OB stars lying between them. However, all modern studies of NGC~1893 that make use of its upper main-sequence obtain distances $\sim5$~kpc \citep{fitzsimmons1993, massey1995, marco2002, negueruela2007}. Other works using fits to the low-mass star sequence tend to give lower distances \citep[around 3.5~kpc;][]{sharma2007,prisinzano2011, lim2014}. The reasons for this discrepancy are unclear. \citet{lim2014} attribute it to the inaccuracies in the spectroscopic parallaxes. However, given the almost universal (hidden) multiplicity of OB stars, spectroscopic parallaxes based on calibrations will most likely underestimate (rather than overestimate) the distance. Whatever the case, the distances to NGC~1893 and Stock~8 are not compatible, and the radial velocities of their associated \ion{H}{ii} regions are quite different. Even at 3.5~kpc, NGC~1893 would be too far away to belong to the Perseus arm as traced in all modern models.

In this paper we have shown that three small clusters located close to the \ion{H}{ii} region Sh2~234 have the same photometric distance and that there is a large population of OB stars in the same area whose spectroscopic distances are compatible. However, the picture of the region of the Galactic Plane around $l=173\degr$ is complex. Several \ion{H}{ii} regions have distances ranging from 1.3~kpc to 2.8~kpc. Their radial velocities range from $+1.4$ to $-25.7\:{\rm km}\,{\rm s}^{-1}$ \citep{fich1990}, suggesting that they are determined mostly by peculiar motions rather than by the Galactic rotation curve (which predicts velocities between $-5$ and $-8\:{\rm km}\,{\rm s}^{-1}$ for this range of distances). Therefore we are likely seeing a number of star forming complexes projected over most of the expected width of the Perseus arm.

Are there any other open clusters placed between the limits of Aur~OB2 defined by \citet{humphreys1978}, i.e., between $l=172\degr$ and $l=174\degr$? NGC~1912 ($l=172\fdg3$; $b=+0\fdg7$) seems to be an intermediate-age cluster with an age around $\log t=8.5$ \citep{subramaniam1999,pandey2007}. The nearby NGC~1907 ($l=172\fdg6$; $b=+0\fdg3$) has a similar age, and a distance of $\sim1.8$~kpc \citep{subramaniam1999,pandey2007}. The distance to NGC~1912 is similar or slightly shorter. In any case, none of these two clusters is a spiral-arm tracer. We do not find any young cluster in addition to those already discussed.

Apart from the stars observed in this paper, there are a few objects in the area with spectroscopic distances compatible with our value for Stock~8, among them, LY~Aur (mentioned above, O9\,II+O9.5\,III), HD~36212 (B2.5\,II) or HD~36280 (B0.5\,IVn), as well as a moderate number of stars from the LS catalog that could also be associated. However, the high concentration of OB stars studied here does not extend significantly beyond the limits of the area that we have studied. The few cataloged OB stars lying between our area of study and NGC~1893 are likely foreground objects. This spatial concentration gives strong support to the association of the diffuse population with the three clusters studied in this paper. Therefore, the area studied in this paper seems to be a closed group, with an age around 5~Ma and only residual present-day star formation. 

\section{Conclusions}

These are the main conclusions of our work:

\begin{enumerate}

\item We have obtained a distance of $2.80^{+0.27}_{-0.24}$~kpc for the open cluster Stock~8 using accurate photometry and spectroscopy  

\item We have found two new open clusters located $\sim20\arcmin$ north of Stock~8, named Alicante~11 and Alicante~12, which are placed at the same distance that Stock~8. 

\item We have calculated spectroscopic distances for 14 other early-type stars in the area, finding that all of them could be located at the same distance as the 3 open clusters. 

\item We have estimated an age for Stock~8 between 4~Ma and 6~Ma using pre-main sequences in the $K_{\textrm{S}}/(J-K_{\textrm{S}})$ diagram. There is no evidence for any of the other clusters or the diffuse population around them to have a significantly different age.

\item We have detected a pre-main sequence population associated with the likely B-type members lying on the main sequence. From analysis of WISE and 2MASS data, we find that many of them present disks, also showing an excess in the $(J-K_{{\textrm S}})$ colour. 

\item Star LS\,V $+34\degr$23 (HD~35633), located  NW of Stock~8, has spectral type O8\,II(f), and is likely the main source of ionization in the nebula in the cluster region.

\item The picture that emerges is that of an area of recent star formation, where the \ion{H}{ii} region Sh~2-234 represents the last remnant of the cloud, and there is only some residual ongoing star formation, concentrated towards the Nebulous Stream and the small obscured cluster CC~14. 

\item The classical picture of Aur~OB2, as a concentration of OB stars extending between Stock~8 and NGC~1893 cannot be held, as NGC~1893 is clearly more distant. Stock~8 and the surrounding association are likely located on the Perseus arm, as defined by \citet{choi2014}. Other nearby \ion{H}{ii} regions, such as Sh~2-237 or the Sh~2-231\,--\,235 complex are also likely located on the Perseus arm, but their differing radial velocities and distances do not support a close connection with Stock~8 and the surrounding association.

\end{enumerate}

\section*{Acknowledgements}
We thank the referee for the careful reading of the manuscript.

This research is partially supported by the Spanish Ministerio de
Ciencia e Innovaci\'on (MICINN) under
grants AYA2012-39364-C02-02 and AYA2015-68012-C2-2-P, and by the Generalitat Valenciana (ACOMP/2014/129). AM acknowledges support from the Generalitat Valenciana through the grant BEST/2015/242 and from the Ministerio de Educaci\'on, Cultura y Deporte through the grant PRX15/00030.

This work is based in part on observations obtained with the Jacobus Kapteyn
Telescope operated on the island of La Palma by the Isaac
Newton Group, in the Spanish Observatorio Roque de los Muchachos of the Instituto de Astrof\'{\i}sica de Canarias. 

This work is based in part on observations made with the Italian Telescopio Nazionale Galileo (TNG) operated on the island of La Palma by the Fundaci\'on Galileo Galilei of the INAF (Istituto Nazionale di Astrofisica) at the
Spanish Observatorio del Roque de los Muchachos of the Instituto de Astrof\'{\i}sica de Canarias.

Based on observations made with the Nordic Optical Telescope, operated
on the island of La Palma jointly by Denmark, Finland, Iceland,
Norway, and Sweden, in the Spanish Observatorio del Roque de los
Muchachos of the Instituto de Astrof\'{\i}sica de Canarias.  

Some of the data presented here have been taken using ALFOSC, which is owned by the Instituto de Astrofisica de Andalucia (IAA) and operated at the Nordic Optical Telescope under agreement between IAA and the NBIfAFG of the Astronomical Observatory of Copenhagen.

Based in part on observations made at Observatoire de Haute Provence (CNRS), France.

This research has made use of the Simbad database, operated at CDS,
Strasbourg (France) and of the WEBDA database, operated at the
Institute for Astronomy of the University of Vienna. This publication
makes use of data products from 
the Two Micron All Sky Survey, which is a joint project of the University of
Massachusetts and the Infrared Processing and Analysis
Center/California Institute of Technology, funded by the National
Aeronautics and Space Administration and the National Science
Foundation.







\newpage
\section{Photometric Tables}

\begin{table*}
\centering
\caption{Coordinates in J2000 for stars with optical photometry in the cluster Stock~8. The second column indicates its number in Table~\ref{Near_IR}, where we present our TNG photometry. The 2MASS photometry values are given for stars with no measurement in Table~\ref{Near_IR}. Material on-line\label{coorSTOCK8}}
\begin{tabular}{llllrrrrrr}
\hline
 Name & Number&RA(J2000)& DEC(J2000)&$J$&$\sigma_{J}$&$H$&$\sigma_{H}$&$K_{S}$&$\sigma_{K_{S}}$\\
\hline\hline
ST1   &		& 05:27:51.82  &   +34:30:48.1&9.075  & 0.027 &  8.441 &  0.020 &  8.219  & 0.018 \\
ST2   &		& 05:28:06.39  &   +34:30:40.3&13.820 & 0.024 &  13.501&  0.027 &  13.419 & 0.036 \\		
ST3   & 	& 05:27:49.91  &   +34:30:43.7&12.767 & 0.019 &  12.592&  0.019 &  12.492 & 0.023 \\
ST4   &  	& 05:28:15.51  &   +34:30:36.3&14.248 & 0.026 &  13.876&  0.036 &  13.777 & 0.041 \\
ST5   &  	& 05:28:04.60  &   +34:30:36.6&14.359 & 0.023 &  13.979&  0.031 &  13.743 & 0.040 \\
ST6   &  	& 05:28:09.92  &   +34:30:32.5&13.820 & 0.023 &  13.210&  0.025 &  13.057 & 0.029 \\
\hline
\end{tabular}
\end{table*}

\begin{table*}
\centering
\caption{Near-IR photometry obtained by us for stars in the cluster Stock~8. Material on-line\label{Near_IR}}
\begin{tabular}{cccrrrrrr}
\hline
Name&R.A(J2000)&DEC(J2000)&$J$&$\sigma_{J}$&$H$&$\sigma_{H}$&$K_{S}$&$\sigma_{K_{S}}$\\
\hline\hline
1&05:28:00.83&+34:22:44.7  &17.168&0.013&16.393&0.016&15.986&0.013\\
2&05:28:15.89&+34:22:47.2  &15.499&0.012&14.513&0.010&14.261&0.010\\
3&05:28:03.07&+34:22:45.5  &14.585&0.010&14.168&0.009&14.049&0.009\\
4&05:28:13.87&+34:22:49.9  &14.910&0.011&14.118&0.009&13.893&0.010\\
5&05:28:13.76&+34:22:50.4  &15.220&0.013&14.380&0.011&14.131&0.011\\
\hline
\end{tabular}
\end{table*}

\begin{table*}
\centering
\caption{Optical photometry for stars in the open cluster Stock~8. The values with label $^{*}$ are not considered in the analysis because their photometric errors are around $0.1$. Material on-line\label{strom-stock8}}
\begin{tabular}{lrlrlrllrllrll}
\hline
Name&$V$&$\sigma_{V}$&$(b-y)$&$\sigma_{(b-y)}$&$c_{1}$&$\sigma_{c_{1}}$&$N$&$m_{1}$&$\sigma_{m_{1}}$&$N$&$\beta$&$\sigma_{\beta}$&$N$\\
\hline\hline
ST1  &11.435&0.021&0.964&0.027&$-$0.250&0.040   &1&0.574	&  0.039   &1&2.597&  0.032  &1\\
ST2  &15.131&0.004&0.525&0.010&$-$0.135&0.036&3&0.252	&0.060&3&2.616&0.052&2\\
ST3  &13.568&0.034&0.353&0.008&0.367   &0.037&6&0.347	&0.025&5&2.803&0.029&5\\
ST4  &15.859&0.008&0.629&0.038&0.124   &0.111$^*$&2&0.134	&0.097$^*$&2&2.603&0.040&2\\
ST5  &15.947&0.045&0.678&0.009&$-$0.046 &0.119$^*$&2&0.179	&0.075&2&2.597&0.061&2\\
ST6  &15.864&0.002&0.762&0.013&0.031   &0.200$^*$&2&0.227	&0.078&2&2.578&0.064&2\\
\hline
\end{tabular}
\end{table*}

\begin{table*}
\centering
\caption{Coordinates in J2000 and 2MASS photometry for stars with optical photometry in the new cluster Alicante~11. Material on-line\label{coor-Alicante11}}
\begin{tabular}{lllrrrrrr}
\hline
Name & RA(J2000)& DEC(J2000)&$J$&$\sigma_{J}$&$H$&$\sigma_{H}$&$K_{S}$&$\sigma_{K_{S}}$\\
\hline\hline
Alicante~11-1   &	 05:27:55.48& +34:50:21.3&     10.858 & 0.018 &  10.234 & 0.016 &  10.062 & 0.017\\
Alicante~11-2   &	 05:27:57.72& +34:49:25.5&     12.051 & 0.019 &  11.788 & 0.018 &  11.683 & 0.020\\
Alicante~11-3   &	 05:27:24.26& +34:49:26.0&     11.243 & 0.022 &  11.113 & 0.027 &  11.003 & 0.019\\
Alicante~11-4   &	 05:27:27.63& +34:48:53.7&     12.208 & 0.023 &  11.886 & 0.030 &  11.812 & 0.023\\
Alicante~11-5   &	 05:27:54.47& +34:48:43.7&     12.982 & 0.021 &  12.815 & 0.022 &  12.749 & 0.023\\
\hline
\end{tabular}
\end{table*}

\begin{table*}
\centering
\caption{Optical photometry for stars in the new open cluster Alicante~11. The values with label $^*$ are not considered in the analysis because of their error are around $0.1$. Material on-line\label{strom-Alicante11}}
\begin{tabular}{lllrlrlrlrll}
\hline
Name&$V$&$\sigma_{V}$&$(b-y)$&$\sigma_{(b-y)}$&$m_{1}$&$\sigma_{m_{1}}$&$c_{1}$&$\sigma_{c_{1}}$&$\beta$&$\sigma_{\beta}$&$N$\\
\hline\hline
Alicante~11-1 &13.225& 0.029    &0.863&	0.047  &0.308   &0.071	 &0.292   &0.066	&2.689& 0.045    &1\\
Alicante~11-2 &13.334&0.028&0.425&0.019&0.042   &0.030&0.899   &0.030&2.840&0.023&3\\
Alicante~11-3 &12.201&0.007&0.313&0.017&0.074   &0.026&0.476   &0.018&2.790&0.022&5\\
Alicante~11-4 &13.636&0.012&0.470&0.005&0.115   &0.016&0.444   &0.025&2.712&0.007&5\\
Alicante~11-5 &13.860&0.023&0.327&0.035&$-$0.007&0.050&0.909   &0.055&2.821&0.049&3\\
\hline
\end{tabular}
\end{table*}

\begin{table*}
\centering
\caption{Coordinates in J2000 and 2MASS photometry for stars with optical photometry in the new cluster Alicante~12. Material on-line\label{coor-Alicante12}}
\begin{tabular}{lllrrrrrr}
\hline
Name & RA(J2000)& DEC(J2000)&$J$&$\sigma_{J}$&$H$&$\sigma_{H}$&$K_{S}$&$\sigma_{K_{S}}$\\
\hline\hline
Alicante~12-1   &	  05:28:17.33&     +34:51:57.0&13.027& 0.021&  12.689& 0.019&  12.590 &0.025  \\      
Alicante~12-2   &	  05:28:15.50&     +34:51:51.4&12.557& 0.019&  12.198& 0.018&  12.098 &0.023  \\
Alicante~12-3   &	  05:28:05.04&     +34:51:41.2&11.829& 0.019&  11.672& 0.019&  11.562 &0.022  \\
Alicante~12-4   &	  05:28:13.83&     +34:51:24.1&11.504& 0.019&  11.173& 0.018&  11.108 &0.019  \\	       
Alicante~12-5   &	  05:28:28.12&     +34:50:34.6&11.743& 0.018&  11.551& 0.016&  11.466 &0.019  \\      
\hline
\end{tabular}
\end{table*}

\begin{table*}
\centering
\caption{Optical photometry for stars in the open cluster Alicante~12. The values with label $^*$ are not considered in the analysis because of their error are around $0.1$. Material on-line\label{strom-Alicante12}}
\begin{tabular}{lllrlrlrlrll}
\hline
Name&$V$&$\sigma_{V}$&$(b-y)$&$\sigma_{(b-y)}$&$m_{1}$&$\sigma_{m_{1}}$&$c_{1}$&$\sigma_{c_{1}}$&$\beta$&$\sigma_{\beta}$&$N$\\
\hline\hline
Alicante~12-1& 14.580&0.048&0.586&0.050&0.051   &0.093$^*$&0.423   &0.040&2.732&0.097$^*$&3\\
Alicante~12-2&14.250&0.019&0.607&0.011&0.026   &0.041	 &0.511  &0.008	&2.740& 0.046&2\\
Alicante~12-3 &12.730&0.043&0.357&0.034&0.010   &0.010&0.221   &0.037&2.743&0.073&3\\
Alicante~12-4 &12.855&0.036&0.486&0.032&0.106   &0.066&0.520   &0.047&2.716&0.057&3\\
Alicante~12-5& 12.868&0.027&0.367&0.037&0.079   &0.055&0.920   &0.016&2.835&0.038&3\\
\hline
\end{tabular}
\end{table*}


\bsp	
\label{lastpage}

\begin{thebibliography}{99}
\bibitem[\protect\citeauthoryear{Alves}{2000}]{alves2000} Alves, D.~R.\ 2000, \apj, 539, 732 
\bibitem[\protect\citeauthoryear{Bik et al.}{2012}]{bik12}
Bik, A., Henning, Th., Stolte, A., 
et al. 2012, ApJ, 744, 87
\bibitem[\protect\citeauthoryear{Blaauw}{1964}]{blaauw1964} Blaauw,  A. 1964, ARA\&A, 2, 213
\bibitem[\protect\citeauthoryear{Blitz et al.}{1982}]{blitz1982} Blitz, L., Fich, M., \& Stark, A.A. 1982, \apjs, 49, 183
\bibitem[\protect\citeauthoryear{Bobylev \& Bajkova}{2010}]{bobylev2010} Bobylev, V.V., \& Bajkova, A.T. 2010, MNRAS 408, 1788
\bibitem[\protect\citeauthoryear{Cabrera-Lavers et al.}{2005}]{cabrera2005} Cabrera-Lavers, A., Garz{\'o}n, F., \& Hammersley, P.~L.\ 2005, \aap, 433, 173 
\bibitem[\protect\citeauthoryear{Chen}{2005}]{chen2005} Chen, Y., Yao, Y., Yang, J., Zeng, Q., \& Sato, S. 2005, ApJ, 629, 288
\bibitem[\protect\citeauthoryear{Choi et al.}{2014}]{choi2014} Choi, Y.K., Hachisuka, K., Reid, M.J., 
et al. 2014, ApJ, 790, 99
\bibitem[\protect\citeauthoryear{Churchwell et al.}{2009}]{churchwell2009} Churchwell, E., Babler, B. L., Meade, M. R., et al. 2009,PASP,121, 213
\bibitem[\protect\citeauthoryear{Costa}{2015}]{costa2015} Costa, E., Moitinho, A., Radiscz, M., et al. 2015, A\&A, 580, A4
\bibitem[\protect\citeauthoryear{Crawford \& Mander}{1966}]{crawford1966} Crawford, D. L., \& Mander, J. 1966, AJ, 71, 114
\bibitem[\protect\citeauthoryear{Crawford \& Barnes}{1970a}]{crawford1970a}Crawford, D. L., \& Barnes, J. V. 1970a, AJ, 75, 978
\bibitem[\protect\citeauthoryear{Crawford et al.}{1970b}]{crawford1970b} Crawford, D. L., Glaspey, J. W., \& Perry, C. L. 1970b, AJ, 75, 822
\bibitem[\protect\citeauthoryear{Crawford}{1975}]{crawford1975} Crawford, D. L. 1975, PASP, 87, 481
\bibitem[\protect\citeauthoryear{Draper et al.}{2000}]{Draper2000} Draper, P.W., Taylor, M., \& Allan, A. 2000, Starlink User Note 139.12, R.A.L.
\bibitem[\protect\citeauthoryear{Drechsel et al.}{1994}]{dre94} Drechsel, H., Haas, S., Lorenz, R., \& Mayer, P. 1994, A\&A, 284, 853
\bibitem[\protect\citeauthoryear{Evans \& Blair}{1981}]{evans1981} Evans, N. J., II, \& Blair, G. N. 1981, ApJ, 246, 394
\bibitem[\protect\citeauthoryear{Fich et al.}{1990}]{fich1990} Fich, M., Treffers, R.R., \& Dahl, G.P. 1990, AJ, 99, 622
\bibitem[\protect\citeauthoryear{Fitzgerald}{1970}]{fitzgerald1970} FitzGerald, M.P. 1970, A\&A, 4, 234
\bibitem[\protect\citeauthoryear{Fitzsimmons}{1993}]{fitzsimmons1993} Fitzsimmons, A. 1993, A\&AS, 99, 15
\bibitem[\protect\citeauthoryear{Foster \& Brunt}{2015}]{foster15} Foster, T., \& Brunt, C.M. 2015, AJ, 150, 147
\bibitem[\protect\citeauthoryear{France et al.}{2004}]{france04} France, K., McCandliss, S.R., Burgh, E.B., \& Feldman, P.D. 2004, ApJ, 616, 257
\bibitem[\protect\citeauthoryear{Froebrich et al.}{2007}]{froebrich2007} Froebrich, D., Scholz, A., \& Raftery, C. L. 2007, MNRAS, 374, 399
\bibitem[\protect\citeauthoryear{Garmany \& Conti}{1984}]{garmany84} Garmany, C. D., \& Conti, P. S. 1984, ApJ, 284, 705
\bibitem[\protect\citeauthoryear{Gillet et al.}{1994}]{gil94}Gillet, D., Burnage, R., Kohler, D., et al. 1994, A\&AS, 108, 181
\bibitem[\protect\citeauthoryear{Gonz{\'a}lez-Fern{\'a}ndez et al.}{2015}]{gonzalez2015} Gonz{\'a}lez-Fern{\'a}ndez, C., Dorda, R., Negueruela, I., \& Marco, A.\ 2015, \aap, 578, A3 
\bibitem[\protect\citeauthoryear{Hardorp et al.}{1965}]{hardorp1965} Hardorp, J., Theile, I., \& Voigt, H.H. 1965, Hamburger Sternw., Warner \& Swasey Obs., 5
\bibitem[\protect\citeauthoryear{Hiltner}{1956}]{hiltner1956} Hiltner, W. A. 1956, ApJS, 2, 389
\bibitem[\protect\citeauthoryear{Hindson et al.}{2013}]{hindson13} Hindson, L., Thompson, M. A., Urquhart, J.S., 
et al. 2013, MNRAS, 435, 2003
\bibitem[\protect\citeauthoryear{Honma et al.}{2012}]{honma2012} Honma, M., Nagayama, T., Ando, K., et al. 2012, PASJ, 64, 136
\bibitem[\protect\citeauthoryear{Howarth et al.}{1998}]{Howarth1998} Howarth, I., Murray, J., Mills, D., \& Berry, D.S. 1998, Starlink User Note 50.21, R.A.L.
\bibitem[\protect\citeauthoryear{Humphreys}{1978}]{humphreys1978} Humphreys, R. M. 1978, ApJS, 38, 309
\bibitem[\protect\citeauthoryear{Ivanov et al.}{2005}]{ivanov05} Ivanov, V. D., Borissova, J., Bresolin, F., \& Pessev, P. 2005, A\&A, 435, 107 
\bibitem[\protect\citeauthoryear{Jarrett et al.}{2011}]{jarret2011} Jarrett, T. H., Cohen, M., Masci, F., et al. 2011, ApJ, 735, 112
\bibitem[\protect\citeauthoryear{Johnson \& Morgan}{1953}]{johnson1953} Johnson, H.L., \& Morgan, W.W. 1953, ApJ, 117, 313
\bibitem[\protect\citeauthoryear{Jose et al.}{2008}]{jose2008} Jose, J., Pandey, A.K., Ojha, D.K., 
et al. 2008, MNRAS, 384, 1675
\bibitem[\protect\citeauthoryear{Kang et al.}{2012}]{kang2012} Kang, J.H., Koo, B.C., \& Salter, C. 2012, AJ, 143, 75
\bibitem[\protect\citeauthoryear{Kiminki et al.}{2015}]{kiminki15} Kiminki, M.M., Kim, J.S., Bagley, M.B., Sherry, W.H., \& Rieke, G.H. 2015, ApJ, 813, 42
\bibitem[\protect\citeauthoryear{Koenig et al.}{2012}]{koenig2012} Koenig, X. P., Leisawitz, D.T., Benford, D.J.,  
et al. 2012, ApJ 744, 130
\bibitem[\protect\citeauthoryear{Koenig \& Leisawitz}{2014}]{koenig2014} Koenig, X. P., Leisawitz, D. T. 2014, ApJ, 791, 131
\bibitem[\protect\citeauthoryear{Kronberger et al.}{2006}]{kronberger2006} Kronberger, M., Teutsch, P., Alessi, B., 
et al. 2006, A\&A, 447, 921
et al. 1971, GCVS3 
\bibitem[\protect\citeauthoryear{Lim et al.}{2014}]{lim2014} Lim, B., Sung, H., Kim, J.S., Bessel, M.S., \& Park, B.-G. 2014, MNRAS, 443, 454
\bibitem[\protect\citeauthoryear{Lim et al.}{2015}]{lim2015} Lim, B., Sung, H., Bessel, M.S., et al. 2015, AJ, 149, 127
\bibitem[\protect\citeauthoryear{McMillan \& Binney}{2010}]{mcmillan2010} McMillan, P.J., \& Binney, J.J. 2010, MNRAS, 402, 934
\bibitem[\protect\citeauthoryear{Malysheva}{1990}]{malysheva1990} Malysheva, L. K. 1990, Soviet Astronomy, 34, 122
\bibitem[\protect\citeauthoryear{Manfroid}{1993}]{manfroid1993} Manfroid, J. 1993, A\&A, 271, 714
\bibitem[\protect\citeauthoryear{Marco et al.}{2001}]{marco2001} Marco A., Bernabeu, G., \& Negueruela I. 2001, AJ, 121, 2075
\bibitem[\protect\citeauthoryear{Marco \& Negueruela}{2002}]{marco2002} Marco A., \& Negueruela I. 2002, A\&A, 393, 195
\bibitem[\protect\citeauthoryear{Marco \& Negueruela}{2013}]{marco2013} Marco A., \& Negueruela I. 2013, A\&A, 552, 92
\bibitem[\protect\citeauthoryear{Massey et al.}{1995}]{massey1995} Massey, P., Johnson, K.E., \& DeGioia-Eastwood, K. 1995, ApJ, 454, 151
\bibitem[\protect\citeauthoryear{Mayer}{1964}]{mayer1964} Mayer, P. 1964, Acta Univ. Carol. Math. Phys., 1, 25M
\bibitem[\protect\citeauthoryear{Mayer \& Macak}{1971}]{mm71} Mayer, P., \& Macak, P. 1971, Bull. Astron. Inst. Czech., 22, 46
\bibitem[\protect\citeauthoryear{Mayer}{2013}]{mayer2013} Mayer, P., Drechsel, H., Harmanec, P., Yang, S., \& Slechta, M. 2013, A\&A, 559, A22
\bibitem[\protect\citeauthoryear{Messineo et al.}{2012}]{messineo2012} Messineo, M., Menten, K.~M., Churchwell, E., \& Habing, H.\ 2012, \aap, 537, A10
\bibitem[\protect\citeauthoryear{Messineo et al.}{2015}]{messineo15} Messineo, M., Clark, J.S., Figer, 
et al. 2015, ApJ, 805, 110
\bibitem[\protect\citeauthoryear{Morgan et al.}{1953}]{morgan53} Morgan, W.W., Whitford, A.E., \& Code, A.D. 1953, ApJ, 118, 318
\bibitem[\protect\citeauthoryear{Morgan et al.}{1955}]{morgan55} Morgan, W.W., Code, A.D., \& Whitford, A.E. 1955, ApJS, 2, 41
\bibitem[\protect\citeauthoryear{Negueruela \& Marco}{2003}]{negueruela2003} Negueruela I., \& Marco A. 2003, A\&A, 406, 119
\bibitem[\protect\citeauthoryear{Negueruela et al.}{2007}]{negueruela2007} Negueruela I., Marco A., Israel, G.L., \& Bernabeu G. 2007, A\&A, 471, 485
\bibitem[\protect\citeauthoryear{Netopil et al.}{2012}]{netopil2012} Netopil, M.,  Paunzen, E., \&  Stütz, Christian., 2012, ASSP, 29, 53
\bibitem[\protect\citeauthoryear{Pandey et al.}{2007}]{pandey2007} Pandey, A.K., Sharma, S., Upadhyay, K., 
et al. 2007, PASJ, 59, 547
\bibitem[\protect\citeauthoryear{Pandey et al.}{2013}]{pandey2013} Pandey, A.K., Eswaraiah, C., Sharma, S.,et al. 2013, ApJ, 764, 172
\bibitem[\protect\citeauthoryear{Perry et al.}{1978}]{perry1978}Perry, C. L., Lee, P. D., \& Barnes, J. V. 1978, PASP, 90, 73
\bibitem[\protect\citeauthoryear{Perry et al.}{1987}]{perry1987} Perry, C. L., Olsen, E. H., \& Crawford, D. L. 1987, PASP, 99, 1184
\bibitem[\protect\citeauthoryear{Preibisch \& Zinnecker}{2007}]{preibisch2007} Preibisch, T., \& Zinnecker, H. 2007, Triggered Star Formation in a Turbulent ISM, Edited by B. G. Elmegreen and J. Palous. Proceedings of the International Astronomical Union 2, IAU Symposium 237, held 14-18 August, 2006 in Prague, Czech Republic. Cambridge: Cambridge University Press, 2007., pp.270-277
\bibitem[\protect\citeauthoryear{Preibisch et al.}{2011}]{preibisch11} Preibisch, T., Ratzka, T., Kuderna, B., 
et al. 2011, A\&A, 530, 34
\bibitem[\protect\citeauthoryear{Prisinzano et al.}{2011}]{prisinzano2011} Prisinzano, L., Sanz-Forcada, J., Micela, G., et al. 2011, A\&A, 527, A77
\bibitem[\protect\citeauthoryear{Reid et al.}{2014}]{reid2014} Reid, M.J., Menten, K.M., Brunthaler, A., 
et al. 2014, ApJ, 783, 130
\bibitem[\protect\citeauthoryear{Rom\'an-Z\'u\~niga et al.}{2015}]{roman15} Rom\'an-Z\'u\~niga, C.G., Ybarra, J.E., Meg\'{i}as, G.D., 
et al. 2015, AJ, 150, 80
\bibitem[\protect\citeauthoryear{Sakai et al.}{2012}]{sakai2012} Sakai, N., Honma, M., Nakanishi, H., et al. 2012, PASJ, 64, 108
\bibitem[\protect\citeauthoryear{Sharma et al.}{2007}]{sharma2007} Sharma, S., Pandey, A.K., Ojha, D.K., 
et al. 2007, MNRAS, 380, 1141 
\bibitem[\protect\citeauthoryear{Shortridge et al.}{1997}]{Shortridge1997} Shortridge, K., Meyerdicks, H., Currie, M., et al. 1997, Starlink User Note 86.15, R.A.L.
\bibitem[\protect\citeauthoryear{Siess et al.}{2000}]{siess2000} Siess L., Dufour E., \& Forestini M. 2000, A\&A, 358, 593
\bibitem[\protect\citeauthoryear{Sim\'on-D\'{i}az \& Herrero}{2014}] {sergio14} Sim\'on-D\'{i}az, S. \& Herrero, A. 2014, A\&A, 562, 135
\bibitem[\protect\citeauthoryear{Skrutskie et al.}{2006}]{Skrutskie2006} Skrutskie, M.F., Cutri, R.M., Stiening, R., et al. 2006, AJ, 131, 1163
\bibitem[\protect\citeauthoryear{Sota et al.}{2011}]{sota11} Sota, A., Ma\'{i}z Apell\'aniz, J., Walborn, N.R., 
et al. 2011, ApJS, 193, 24
\bibitem[\protect\citeauthoryear{Sota et al.}{2014}]{sota14} Sota, A., Ma\'{i}z Apell\'aniz, J., Morrell, N.I., 
et al. 2014, ApJS, 211, 10
\bibitem[\protect\citeauthoryear{Stetson}{1987}]{stetson1987} Stetson, P. B. 1987, PASP, 99, 191
\bibitem[\protect\citeauthoryear{Strai{\v z}ys et al.}{2010}]{straizys2010} Strai{\v z}ys, 
V., Drew, J.~E., \& Laugalys, V.\ 2010, Baltic Astronomy, 19, 169 
\bibitem[\protect\citeauthoryear{Subramaniam \& Sagar}{1999}]{subramaniam1999} Subramaniam A., \& Sagar R. 1999, AJ, 117, 937 
\bibitem[\protect\citeauthoryear{Turner}{1980}]{turner1980} Turner, D. 1980, ApJ, 240, 137
\bibitem[\protect\citeauthoryear{Vall\'ee}{2015}]{vallee2015} Vall\'ee, J.P. 2015, MNRAS, 450, 4277
\bibitem[\protect\citeauthoryear{Walborn}{1971}]{walborn1971} Walborn, N.R. 1971, ApJS, 23, 257
\bibitem[\protect\citeauthoryear{Walborn \& Fitzpatrick}{1990}]{walborn1990} Walborn, N.~R., \& Fitzpatrick, E.~L.\ 1990, \pasp, 102, 379 
\bibitem[\protect\citeauthoryear{V\'azquez et al.}{2008}]{vazquez2008} V\'azquez, R., May, J., Carraro, G., et al. 2008, ApJ, 672, 930
\bibitem[\protect\citeauthoryear{Wouterloot \& Brand}{1989}]{wouterloot1989} Wouterloot, J.G.A., \&  Brand, J. 1989, A\&AS, 80, 149
\bibitem[\protect\citeauthoryear{Wright et al.}{2010}]{wright2010} Wright, E.L., Eisenhardt, P.R.M., \& Mainzer, A.K. et al. 2010, AJ, 140, 1868
\bibitem[\protect\citeauthoryear{Wright et al.}{2014}]{wright14} Wright, N.J., Parker, R.J., Goodwin, S.P., \& Drake, J.J. 2014, MNRAS, 438, 639
\bibitem[\protect\citeauthoryear{Xu et al.}{2013}]{xu2013} Xu, Y., Li, J. J., Reid, M. J. et al. 2013, ApJ, 769, 15

\end{thebibliography}
\end{document}